\documentclass{aastex63}
\usepackage{subfigure}

\newcommand\ha{H$\alpha$}
\newcommand\pb{Pa$\beta$}

\newcommand{\chisq}{$\chi^2_{red.}$}
\newcommand{\ebv}{$\rm E(B-V)$}

\received{\today}
\revised{}
\accepted{}
\submitjournal{ApJ}

\shorttitle{Looking for obscured YSCs in NCG 1313}
\shortauthors{Messa et al.}
\graphicspath{{./}{figures/}}

\begin{document}

\title{Looking for obscured young star clusters in NCG 1313}

\correspondingauthor{Matteo Messa}\email{matteo.messa@unige.ch}

\author[0000-0003-1427-2456]{Matteo Messa}
\affiliation{Department of Astronomy, University of Massachusetts Amherst, 
710 North Pleasant Street, Amherst, MA 01003, USA}

\author[0000-0002-5189-8004]{Daniela Calzetti}
\affiliation{Department of Astronomy, University of Massachusetts Amherst, 
710 North Pleasant Street, Amherst, MA 01003, USA}

\author[0000-0002-8192-8091]{Angela Adamo}
\affiliation{Department of Astronomy, Oscar Klein Centre, Stockholm University,  AlbaNova, Stockholm SE-106 91, Sweden}

\author[0000-0002-3247-5321]{Kathryn Grasha}
\affiliation{Research School of Astronomy and Astrophysics, 
Australian National University, 
Weston Creek, ACT 2611, Australia}
\affiliation{ARC Centre of Excellence for All Sky Astrophysics in 3 Dimensions (ASTRO 3D), 
Australia}

\author[0000-0001-8348-2671]{Kelsey E. Johnson}
\affiliation{Department of Astronomy, University of Virginia, Charlottesville, VA 22904-4325, USA}

\author[0000-0003-2954-7643]{Elena Sabbi}
\affiliation{Space Telescope Science Institute, 3700 San Martin Drive, Baltimore, MD 2121, USA}

\author[0000-0002-0806-168X]{Linda J. Smith}
\affiliation{Space Telescope Science Institute and European Space Agency, 3700 San Martin Drive, Baltimore, MD 21218}

\author{Varun Bajaj}
\affiliation{Space Telescope Science Institute, 3700 San Martin Drive, Baltimore, MD 2121, USA}

\author[0000-0001-9338-2594]{Molly K. Finn}
\affiliation{Department of Astronomy, University of Virginia, Charlottesville, VA 22904-4325, USA}

\author[0000-0001-8078-3428]{Zesen Lin}
\affiliation{Key Laboratory for Research in Galaxies and Cosmology, Department of Astronomy, University of Science and Technology of China,
Hefei 230026, China}




\begin{abstract}
Using recently acquired HST NIR observations (J, \pb\ and H bands) of the nearby galaxy NGC1313, we investigate the timescales required by a young star cluster to emerge from its natal cloud. We search for extincted star clusters, potentially embedded in their natal cloud as either: 1. compact sources in regions with high \ha/\pb\ extinctions; 2. compact HII regions that appear as point-like sources in the \pb\ emission map.
The NUV--optical--NIR photometry of the candidate clusters is used to derive their ages, masses and extinctions via a least$-\chi^2$ SED broad and narrow--band fitting process. 
The 100 clusters in the final samples have masses in the range $\rm \log_{10}(M/M_\odot)=2.5-3.5$ and moderate extinctions, $\rm E(B-V)\lesssim1.0$ mag. 
Focusing on the young clusters ($0-6$ Myr) we derive a weak correlation between extinction and age of the clusters. Almost half of the clusters have low extinctions, $\rm E(B-V)<0.25$ mag, already at very young ages ($\le3$ Myr), suggesting that dust is quickly removed from clusters. A stronger correlation is found between the morphology of the nebular emission (compact, partial or absent, both in \ha\ and \pb) and cluster age. 
Relative fractions of clusters associated with a specific nebular morphology is used to estimate the typical timescales for clearing the natal gas cloud, resulting between 3 and 5 Myr, $\sim1$ Myr older than what estimated from NUV--optical--based cluster studies. 
This difference hints to a bias for optically--only based studies, which JWST will address in the coming years.
\end{abstract}

\keywords{Young star clusters --- Compact H II region --- Star formation --- Interstellar dust extinction --- Spiral galaxies}


\section{Introduction} \label{sec:intro}
Recent studies comparing the spatial distribution within galaxies of giant molecular clouds (GMCs) and HII regions suggest that GMCs spend most of their lifetime in an inert state. From the moment they begin to host star formation, their dense cores are quickly dissolved, on timescales of a few million years \citep{kobulnicky1999,matthews2018,kruijssen2019}. Such short timescales, suggesting very efficient feedback from young stars, have been found with the same technique in various nearby galaxies and at different galactocentric distances \citep{kruijssen2019,chevance2020a}. According to these results, the star formation process is rapid and inefficient because feedback acts on very short timescales, resulting in only a small fraction of the gas being converted into stars \citep[e.g.][]{matthews2018,chevance2020b}.

Short feedback timescales have been typically derived in recent years by comparing tracers of different stages of the SF process. \citet{corbelli2017} studied GMCs, embedded clusters (in the mid-infrared, MIR, where the extinction effect is lower) and exposed clusters (in \ha\ and UV bands) in the nearby galaxy M33, finding that young clusters remain in an embedded phase only for $\sim2$ Myr on average. 
Similarly, \citet{grasha2018,grasha2019} found median ages of $2-4$ Myr for clusters associated to GMCs in NGC7793 and NGC5194. This also suggests that star clusters are able to emerge from their natal clouds long before the timescale required for clouds to disperse.

Star clusters play a fundamental role in the SF process, as most of the SF takes place in clustered environments \citep{lada2003} and typically $\sim5-20\%$ of the stars formed will evolve in bound systems (e.g. \citealp{bastian2012,adamo2015,chandar2015,johnson2016,messa2018a,messa2018b} and the review by \citealp{krumholz2019}). 
Clustered star formation hosts most of the massive stars, responsible for the feedback regulating SF processes. The very  early stages of  cluster evolution can thus reveal, via the time needed to clear the gas and dust cloud around them, which feedback mechanism is mostly responsible for stopping star formation and therefore determining its efficiency. \citet{whitmore2011} and \citet{hollyhead2015} studied the morphologies of \ha\ emission associated to young clusters in the nearby galaxy M83, revealing that clusters are no longer embedded in their natal gas clouds by ages $<4$ Myr, in agreement with the aforementioned studies of GMCs-clusters association. \citet{hannon2019} measured a progression of the median ages when going from concentrated HII regions (median age $\sim3$ Myr), to partially exposed ($\sim4$ Myr) to no \ha\ emission ($>5$ Myr). The same authors suggest that the typical timescale for gas clearing can be very short, on the order of $\sim2$ Myr. 

Almost all extragalactic young cluster studies are based on NUV-optical bands and therefore could be biased against extincted clusters (with extinctions higher than $\rm A_V\gtrsim1.0$ mag). 
Embedded clusters are likely extincted in the optical bands and can be revealed with longer wavelength emission. Radio observations can be used to directly detect the free-free emission from the gas ionized by newly formed massive stars \citep[e.g.][]{kobulnicky1999,johnson2001,johnson2004,johnson2009,johnson2003,turner2004,aversa2011,kepley2014}.
The low detection rate of compact thermal radio sources is consistent with this embedded phase being short-lived \citep[e.g.][]{tsai2009,aversa2011}. Millimeter (and sub-millimeter) observations enable us to probe the molecular material in the vicinity of young star clusters \citep[e.g.][]{finn2019,johnson2015,johnson2018}.
Infrared emission can be used to trace the warm dust cocoons surrounding young clusters \citep[e.g.][]{vacca2002,johnson2004,corbelli2017}, although the extent to which the infrared luminosities and colors can be used as diagnostics is highly dependent on the physical distribution of dust \citep[e.g.][]{whelan2011}. 
Fully constraining the characteristics of embedded clusters requires this full wavelength coverage (from the radio to the infrared).

Typical mid-infrared (MIR) resolution ($\rm PSF-FWHM\gtrsim1.9''$ for \textit{Spitzer}) allows to resolve single clusters only for galaxies at close distances $\lesssim 1$ Mpc, making it difficult to build statistical samples. Already at a few Mpc, the SF clumps observable in the MIR bands may contain several star clusters \citep[see e.g.][]{lin2020}. Studying MIR clumps on such scales ($\sim50$ pc) reveal sources with foreground extinctions up to $\rm A_V\sim15$ mag \citep{elmegreen2019} and are useful to study the collapse of gas clouds \citep{elmegreen2018,elmegreen2019,elmegreen2020}, but cannot trace the effect of feedback on the $\rm \sim pc$ scale. The main unknown of the current studies on the gas clearing timescales is the possible effect of young embedded (and therefore possibly extincted) clusters, which are consistently neglected in studies of cluster populations focused on the optical-NUV bands \citep[e.g.][]{whitmore2011,hollyhead2015,hannon2019}. With the present work we are filling this gap, by targeting the young and embedded clusters. 

One way of combining high spatial resolution with the study of embedded star clusters is to rely on the infrared bands of HST.
It is expected that newly formed clusters be surrounded by a dense compact (with radii of $\sim1$ pc) cloud of gas and dust, that makes them extincted in the optical and NUV bands.
\citet{calzetti2015b} already demonstrated how the use of near-infrared (NIR) HST band, and in particular the presence of NIR narrow-band filters, helps characterizing a highly extincted cluster (with foreground extinction $A_V\sim2$ mag and mixed attenuation from a cloud of $A_V\sim49$ mag in total), detected in the optical spectrum only in the I band, in the centre of NGC 5253. 

In this work, we extend the NIR study of embedded clusters in the nearby galaxy NGC1313, a mildly-inclined ($i=40^\circ.7$) SBd galaxy at a distance of $4.39\pm0.04$ Mpc \citep{jacobs2009}.
Similarly to the study of \citet{calzetti2015b} we will use newly acquired HST-WFC3 NIR data, namely  F110W, F128N and F160W, corresponding to J, Pa$\beta$, and H bands.
NGC1313 has a stellar mass of $ \rm 2.6\times10^9\ M_\odot$, an extinction--corrected $\rm SFR_{UV}=1.15\ M_\odot/yr$ \citep{calzetti2015_legus} and has been suggested to be currently in interaction with a satellite galaxy, that has produced a loop of HI gas around the galaxy \citep{peters1994} and a recent increase (on a timescale of 100 Myr) of the SFR in the south--west arm  \citep{silva-villa2012}. Due to both its physical and morphological properties, with the presence of a bar and a rather irregular appearance, NGC 1313 has been compared to the Large Magellanic Cloud \citep{de-vaucouleurs1963}. The star cluster population of NGC 1313 is quite numerous, with 673 clusters with ages $<300$ Myr \citep{grasha2017b} of which 195 have compact morphology, ages $\le200$ Myr and masses $\rm \ge5000\ M_\odot$ \citep{ryon2017}. The proximity of the galaxy, combined with the HST resolution allowed the study of its cluster population on physical scales $\lesssim1$ pc and \citet{ryon2017} finds a median cluster size of $2.3$ pc. Using HST narrow--band observations of NGC 1313 and of two galaxies at similar distance, \citet{hannon2019} studied the morphology of \ha\ regions associated with the star clusters, estimating the typical timescales for a compact gas nebula to disperse ($<5$ Myr). We will, in the current work, complement the \citet{hannon2019} study with our new NIR based analysis.

The paper is organized as follows: in Section~\ref{sec:data} we present the HST observations used in this study, while in Section~\ref{sec:catalogs} we describe the methodology used to extract the catalogs of the young cluster candidates. In Section~\ref{sec:analysis} and \ref{sec:results} we present the analysis performed and the relative results. We discuss the results of our analysis in Section~\ref{sec:discussion} and summarize our findings in Section~\ref{sec:summary}.

\section{Data} \label{sec:data}
Recently acquired near infrared (NIR) photometry of NGC 1313 with the WFC3-IR camera, on board HST, covers two pointings of the galaxy with two broad-band (F110W and F160W) and one narrow-band (F128N) filters (GO program 15330; PI D. Calzetti). The narrow-band filter, centered at 1.283 $\rm \mu m$, covers the wavelengths of the Paschen--$\beta$ (\pb) hydrogen recombination line emission. 
For each filter individual exposures were corrected for bias, dark and flat-field using the standard pipeline CALWF3 version 3.4.2. 
Images were aligned to the Gaia DR2 \citep{gaia2018} reference frame using TweakReg, and mosaicked together using AstroDrizzle to a pixel scale of 0.08 arcsec/pixel.
Proper motions were not applied in the alignment, as the high uncertainties of proper motions of extragalactic sources in the DR2 catalog would yield worse alignments, and poor subsampling in the mosaics.
In this study we take advantage of the broad archival HST wavelength coverage of NGC 1313; in detail we use observations in five filters from the Legacy ExtraGalactic UV Survey \citep[LEGUS,][]{calzetti2015_legus} covering the NUV-optical range with broad-band filters (F275W, F336W, F435W, F555W and F814W) and in two additional filters from the LEGUS--H$\alpha$ follow-up survey (GO-13773; PI R. Chandar) observing the galaxy with a narrow filter covering the \ha\ emission-line (F657N) and a medium-band filter sampling the line-free continuum (F547M). In total we have 10 filters spanning the wavelength range NUV-NIR, including two filters for the observation of the two hydrogen recombination lines \ha\ and \pb ; details of the observations are summarized in Tab.~\ref{tab:data}.
Fig.~\ref{fig:maps_ext_spitzer} (top left) shows a RGB--composite with \ha, F555W and \pb\ in the blue, green and red channels, respectively.
We study the cluster population in the area delimited by the field of view of the NIR pointings, outlined in the same figure.

\begin{table}[]
    \centering
    \begin{tabular}{llrrl}
    \hline
        Instrument  & Filter & $\rm P_\lambda$ [\AA] & Expt [s] & Program/PI \\
    \hline
    \hline
        WFC3-IR     & F160W & 15369  & 3594  & GO 15330/Calzetti \\
        WFC3-IR     & F128W & 12832  & 3594  & GO 15330/Calzetti \\
        WFC3-IR     & F110W & 11534  & 1994  & GO 15330/Calzetti \\
        ACS-WFC     & F814W & 8047   & 4569  & GO 9796/Miller \\
        WFC3-UVIS   & F657N & 6567   & 3090  & GO 13773/Chandar \\
        WFC3-UVIS   & F547M & 5447   & 1108  & GO 13773/Chandar \\
        ACS-WFC     & F555W & 5360   & 5600  & GO 9796/Miller \\
        ACS-WFC     & F435W & 4329   & 4560  & GO 9796/Miller \\
        WFC3-UVIS   & F336W & 3355   & 4818  & GO 13364/Calzetti \\
        WFC3-UVIS   & F275W & 2707   & 5058  & GO 13364/Calzetti \\
    \hline
    \end{tabular}
    \caption{Summary of the HST observations used in this work, along with the pivot wavelength of each filter ($\rm P_\lambda$) and the exposure times (Expt).}
    \label{tab:data}
\end{table}

\section{Source catalogs} \label{sec:catalogs}
We look for young dust-embedded star clusters following two separate approaches: (1) we define regions of high extinction, by constructing an extinction map from the combination of \ha\ and \pb\ emission line maps and looking for NIR compact sources within such regions; (2) we select, within the entire galaxy, sources with compact \pb\ line emission.
In addition to the aforementioned source selections, we also consider the cluster catalog produced by LEGUS \citealp{adamo2017}, based on HST observation in 5 NUV--optical broadband filters. 

\subsection{Sources in high-extinction regions} \label{sec:catalogs_highext}
We make use of the observations in the UVIS-F657N and IR-F128N filters to construct \ha\ and \pb\ nebular emission-line maps. 
In order to estimate the continuum emission at the central wavelength of the F657N filter, we use two nearby filters, F547M and F814W. The former is a medium--band filter not containing nebular lines, the latter instead does contain emission lines, but its large bandwidth causes the flux to be greatly dominated, even for very young sources, by the stellar continuum; the nebular lines play only a minimal contribution. We linearly interpolate the observed flux on  logarithmic scale in those two filters on a pixel--by--pixel basis, and we find, for every pixel of the data, the continuum corresponding to the pivot wavelength of the F657N filter, $\rm \lambda_{F657N}=6566.6$ \AA. This approach is made possible by the comparable widths of the PSF in the various HST optical filters, ensuring that the contribution to the flux in every pixel is coming from the same region in every observed band. 
Having subtracted the continuum, we are left with the combined emission of the \ha\ line and the [$\rm N_{II}$] doublet ($\lambda\lambda\ 6548,6584$ \AA). In order to remove the contribution of the [$\rm N_{II}$] lines from the emission, we use their relative ratio to the \ha\ line as given by \citet{kennicutt2008} (for NGC 1313, $\rm [N_{II}]/H\alpha=0.34$).

We create a map of the \pb\ emission in a similar way, using F110W and F160W as nearby filters. We interpolate the flux in those two filters to estimate the value of the stellar continuum at the pivot wavelength of the F128N filter, $\rm \lambda_{F128N}=12 831.8$ \AA. We point out that the F110W filter contains the \pb\ emission and therefore we use the first estimate of the \pb\ line emission map to create a line--free F110W map and then repeat the interpolation process. The second estimate of \pb\ emission line map differs from the first one by $5\%$ at maximum. Further iterations of this ``cleaning'' process differ from the previous ones by less than $0.1\%$ and quickly converge. 

We convert the \ha\ and \pb\ emission maps to a common physical resolution by degrading the pixel scale of the \ha\ map from $0.04$ to $0.08$ arcsec/pix and then convolve each map to a gaussian kernel with $\sigma=3$ px ($\sim5$ pc at the distance of NGC 1313), in order to smooth possible pixel--scale inaccuracies of the maps.
We calculate the ratio \ha/\pb, correcting for the Galactic reddening:
\begin{equation}
\rm R_{dered.}\equiv\left(\frac{H\alpha}{Pa\beta}\right)_{dered.} = \left(\frac{H\alpha}{Pa\beta}\right)_{observed} 10^{0.4\cdot[k(H\alpha)-k(P\beta)]\cdot E(B-V)}
\end{equation}
where, for the Milky Way extinction law, $\rm k(H\alpha)-k(P\beta)=1.69$ (according to the parametrization of \citealp{fitzpatrick1999}, with total-to-selective extinction value $\rm R_V=3.1$) and we used $\rm E(B-V)=0.096$ mag\footnote{value taken from the NASA Extragalactic Database (NED).} for the foreground extinction correction in the direction of NGC 1313.
Finally, we convert the line ratio into an expected extinction value. Taking the intrinsic line ratio $\rm R_{intrinsic} = 17.57$\footnote{appropriate for $\rm H_{II}$ regions with electron temperature $T_e\sim11,500$ K}, we assume that the different ratios observed over the different sub-regions of the galaxies are due to variations in the nebular extinction:
\begin{equation}
\rm A_V = \frac{R_V}{-0.4\cdot[k(H\alpha)-k(P\beta)]}\cdot \log_{10}\left(\frac{R_{dered.}}{R_{intrinsic}}\right) 
\end{equation}
with $\rm R_V=3.1$. In constructing the extinction map we limit our selection to regions where both \ha\ and \pb\ line emissions have a S/N$>3$.

\begin{figure*}
\centering
\subfigure{\includegraphics[width=0.47\textwidth]{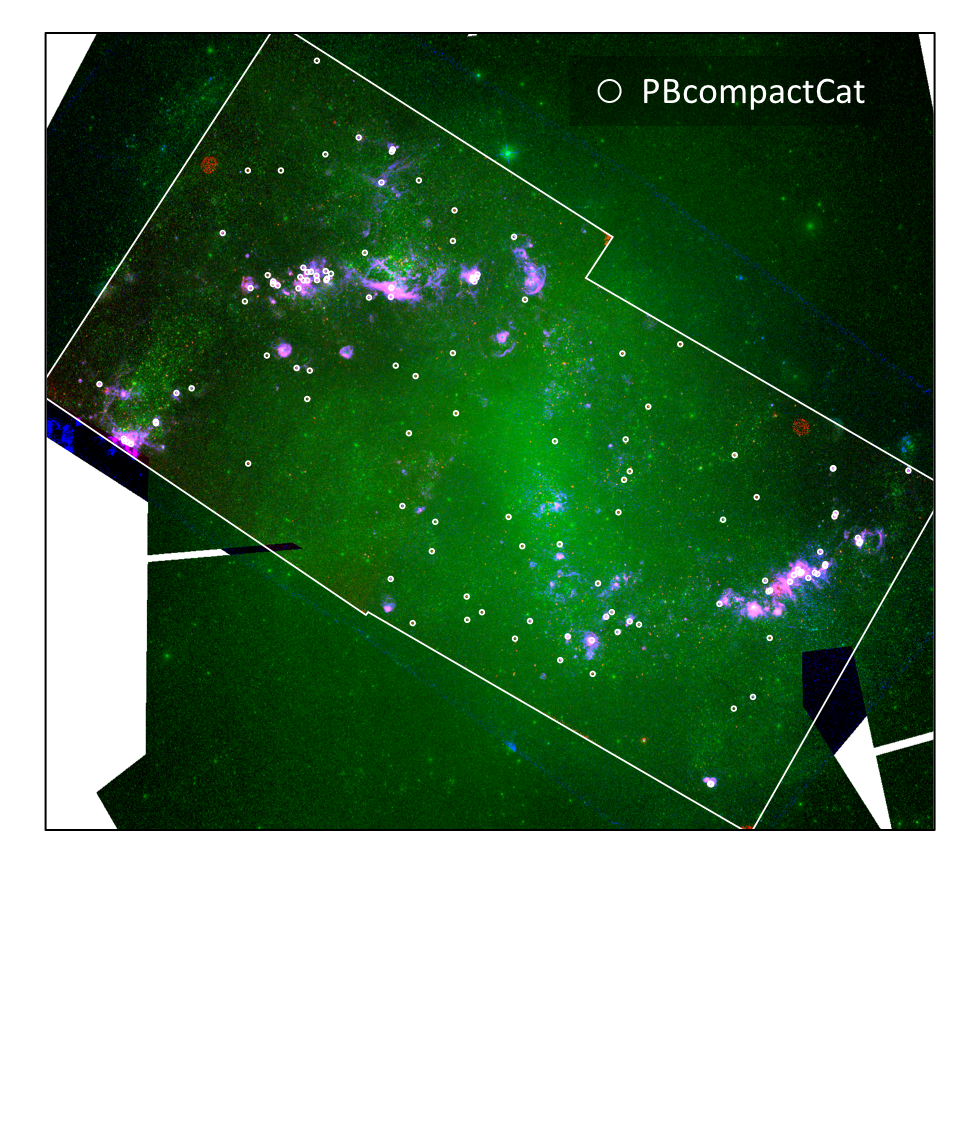}}
\subfigure{\includegraphics[width=0.47\textwidth]{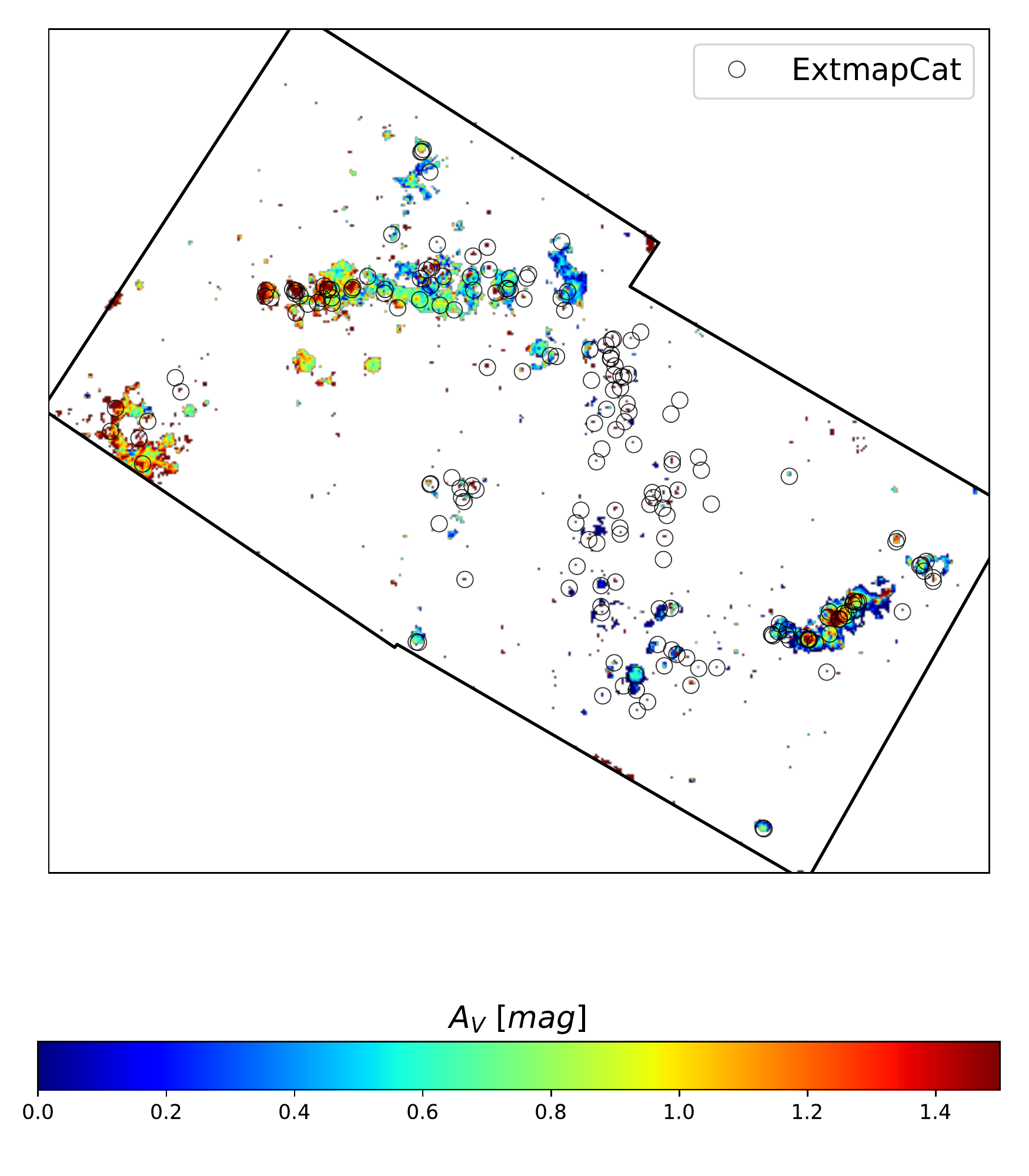}}
\subfigure{\includegraphics[width=0.47\textwidth]{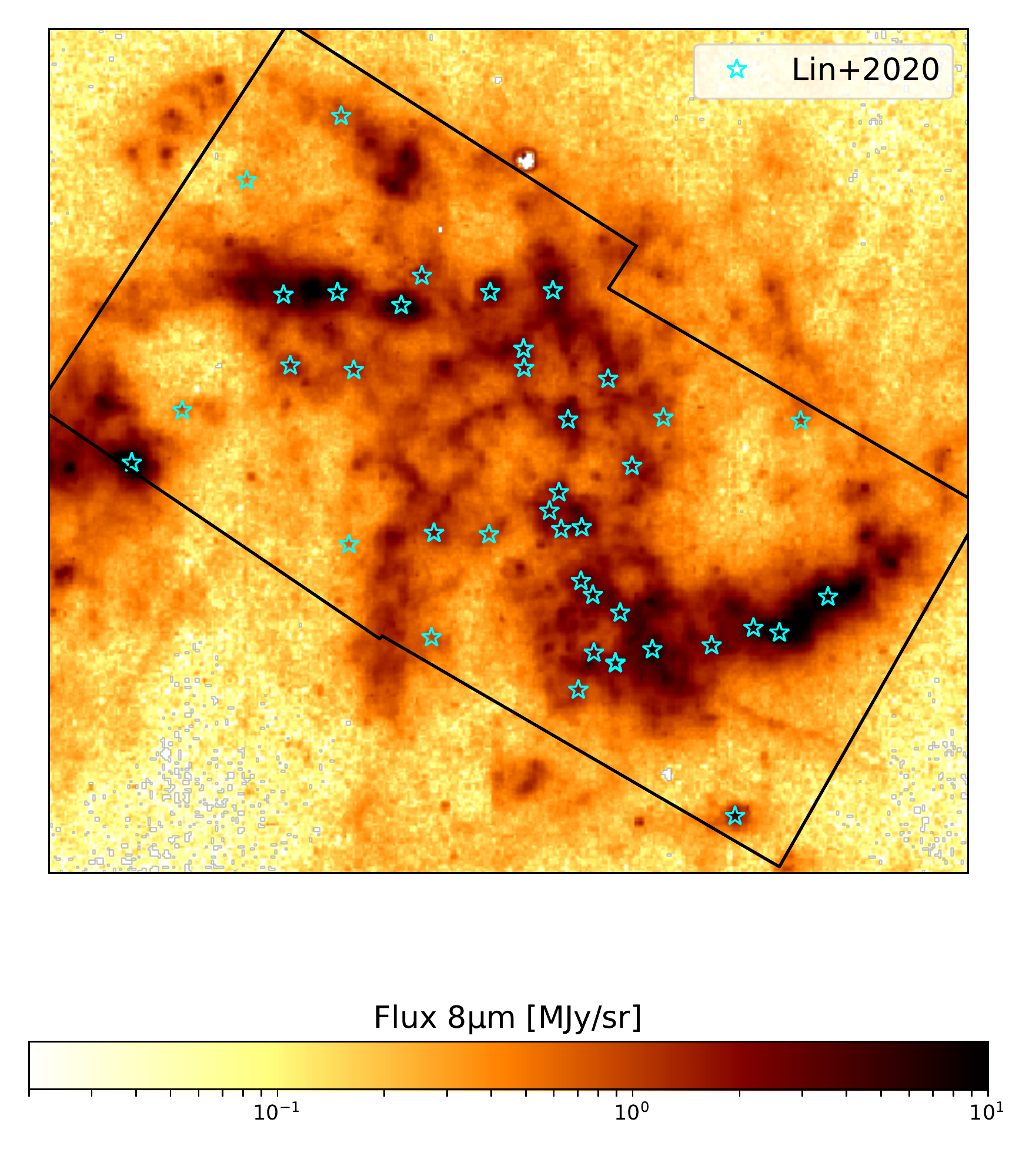}}
\subfigure{\includegraphics[width=0.47\textwidth]{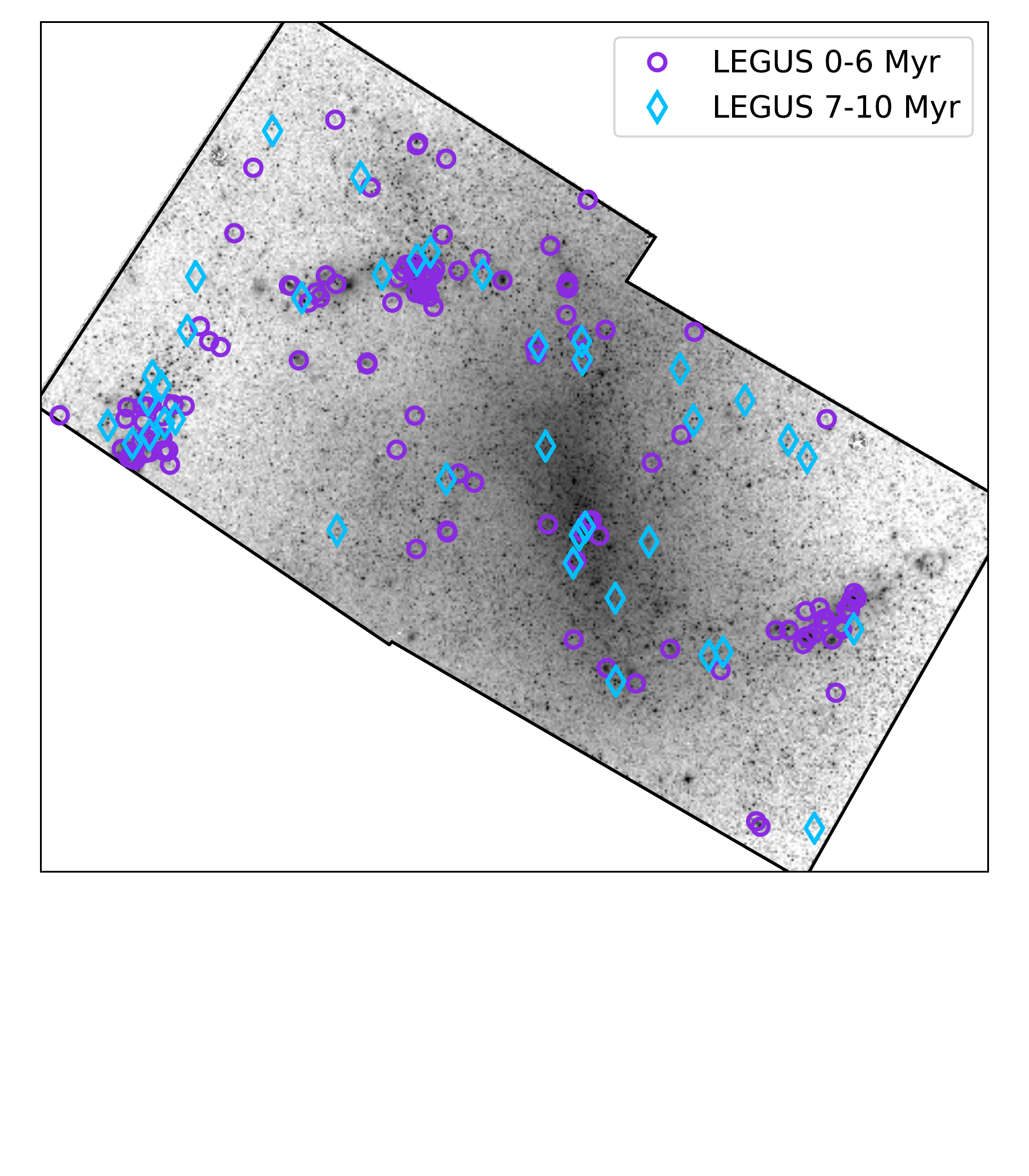}}
\caption{(Top left): RGB composite of NGC1313 showing continuum--subtracted \ha\ in the blue channel, F555W in green and continuum--subtracted \pb\ in red. The coordinates of the sources from the \textit{PBcompactCat} are shown with white circles The thick white contour marks the FoV of the HST NIR observations. (Top right): map of the \ha/\pb\ ratio, converted into an extinction map. The map highlights the regions of recent star formation and suggests the presence of three main extended extincted regions, two in the eastern side of the galaxy and one in the western side, plus smaller compact ones. Position of the sources from the \textit{ExtmapCat} are shown as black circles. (Bottom left): $8\ \mu$m emission from Spitzer, tracing the emission of the warm dust. We note that there is good spatial coincidence between the brightest $8\ \mu$m areas and the nebular extinction map. Cyan empty stars mark the position of the MIR sources studied in \citet{lin2020}.
(Bottom right): position of the cluster of the LEGUS catalog, divided in two age bins (0-6 Myr: purple circles, 7-10 Myr: blue diamonds), plotted over the F555W data. The age division is motivated by the analyses in Section~\ref{sec:avg_ext}.}
\label{fig:maps_ext_spitzer}
\end{figure*}

We show the extinction map in Fig.~\ref{fig:maps_ext_spitzer}. In the same figure, as well as in Fig.~\ref{fig:maps_ext_spitzer_CO}, we compare it to the $\rm 8\ \mu m$ emission from the Spitzer telescope, revealing the dust emission. We find good spatial coincidence of the extincted nebular regions with the regions of brightest emission in $\rm 8\ \mu m$. As expected, the extinction map reveals the extended regions along the spiral arms of NGC 1313 associated to recent star formation. In addition, it reveals compact regions with elevated $\rm A_V$ values, associated with single sources; these latter cases are frequently caused by sources with high \pb\ emission but almost no \ha, mostly located along the spiral arms but with some exceptions. 

Previous studies of clusters in NGC 1313 in the NUV-optical found on average extinctions of $\rm E(B-V)\sim0.25$ ($A_V\sim0.8$) for young clusters. We aim at studying clusters with higher extinction that may be missed in the NUV-optical bands.
We therefore focus on the regions of the extinction map with $A_V\ge0.8$ mag and we visually select all the NIR sources with a clear counterpart in the F814W filter. 
We find a total of 188 cluster candidates. In the rest of the paper we are referring to this source catalog selected on the basis of the extinction map as \textit{ExtmapCat}. We plot the positions of the sources within the galaxy in Fig.~\ref{fig:maps_ext_spitzer} (top right panel).

\subsection{Sources with compact Paschen-$\beta$ emission}
In order to search for sources not detected in \ha\ but emitting in \pb, and therefore missed by the extinction map (limited by the \ha\ detection), we select a second catalogue of sources. 
From the continuum--subtracted \pb\ map described previously, we select sources with detectable (S/N$>3$) \pb\ compact emission (i.e. with a narrow light profile, consistent with the one of stars and clusters, avoiding diffuse emission). 
This approach helps to detect embedded sources that have \pb\ detection but may be missed in \ha; this catalog and \textit{ExtmapCat} are complementary, as they can overlap but are not mutually exclusive. 

We retrieve 124 sources with compact \pb\ emission and for the rest of the paper we refer to this catalog as \textit{PBcompactCat}. Of these, 40 are in common with the \textit{ExtmapCat}.
We plot the coordinates of the sources in this second catalogue in Fig.~\ref{fig:maps_ext_spitzer} (top left panel). While many sources reside in the extended regions of recent star formation outlined previously, some \pb-compact sources reside outside the main spiral arms.

\begin{figure*}
\centering
\includegraphics[width=0.88\textwidth]{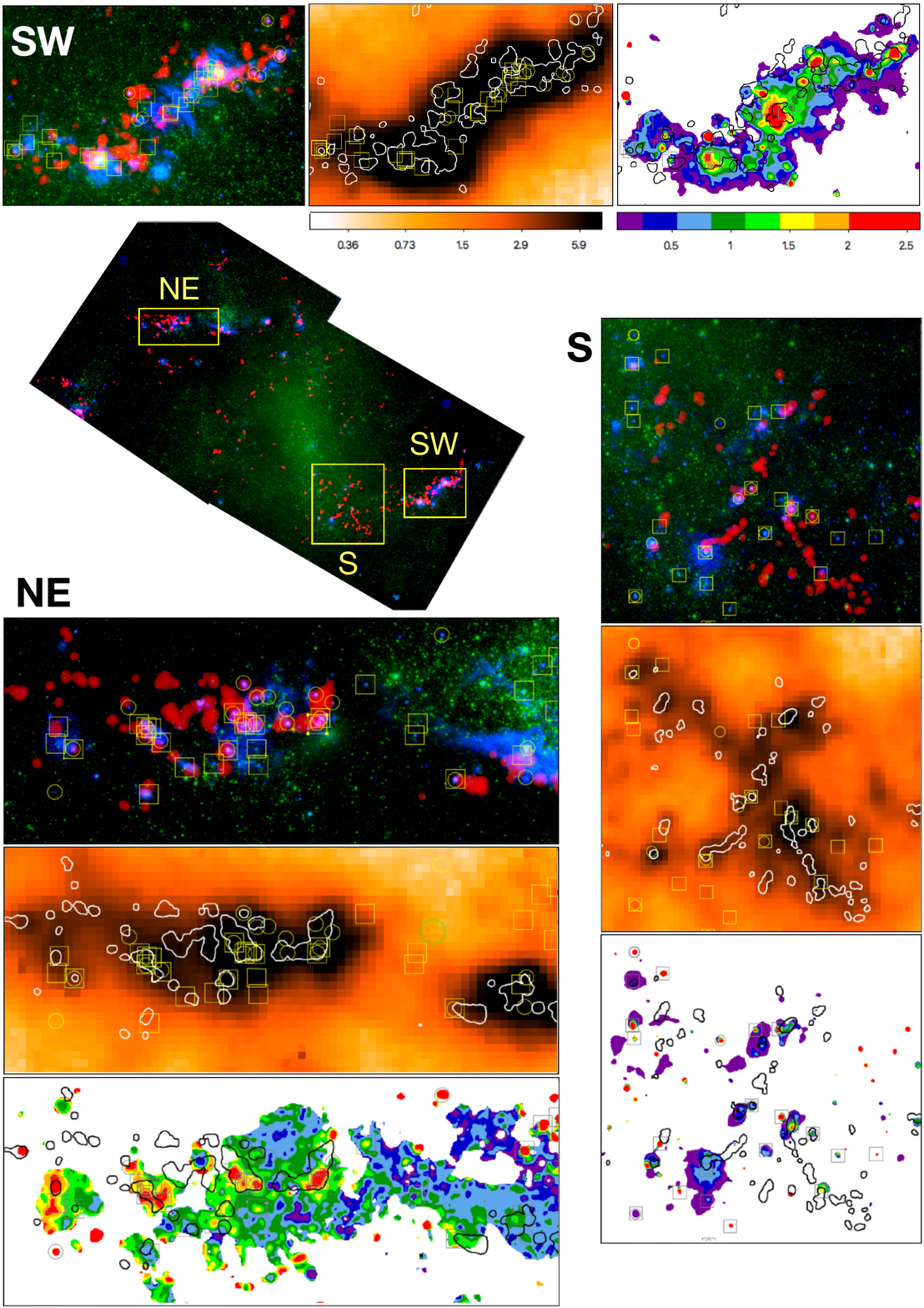}
\caption{Multi--band zoom-ins within NGC 1313. The RGB panels show \pb\ (blue), V--band (green) and CO emission (red). The heatmap panels show the $8\ \mu$m emission from Spitzer (in units of $\rm MJy/sr$), with CO contours over--plotted in white. The multi--colored maps show the distribution of nebular extinction (in units of $\rm A_V\ [mag]$), with CO contours over--plotted in black. In all panels, the positions of sources in \textit{ExtmapCat} and \textit{PBcompactCat} are shown as empty squares and circles, respectively.}
\label{fig:maps_ext_spitzer_CO}
\end{figure*}

In the same Fig.~\ref{fig:maps_ext_spitzer}, we compare the position of sources in our catalogs with the position of MIR compact regions, selected in Spitzer $8 \mu m$ observations by \citet{lin2020}. We remind that $8 \mu m$ observations have a much lower resolution than the HST one, and that sources in the \citet{lin2020} catalog were selected to have some HST cluster counterpart. Overall, the different catalogs cover similar regions of the galaxy; however, some of the \citet{lin2020} sources are not present in our catalogs.
In Fig.~\ref{fig:maps_ext_spitzer_CO} we compare the spatial position of the nebular emission and $\rm 8\ \mu m$ emission with CO clouds found by ALMA (details on the data and of their analysis will be presented in a forthcoming paper, Finn et al., in prep.) focusing on three sub-regions of the galaxy. Also CO emission (which traces the GMCs) has a good spatial coincidence with $\rm 8\ \mu m$ emission and regions of elevated nebular extinction, with some visible exceptions; in the SW portion of the spiral arms, some CO clouds are displaced from the peak of nebular emission, suggesting the presence of gas which is still dark in NIR. The S region is instead mostly missed by the extinction map, or presents only low extinction values, $\rm A_V<0.5$ mag, despite the presence of both CO and $\rm 8\ \mu m$ emission. A possible cause is that the entire S region is in a very early phase of star-formation, still invisible even in the NIR bands. 
This multi--band comparison shows how, while the sources extracted in \textit{ExtmapCat} and \textit{PBcompactCat} are distributed across the entire galaxy, some regions of early star formation may still be so obscured to be inaccessible in the HST NIR bands.

We point out that sources in both \textit{ExtmapCat} and \textit{PBcompactCat} were included in the catalogs without a detailed morphological classification; the two catalogs may therefore contain interlopers, such as bright stars. Photometry and size analyses are performed on all the sources (in Section~\ref{sec:analysis}), and their results will be used in Section~\ref{sec:extmapcat_results} and Section~\ref{sec:pbcompact_results} to clean the samples, producing final catalogs.

\subsection{LEGUS catalogue}
In the course of the current paper we will discuss our results in light of the results achieved by the study of the cluster catalog of NGC 1313 obtained by the LEGUS project. Such catalog was used to study the hierarchical distribution of young clusters \citep{grasha2017a,grasha2017b}, the distribution of their sizes \citep{ryon2017} and the typical evolution timescale associated to their HII regions \citep{hannon2019}. 

The methodology used to extract cluster catalogs in LEGUS is fully described in \citet{adamo2017} and we direct the interested reader to that work for the details. We report here only few main points, relevant for the current work.
The LEGUS cluster catalog of NGC 1313 was build upon 5 broadband filters, covering $NUV,U,B,V$ and $I$ bands. The optical data are the same used by this current analysis. In addition we used the data in the filters F675N and F546M already presented in \citet{hannon2019}.
Ages and masses were derived via SED fitting of the 5 mentioned optical broad-bands only for cluster candidates brighter than $M_V=-6$ mag in the $V$ band and with photometric detections in at least 4 of the bands. A visual morphological classification was used to separate centrally concentrated clusters from multi-peaked compact associations and to separate out contaminating sources (foreground stars, background galaxies, etc...) picked up by the initial source extraction. 
The spatial distribution of the clusters from the LEGUS catalog within the galaxy is shown in Fig.~\ref{fig:maps_ext_spitzer} (bottom right panel). 

\section{Analysis} \label{sec:analysis}
At variance with the standard LEGUS SED analysis, we use in this work the information produced by the combined 10 bands photometry from NUV to NIR (Section~\ref{sec:photometry}) to estimate estimate ages, extinctions and masses of the sources in our catalogs (Section~\ref{sec:fit}).
We also use \pb\ and \ha\ equivalent width measurements as an independent way to estimate cluster ages (Section~\ref{sec:ew}).

\subsection{Photometry} \label{sec:photometry}
For all the catalogs we use a standard aperture photometry approach, using circular apertures with radii of 0.16'' (4 px in the optical filters, 2 px in the NIR ones) in all bands. The sky background is estimated in an annulus centered on the source with radii of 0.20'' and 0.28'' by taking a clipped\footnote{The clipping avoid considering pixels more than $3\sigma$ above the mean of the pixel values in the annulus. This helps eliminating sources from the background estimate.} median. The value of the sky is then normalized by the area in the circular aperture and subtracted from the flux. Previous to the photometry, each source is centered using a centroid algorithm in the F814W filter and its coordinates are then kept fixed in all the other bands. We used F814W as reference because it is the reddest optical filter in our set. The NIR filters have an instrumental PSF which is roughly twice as large (in arcsec) compared to the one of the optical filters. Optical filters are therefore fundamental to discriminate multiple sources, i.e. separated by less than a few pixels transverse to the line of sight. 

An aperture of 0.16'' in radius includes $\sim80\%$ of the flux for PSF-like light profiles, and this fraction decreases for wider light profiles. We apply an aperture correction to account for this missing flux. We consider as \textit{total} flux the one enclosed in a circular aperture with a radius of 0.8''. In order to estimate the aperture correction, we convolve the instrumental PSF\footnote{Obtained from isolated stars in the data.} of each filter with a Moffat profile \citep{elson1987}, considered the most accurate function to describe the light profile of clusters \citep[e.g.][]{elson1987,bastian2013}. We take as reference value for the aperture correction the one obtained from a Moffat profile with an effective radius $\rm R_{eff}=2.5$ pc, as the distribution of cluster sizes is peaked at that value \citep{ryon2015,ryon2017}.
Different choices of the aperture correction would change the normalization of the fluxes, but not the shape of the SED. In the context of broadband SED fitting they would have an effect only on the recovered masses, leaving unchanged ages and extinctions. We discuss different choices of aperture corrections in Section~\ref{sec:masses}.

The radius of the photometric aperture was chosen as a compromise between including most of the sources' fluxes within the aperture and avoiding contamination from nearby sources. Nevertheless, some of the sources suffer from contamination from the light of nearby sources. We consider as \textit{contaminated} the cases where, within an aperture, the flux of nearby sources contributes at least $\sim1/3$ of the total. 61 sources (out of 188, i.e. $32\%$) of the \textit{ExtmapCat}  have contamination in at least 1 filter, while for the \textit{PBcompactCat} catalog the same is true for 16 sources (out of 124, i.e. $13\%$). Most of the contamination takes place in the bluer bands of our filter set, especially in the F275W and F336W filters. The source selection, focused on the red-optical and NIR filters can be a possible cause; many sources of the catalogs are bright in the red filters but their flux in the blue part of the spectrum appear dim, and therefore more easily contaminated by nearby sources.

\subsection{Cluster sizes and alternative photometry} \label{sec:sizes-phot}
The size of stellar clusters in NGC 1313 was studied in details in \citet{ryon2017}. 
The authors showed that the concentration index, CI (defined as the magnitude difference in circular apertures of 1 px and 3 px radii), is able to reproduce the overall distribution of sizes derived by more detailed means such as the software \texttt{GALFIT}. The advantage of using a such a simple metric is that it does not require any supervision, and therefore produces reliable results quickly. 
One possible limitation of the size estimates given by the CI is the presence of crowding, where nearby sources can greatly affect the measure of the CI. 

In order to overcome this problem we implement an alternative measure of the sizes by performing a 2D fitting of the source. This method builds upon the size-photometry script described in \citet{messa2019}; we model the source with the stellar-PSF convolved with a Moffat profile, normalized by the flux and summed to a $1^{st}$--degree polynomial that models the sky background. From each source we produce a cutout of the F814W data with $11\times11$ px size that we fit, comparing it to the source model via a Levenberg-Marquardt minimization. This method was tested to be ineffective in discriminating sizes smaller than $\rm R_{eff}=0.69$ px (corresponding to $\rm R_{eff}=0.59$ pc at the distance of NGC 1313). We therefore consider all sources with $\rm R_{eff}\le0.69$ px as consistent with having a stellar--PSF.

We use the size-photometry method just described also to perform a complementary photometry. In this case we do not rely on an (average) aperture correction, and the results are used to provide an estimate of the total flux of the clusters based on their sizes. For each source we keep fixed the central coordinates and the size retrieved in the F814W filter and then we repeat the Levenberg-Marquardt minimization in all the other filters, leaving only the flux (and the parameters describing the background) as free parameters. More details on this method are given in Appendix~\ref{sec:app_sizeflux}. The photometric results obtained in this way are used in Section~\ref{sec:masses} to discuss the mass distribution of the cluster catalogs.

\begin{figure}
\centering
\includegraphics[width=0.8\textwidth]{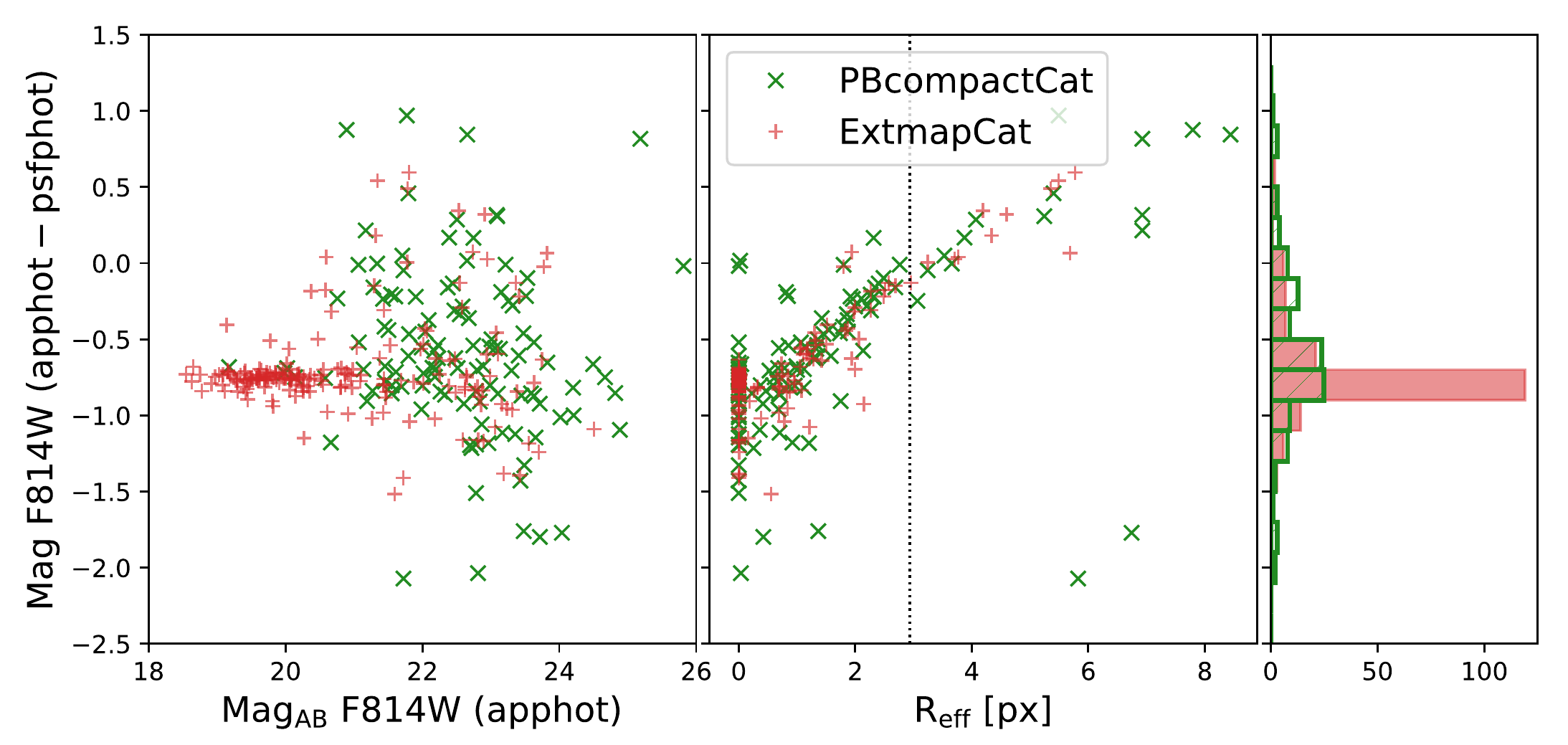}
\caption{Comparison between the photometry in the F814W filter obtained via the aperture photometry process (apphot) and via the alternative size--photometry (psfphot), in function of the F814W photometry itself (left panel) and of the recovered effective radius, $\rm R_{eff}$ (central panel). The distribution of the photometry differences (right panel) shows a peak around $0.8$ mag, due to the presence of many sources with PSF--like light profiles. The black dotted vertical line in the central panel marks $\rm R_{eff}=2.9$ px, corresponding to the value $\rm R_{eff}=2.5$ pc used as reference for estimating the average aperture correction (see Section~\ref{sec:photometry}).}
\label{fig:comparison_phot}
\end{figure}

The comparison between the photometry obtained with this method and the one obtained via the aperture photometry analysis described in Section~\ref{sec:photometry} is shown in Fig.~\ref{fig:comparison_phot}.
There is a magnitude offset of $\sim0.8$ mag, especially clear for the \textit{ExtmapCat} sources. This offset is caused by the \textit{aperture photometry} method having a fixed aperture correction (calibrated on a typical cluster size, see previous section) as opposed to the size--photometry method. The right panel of Fig.~\ref{fig:comparison_phot} shows how the offset decreases when increasing estimated source sizes, being null at $\rm R_{eff}\sim2.5$ pc.
Combining the information from the two panels in Fig.~\ref{fig:comparison_phot} we notice that the offset is mostly caused by bright sources with narrow profiles; these are probably stars and are excluded from the final samples (see Section~\ref{sec:extmapcat_results} and Section~\ref{sec:pbcompact_results}). The same sources drive the narrow scatter of the distribution for the \textit{ExtmapCat} compared to the \textit{PBcompactCat}. Finally, the large overall scatter of the magnitude difference between the two methods (going from $-2$ to $+1$ mag) can be simply explained by the different sizes of the sources considered. When focusing on the sources at a fixed estimated size, the scatter reduces considerably (Fig.~\ref{fig:comparison_phot}, central panel).

\subsection{SED fitting}\label{sec:fit}
We perform a least--$\chi^2$ fitting of the broadband SED obtained from the photometry in Section~\ref{sec:photometry}, following a very similar approach to the one used in LEGUS (see \citealp{adamo2017}).
We use as spectral models the Yggdrasil single-stellar population (SSP) models\footnote{Available at \url{https://www.astro.uu.se/~ez/yggdrasil/yggdrasil.html}} \citep{zackrisson2011}. In details, we use the modes created from Starburst99 Padova-AGB tracks \citep{leitherer1999,vazquez2005} with a universal \citet{kroupa2001} initial mass function (IMF) in the interval $\rm 0.1-100 M_\odot$. These models assume that the IMF is fully sampled. The stellar models are then used as input to run Cloudy \citep{ferland2013} and obtain the evolution of the nebular continuum and lines emission, produced by the ionized gas surrounding the young clusters. Yggdrasil adopts a spherical gas distribution around the emitting source, with  hydrogen number density $\rm n_H=10^2\ cm^{-3}$ and gas filling factor (describing the porosity of the gas) $f_{fill}=0.01$, typical of HII regions \citep{kewley2002}, and assumes that the gas and the stars form from material of the same metallicity. We consider a sub-solar metallicity ($\rm Z=0.008$) for NGC 1313 as it results in a best fit of the young clusters but we test also models with solar ($\rm Z=0.020$) metallicity (see discussion in Section~\ref{sec:test_metext} and the figures in Appendix~\ref{sec:app_testmet}). 
We choose the models with a gas covering fraction $f_{cov}=0.5$, i.e. only $50\%$ of the Lyman continuum photons produced by the central source ionize the gas. While 
Yggdrasil provides multiple choices for $f_{cov}$, we decide to keep it fixed, as it reasonably describe the case of ionized and partially leaking nebulae around young star clusters. 
Including the nebular emission in the cluster models is fundamental, and its omitting would lead to misleading results \citep{zackrisson2001,reines2010,adamo2010b}. Nevertheless, we do not have enough information to disentangle the gas conditions from our fitting process. We tested fitting the sources using models with $f_{cov}=1.0$ i.e. no leakage of LyC photons, finding no substantial difference in the distribution of recovered cluster ages, masses and extinctions. We therefore kept models with $f_{cov}=0.5$ as the reference one in our analysis.

Starting from the Yggdrasil models, we create a grid for the fitting by combining the spectra at progressive age steps and increasing internal reddening.
We keep the same age steps included in Yggdrasil, which provides a $\rm 1\ Myr$ interval frequency in the range $\rm [1; 15]\ Myr$. The sampling decreases at older ages but we are not affected by it, being interested in the very young sources. 
The models are reddened prior to being fitted using a foreground dust geometry with a grid of extinctions in the range $\rm E(B-V) = [0.0; 2.5]$ with a step of $0.01$ mag. This is the only difference to the LEGUS approach, which used a grid of extinction extending only up to $\rm E(B-V) = 1.5$ mag.
We use as reference the Milky Way extinction law \citep{cardelli1989} but, as a test of the robustness of the results from our fit, we also consider the differential (i.e. the gas emission suffers higher extinction than the stars) starburst attenuation law \citep{calzetti2000}.
The fitting is done using a traditional $\chi^2$ approach on the grid just described. The model spectrum is normalized for each age-extinction combination in order to minimize the reduced $\chi^2$ ($\chi^2_{red.}$), for every source. The best fit for a source is given by the combination of age-extinction that minimize the $\chi^2_{red.}$ value. 
An uncertainty interval is given for the best-fit parameters by considering all the sources within $1\sigma$ from the best-fit values, i.e. the parameters that result in a $\chi^2_{red.}$ within 3.53 from the minimum $\chi^2_{red.}$.

Some of the sources in our catalogs have large photometric errors, while some others are contaminated, in some filters, by the emission of nearby sources, as described in Section~\ref{sec:photometry}. We decide to exclude from the fit the cases where the photometric uncertainty is above 0.3 mag (i.e. $\rm S/N<3$), as sources with an uncertainty above that level have only a partial detection. We also exclude from the fit the cases of contamination, as their photometry would lead to misleading results. 
We fit the sources where at least 5 bands are left after the uncertainty-contamination cut, for a total of 170 sources for the \textit{ExtmapCat} and 98 sources in the case of the \textit{PBcompactCat}.
The results of the fitting process are shown and described in Section~\ref{sec:results}.

\subsection{\ha\ and \pb\ equivalent width analysis}\label{sec:ew}
A direct way of deriving the age of a cluster is to use the equivalent widths, EWs, of the nebular emission lines, defined as:
\begin{equation} \label{eq:EW_definition}
EW_\lambda = \int \frac{F_{obs}(\lambda)-F_c(\lambda)}{F_c(\lambda)}d\lambda    
\end{equation}
where $F_{obs}(\lambda)$ is the observed flux across the emission line at the wavelength $\lambda$, and $F_c(\lambda)$ is the continuum level underneath the emission line. 
We do not have spectra of our sources and we have therefore to rely on a combination of broad and narrow filters to estimate the $EW$. Similarly to what done in Section~\ref{sec:catalogs_highext}, we use the filters adjacent to the narrow one containing the line to estimate the continuum. 
For $EW_{H\alpha}$ we use a logarithmic interpolation of the fluxes in the F547M and F814W filters to find the flux of the continuum emission at the pivotal wavelength of the F657N filter, $F_{C,H\alpha}$. We calculate the flux in the \ha\ line by subtracting $F_{C,H\alpha}$ to the total flux measured in F657N, $F_{L,H\alpha}=F_{F657N}-F_{C,H\alpha}$. From Eq.~\ref{eq:EW_definition} we derive in our case:
\begin{equation} \label{eq:EW_measure}
EW_{H\alpha,obs} = \frac{F_{L,H\alpha}}{F_{C,H\alpha}}\cdot BW_{F657N}    
\end{equation}
where $\rm BW_{F657N}=122$ \AA is the width of the narrow band filter.

In a similar way (using $\rm BW_{F128N}=157$ \AA as the width of the narrow \pb\ filter), we derived EW(\pb) using F110M and F160W to estimate the continuum $F_{C,Pa\beta}$ and consequently to measure the \pb\ line emission at $F_{L,Pa\beta}=F_{F128N}-F_{C,Pa\beta}$, after using the iterative  procedure for the F110W described in Section~\ref{sec:catalogs_highext}.

\begin{figure}
\centering
\includegraphics[width=0.5\textwidth]{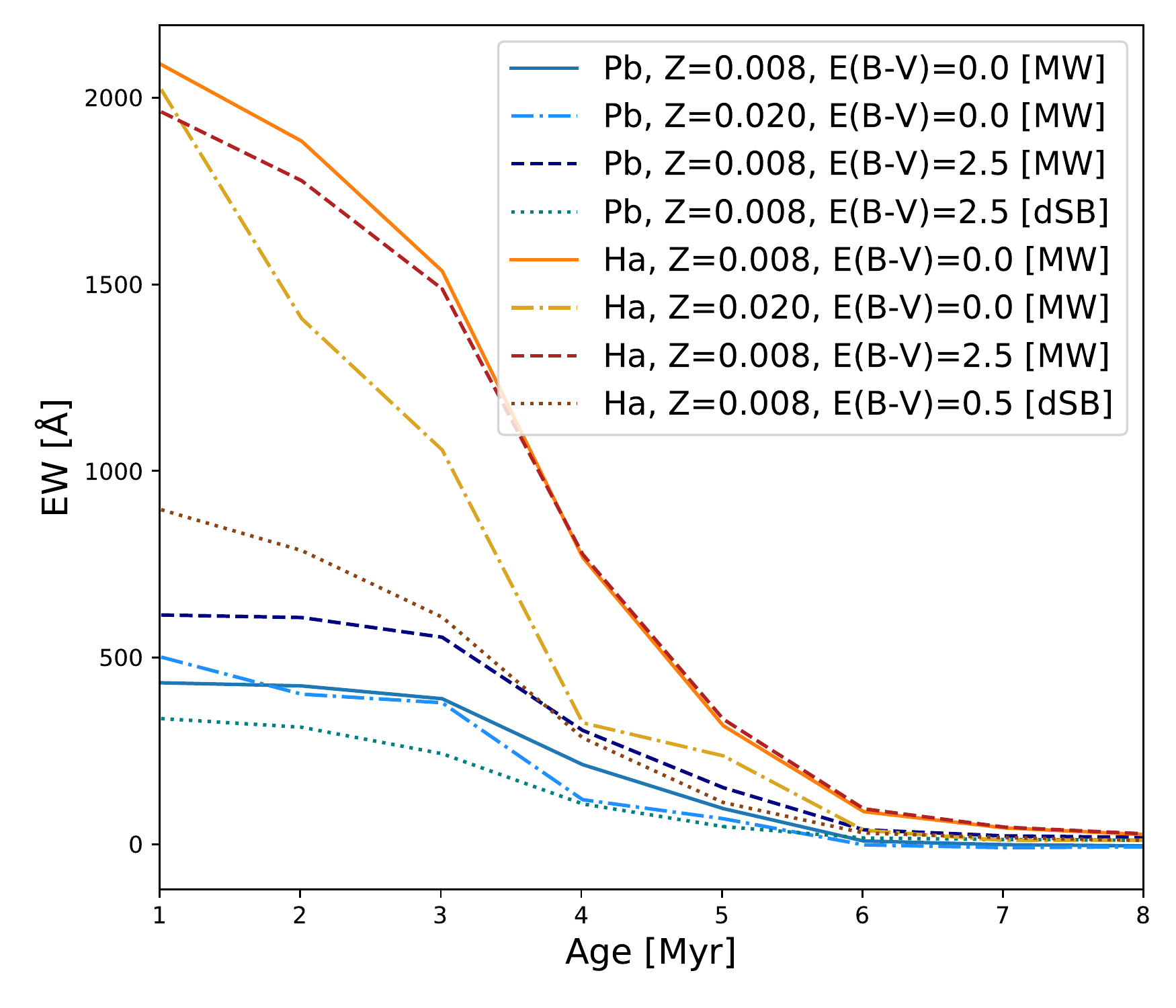}
\caption{Equivalent widths of \ha\ (red/yellow colors) and of \pb\ (blue colors) in function of cluster ages, as measured from the models convolved with the filters used for the fit. Different combinations of metallicity and extinction curves are given (see legend).}
\label{fig:ew_theoretical}
\end{figure}

We use the SSP models described in the previous section to estimate the EW expected at different cluster ages. The EW values that we measure depend on our specific set of filters, and therefore we cannot rely on theoretical values; we prefer instead to use directly the models to create a comparison set for the data. Fig.~\ref{fig:ew_theoretical} shows the evolution of the \ha\ and \pb\ equivalent widths with age of the clusters for different metallicities and extinctions. From the definition in Eq.~\ref{eq:EW_definition}, the EW should not be affected by extinction, if lines and continuum are extincted in the same way. However, since we have to rely on a combination of broad and narrow filters, we see the measured EW being slightly different (typically $<10\%$) at very young ages ($\le 3$ Myr) for different extinction values. 
Differential Starburst extinction is the one with the largest impact, acting more severely on the lines, $\rm E(B-V)_{star} = 0.44\cdot E(B-V)_{gas}$ \citep{calzetti2000} as can be appreciated from Fig.~\ref{fig:ew_theoretical}.

\subsection{Completeness test} \label{sec:completeness_set}
We study the completeness of our samples, both in terms of photometry and of SED fitting. We built a \textit{control} sample of sources with preset ages and masses, adding them to the observational data and going through the entire process of photometric analysis and SED fitting to recover their properties. 
We fix the sizes of the sources at $\rm R_{eff}=1$ px and the masses at $\rm M=10^3\ M_\odot$; these values, according to the results of our analysis presented in the next section (in particular  Fig.~\ref{fig:reffs} and Fig.~\ref{fig:mass_comparison}) are reasonable assumptions. We simulate 100 sources for each combination of ages in the age range [1;6] Myr with steps of 1 Myr and of extinctions in the range [0;2] mag with steps of 0.25 mag, for a total of 30 combinations (3000 simulated sources). 
The galaxy is divided in grids of size $4"\times4"$ and the simulated sources are located only in cells where at least one source from \textit{ExtmapCat} or \textit{PBcompactCat} is found. In this way we measure completeness in the same regions where the real data are located. Only 100 sources per time are inserted in the data--frames, in order to avoid biasing the photometry due to overcrowding. 

We perform aperture photometry on the entire sample of simulated clusters; the almost totality of them have photometric uncertainties below 0.3 mag ($\rm S/N>3$) for all ages and extinctions, in both F814W and F128N, the two filters used as the reference ones for the source extraction (the exact number is collected in Tab.~\ref{tab:completeness_extraction} in Appendix~\ref{sec:app_completeness}). We deduce that we are complete in the detection of the catalog up to a level between $90\%$ and $100\%$. 
Most of the simulated sources have photometric uncertainties below 0.3 mag in at least 5 filters and are therefore fitted. Fig.~\ref{fig:completeness_percentage} in Appendix~\ref{sec:app_completeness} gives the exact number of such sources, at each age and extinction; their fraction stays around $\sim90\%$ for sources up to extinctions of $\rm E(B-V)=1.5$ mag. For higher extinctions it decreases and strongly depends on the age.

We discuss in Section~\ref{sec:completeness} the accuracy we have on the simulated clusters in retrieving the input ages and extinctions.

\section{Results} \label{sec:results}

\subsection{Ages, extinctions and masses of ExtmapCat}\label{sec:extmapcat_results}

\begin{figure*}
\centering
\subfigure{\includegraphics[width=0.49\textwidth]{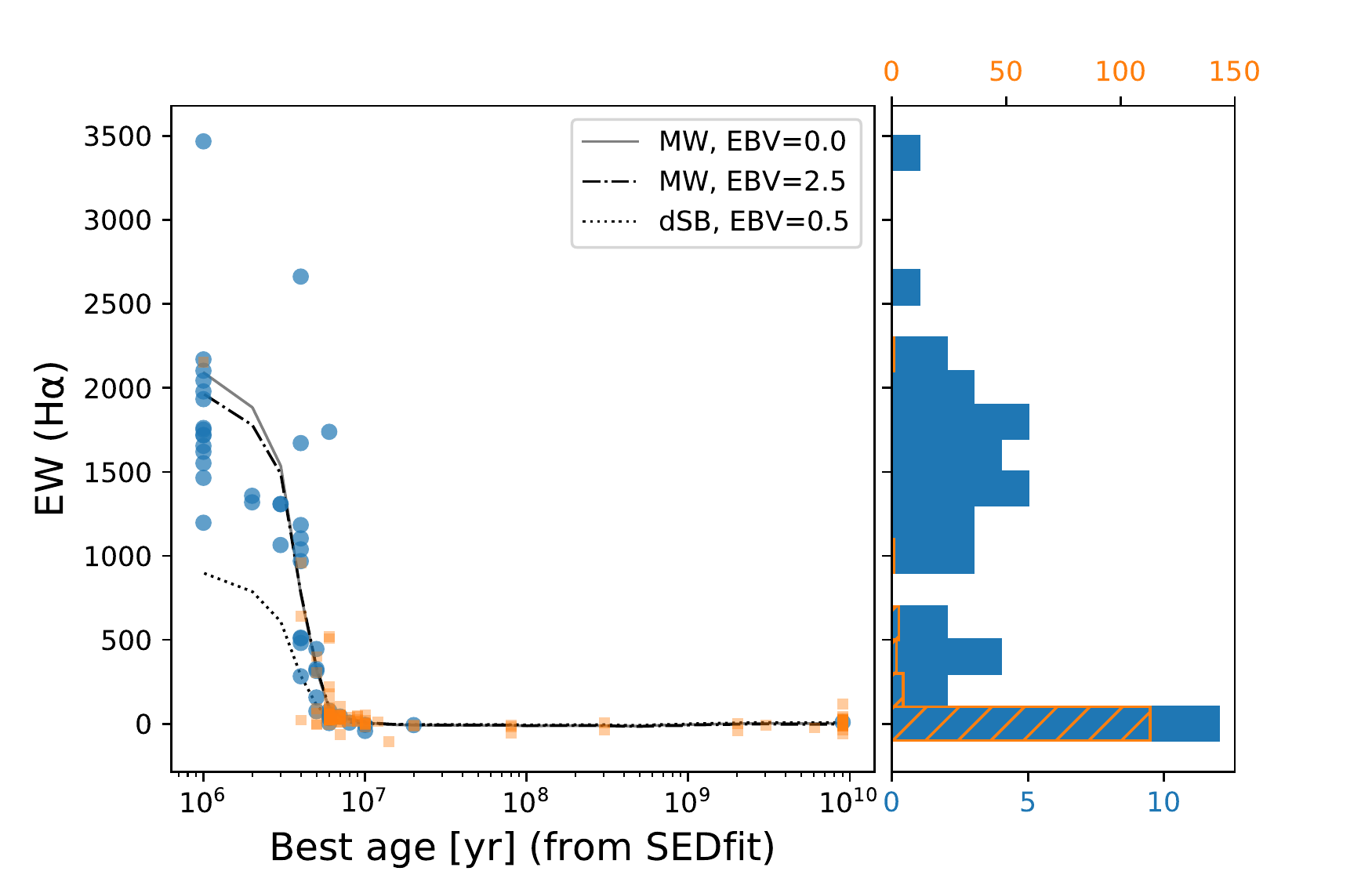}}
\subfigure{\includegraphics[width=0.49\textwidth]{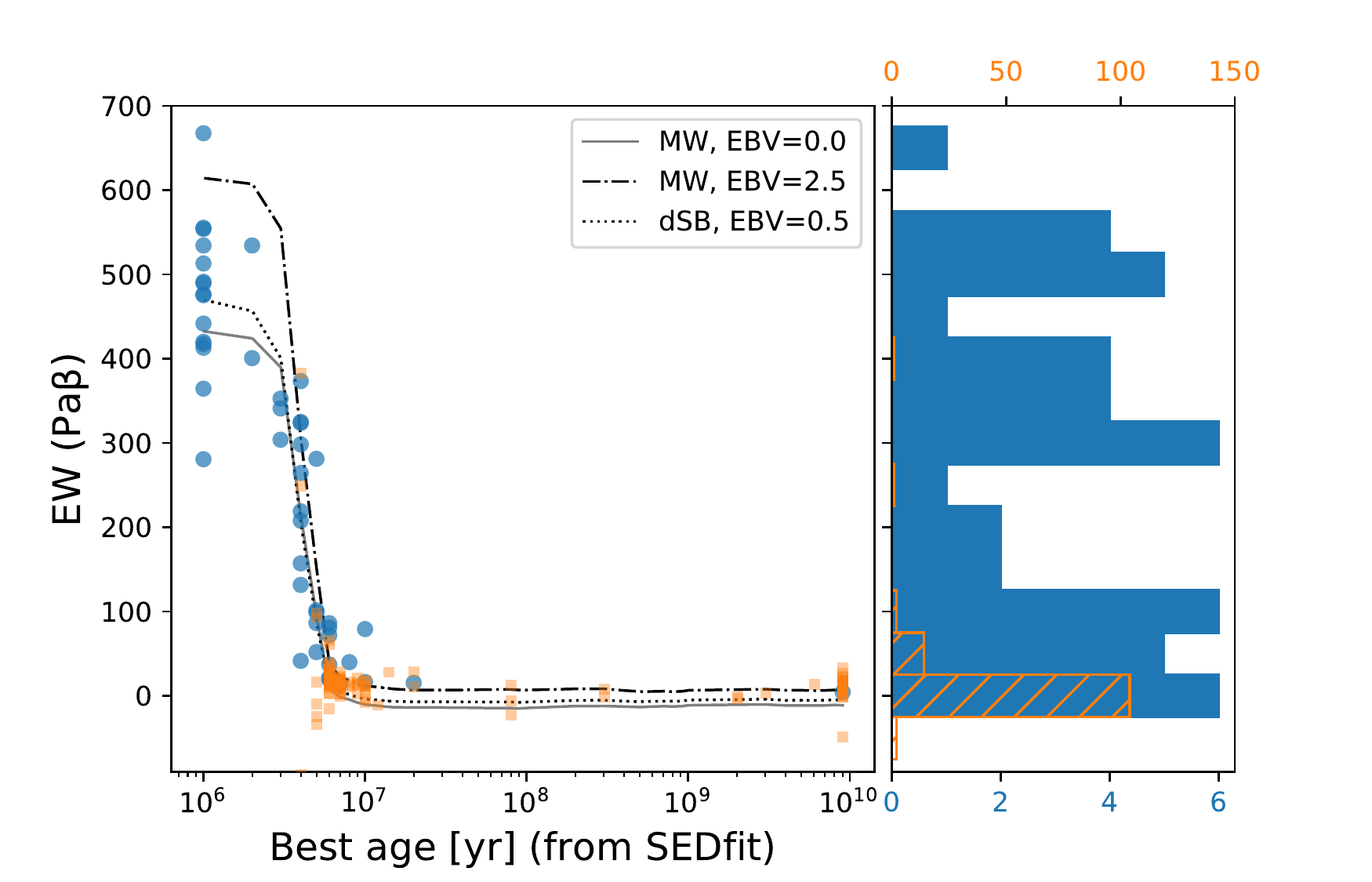}}
\caption{Distribution of estimated \ha\ (left panel) and \pb\ (right panel) equivalent widths and best-fit age values  for the {\em ExtmapCat} sources. The sample is color-coded according to the light profile being either PSF-like ($\rm R_{eff}\le0.59$ pc, orange) or larger ($\rm R_{eff}>0.59$ pc, blue). The solid, dash--dotted and dotted lines are the expected trends, see Fig.~\ref{fig:ew_theoretical}.}
\label{fig:ew_ages_extcat}
\end{figure*}

We show in Fig.~\ref{fig:ew_ages_extcat} the best-fit age values of the sources in the \textit{ExtmapCat}, along with their \ha\ and \pb\ EW values; we color code the sources according to their size, separating the ones with a light profile consistent with the instrumental PSF ($\rm R_{eff}\le0.59$ pc in the F814W filter), and therefore possibly stars, from those with larger size ($\rm R_{eff}>0.59$ pc). Overall we find a good consistency between the ages derived from the SED fit and what could be expected from the EW values. We do not find sources with $\rm age>8$ Myr with high EW in either \ha\ or \pb, confirming the general robustness of our fitting procedure. 
The largest differences of the EW values from the ones predicted (in Section~\ref{sec:ew}) are at the youngest ages of 1 Myr, possibly caused by an inaccuracy in the derived ages. Confirming this hypothesis, we show in Appendix~\ref{sec:app_completeness} how ages in the range $0-3$ Myr are degenerate.
If, instead of the Milky Way extinction curve, we assume the differential starburst one, predictions and observation remain consistent with each other (see Fig.~\ref{fig:ew_ages_extcat}).

\begin{figure*}
\centering
\subfigure{\includegraphics[width=\textwidth]{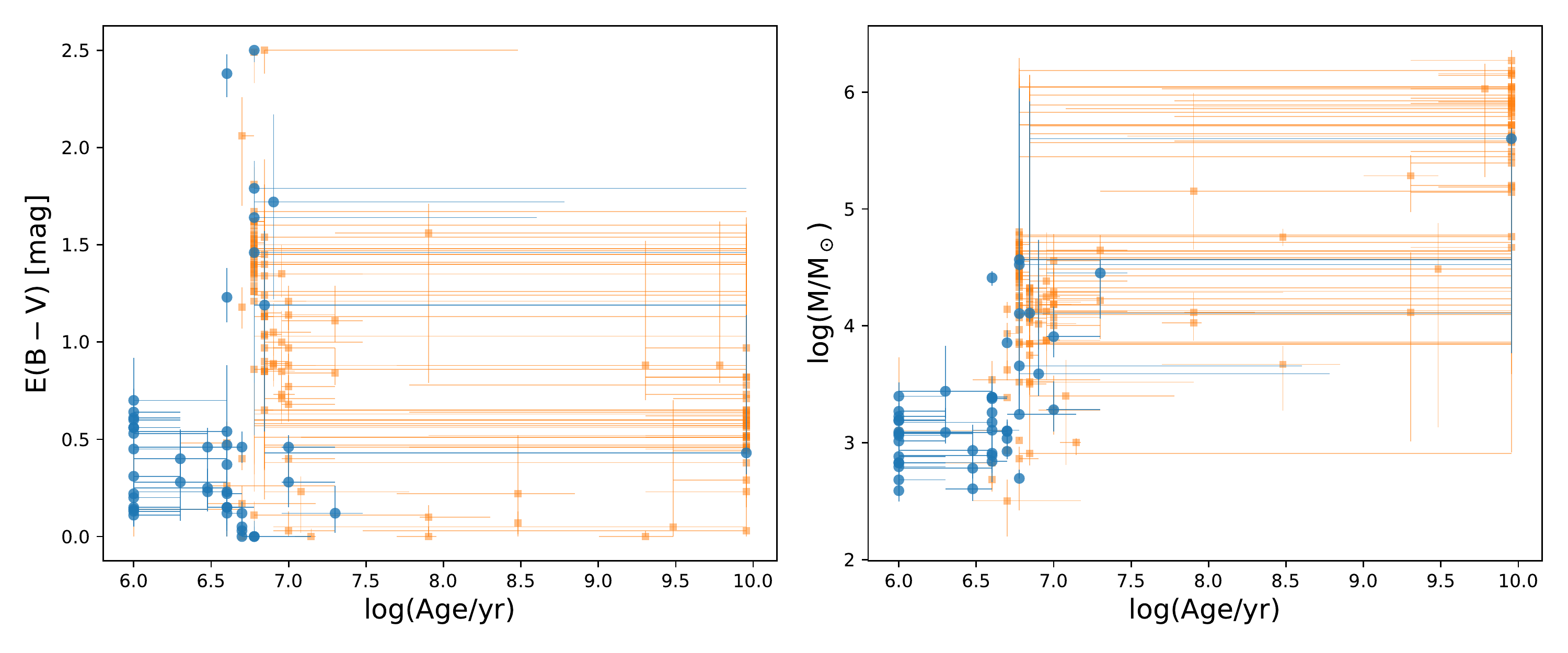}}
\subfigure{\includegraphics[width=\textwidth]{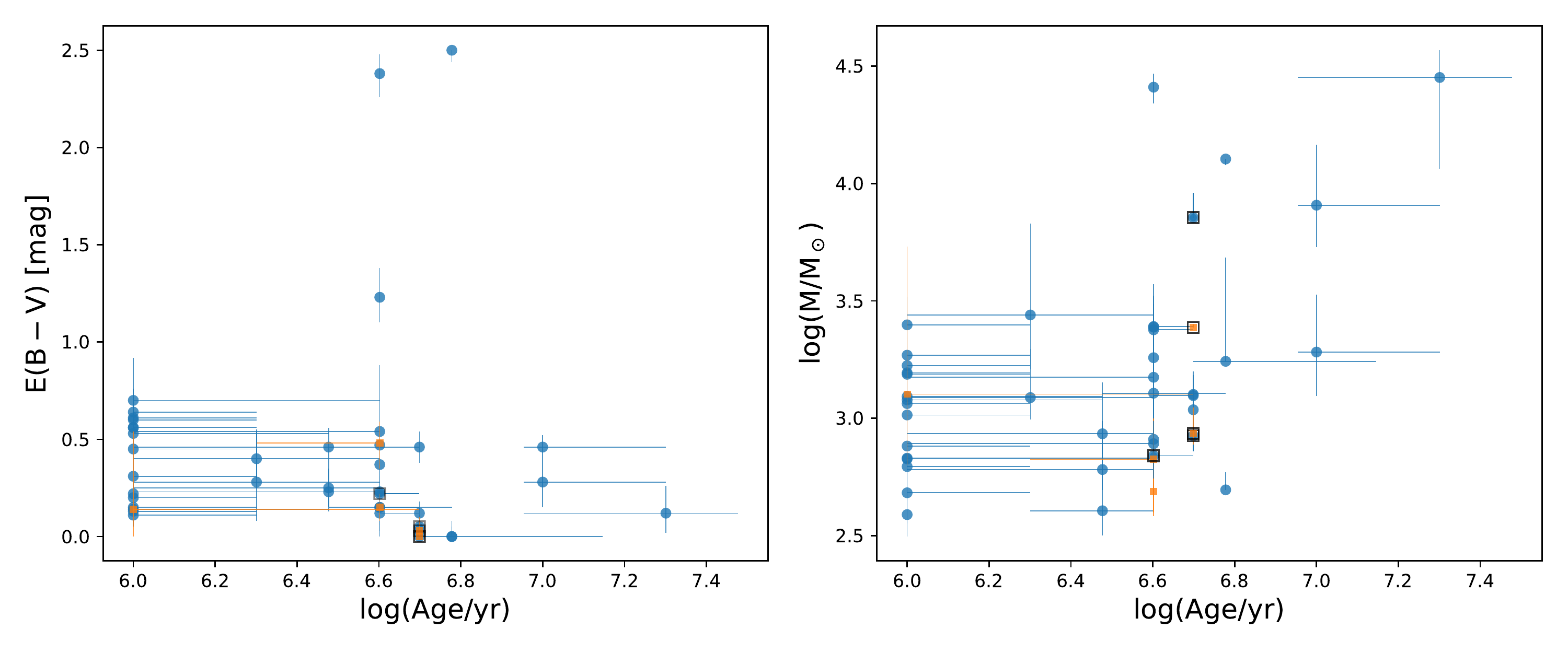}}
\caption{Ages, extinctions (left panels) and masses (right panels) of the \textit{ExtmapCat}, as derived from the broad--band SED fit. The sample is color-coded according to the light profile being either PSF-like ($\rm R_{eff}\le0.59$ pc, orange) or larger ($\rm R_{eff}>0.59$ pc, blue). The bottom row shows the results for the final selection of the sample (\textit{ExtmapCat--final}), see main text. Empty black squares in the bottom plot are used for the 5 sources with blue SED, typical of young sources,but no line emission. The scale of the X--axis in the bottom--row plots was adapted to the narrower range spanned by the data in the final sample.}
\label{fig:age_ebv_mass_extcat}
\end{figure*}

In Fig.~\ref{fig:age_ebv_mass_extcat} we show the same best-fit ages along with extinctions (left panels) and masses (right panels), and their uncertainties.  
We notice that the sources cover, unevenly, the entire age range considered in the model grids. We however point out that almost all of the sources falling in the last age bin considered (age$\sim10$ Gyr) have a PSF-like light profile, consistent with being stars. 
Their masses are on average very high ($\rm >10^5\ M_\odot$) but this is caused by considering them as very old and mildly extincted clusters.
Most of the sources with best age between $\sim10$ Myr and $\sim10$ Gyr have large uncertainties, covering hundreds of Myrs. 
Many of the sources are aggregated around $\sim10$ Myr of age and elevated extinction values; this can be due to degeneracies in the models causing artificial high extinction estimates for dim clusters (due to e.g. the presence of red supergiant stars), as discussed in \citet{hannon2019}.

In order to clean the catalog from sources with un-constrained properties, we leave out the sources with age uncertainties $\rm \Delta \log_{10}(age)>1.5$. In this way all the sources with an age uncertainty covering almost the entire range from $\sim10$ to $\sim10^4$ Myr are discarded (see the comparison between top and bottom panels in Fig.~\ref{fig:age_ebv_mass_extcat}).
We also leave out of the final sample most of the stellar-like sources; among the sources with $\rm R_{eff}<0.59$ pc, we keep only those with \ha\ and \pb\ equivalent width consistent with having ages $\le5$ Myr (i.e. with values of $\rm 317$ \AA and $\rm 95$ \AA in the case of \ha\ and \pb, respectively). We show the remaining sources in Fig.~\ref{fig:age_ebv_mass_extcat} (bottom panels). 
This final selection (\textit{ExtmapCat--final}), counts 46 sources as reported in Tab.~\ref{tab:summary_samples}.

The recovered masses are mostly distributed in the range $\rm [300;3000]\ M_\odot$ (Fig.~\ref{fig:age_ebv_mass_extcat}, right panels). Sources with masses below $\sim1000\ M_\odot$ can be expected to have an uneven sampling of the stellar initial mass function (IMF). This could cause the absence of massive stars able to ionize the gas surrounding the cluster, and, as consequence, the absence of line emission in our photometry. From a visual inspection of the broadband SEDs of our sample, we find that only 5 sources in the entire \textit{ExtmapCat} sample are consistent with having a SED typical of very young (age $\rm\le6\ Myr$) clusters but show no or little line emission in \ha\ and \pb\footnote{Despite the lack of compact line emission, such sources have been selected in the \textit{ExtmapCat} sample because are located in regions of diffuse nebular emission.}. 
The low number of clusters suffering from the lack of ionizing stars, despite the low range of masses recovered, could suggest an underestimation of the masses, due to a poor choice of the average aperture correction values. We discuss this possibility in Section~\ref{sec:masses}.

\begin{figure*}
\centering
\includegraphics[width=0.8\textwidth]{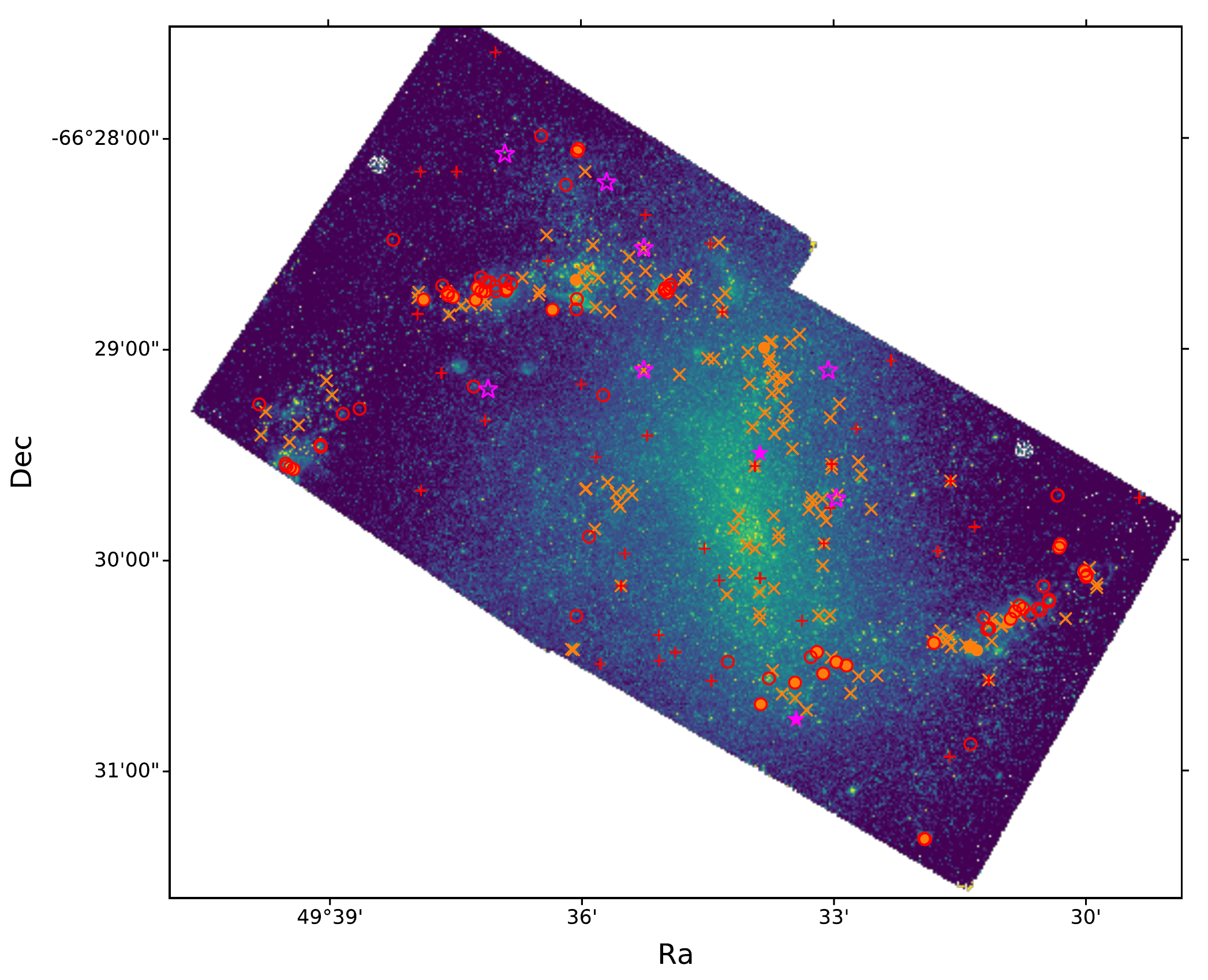}
\caption{Distribution of the sources of the \textit{ExtmapCat} (orange symbols) and \textit{PBcompactCat} (red symbols). The clusters in the final samples have been marked by filled orange circles and empty red circles for the respective catalogs. The positions within the galaxy are plotted over the F128N image data. The clusters from the final catalogs with derived extinctions $\rm E(B-V)>1.5$ are marked by magenta stars (filled for \textit{ExtmapCat}, empty for  \textit{PBcompactCat}) and are discussed in Section~\ref{sec:discuss_single_extincted}.}
\label{fig:coo_dist}
\end{figure*}

We show the position of \textit{ExtmapCat--final} sources within the galaxy in Fig.~\ref{fig:coo_dist}. They are all located in regions of recent star formation, mainly along the  arms of the galaxy. Conversely, the sources discarded from the final selection are more scattered along the entire galaxy, confirming the reliability of our criteria in selecting young clusters.

Only 3 young sources with extinctions $\rm E(B-V)>1.0$ mag remain after the selections. We show the distribution of their $\chi^2_{red.}$ from the fitting analysis, along with the broad--band SED in Appendix~\ref{sec:ind_fit_extincted} (Fig.~\ref{fig:individual_ext1}). Two of them (C93 and C142) have high values for the best--fits $\chi^2_{red.}$, ($\chi^2_{red.}>20$). C10 is the best candidate for being young and highly extincted, with $\rm E(B-V)=1.23$ mag. 

\subsubsection{Testing metallicities and extinction laws}\label{sec:test_metext}
\begin{figure}
\centering
\includegraphics[width=\textwidth]{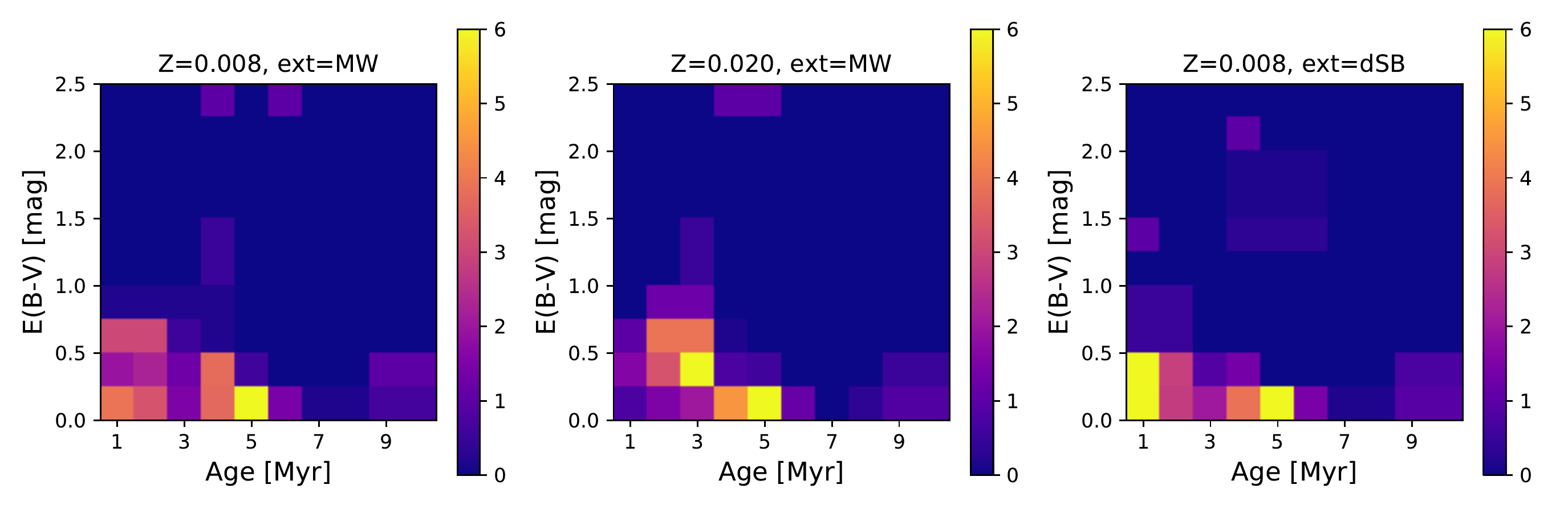}
\caption{Density plots for the \textit{ExtmapCat--final} sources, showing their distribution in the age--extinction plane. The different panels refer to the results using sub-Solar metallicity and Milky-Way extinction (left), Solar metallicity and Milky-Way extinction (centre), sub-Solar metallicity and differential starburst extinction (right).}
\label{fig:age_ebv_extcat}
\end{figure}

We performed the SED fit varying some of the model parameters, namely the metallicity and the extinction law.
In order to better display the distribution of the sources in the age-extinction plane, along with their uncertainties, we use a density plot, based on a fixed grid of ages and extinction intervals (Fig~\ref{fig:age_ebv_extcat}). Each source is considered to cover all cells in the grid included within its upper and lower limits (both in age and E(B-V) values) as given by the face value $\pm$ the uncertainty. Its value in such cells is normalized by the number of the cells covered; in this way the final sum of the cell values over the entire grid is equal to the total number of sources considered. 
We use a grid with a 1 Myr step in ages in the range $\rm [1;10]\ Myr$ and a with a step of $0.25$ mag in the $E(B-V)$ range $\rm [0.0;2.5]\ mag$. 

\begin{figure*}
\centering
\subfigure{\includegraphics[width=0.49\textwidth]{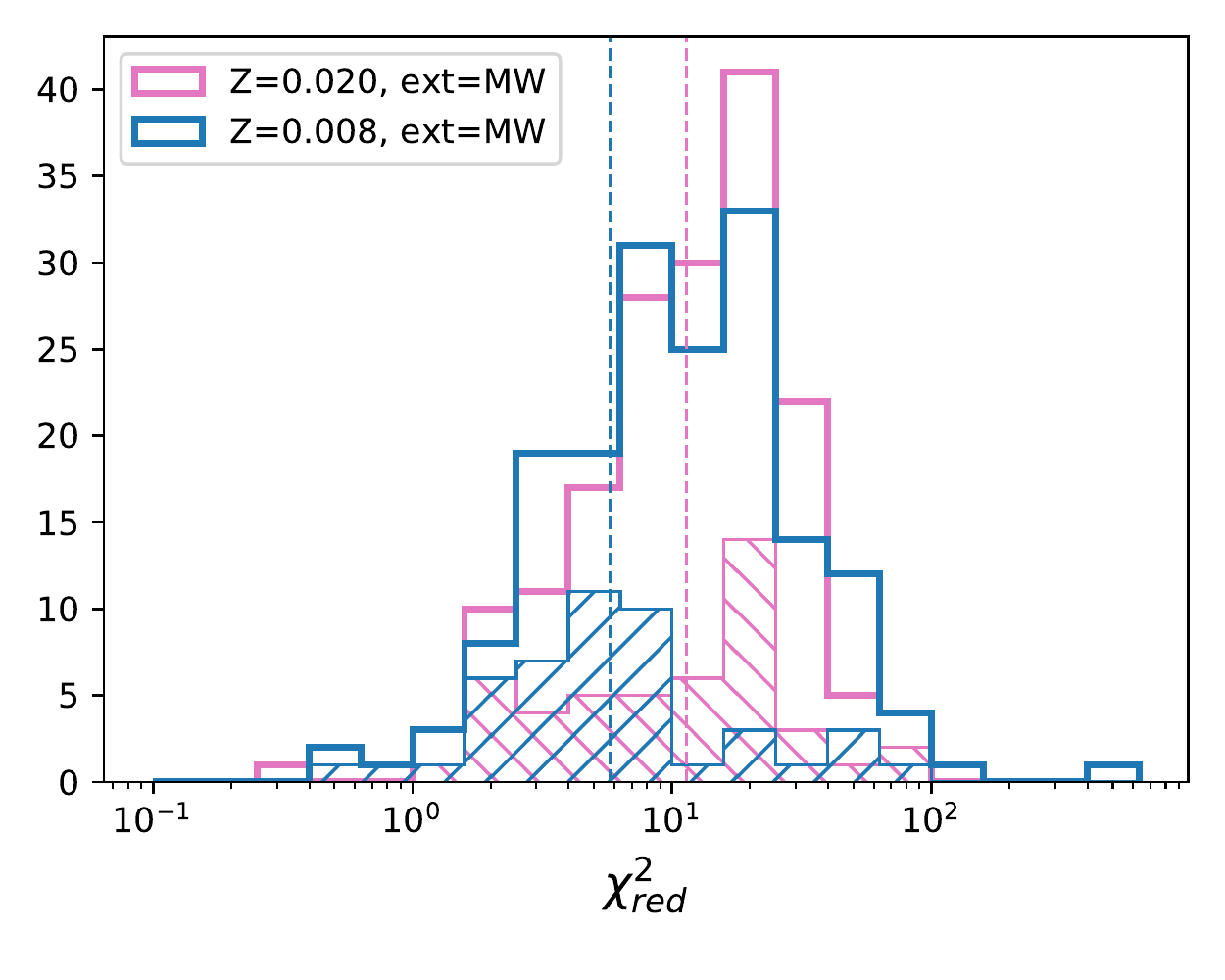}}
\subfigure{\includegraphics[width=0.49\textwidth]{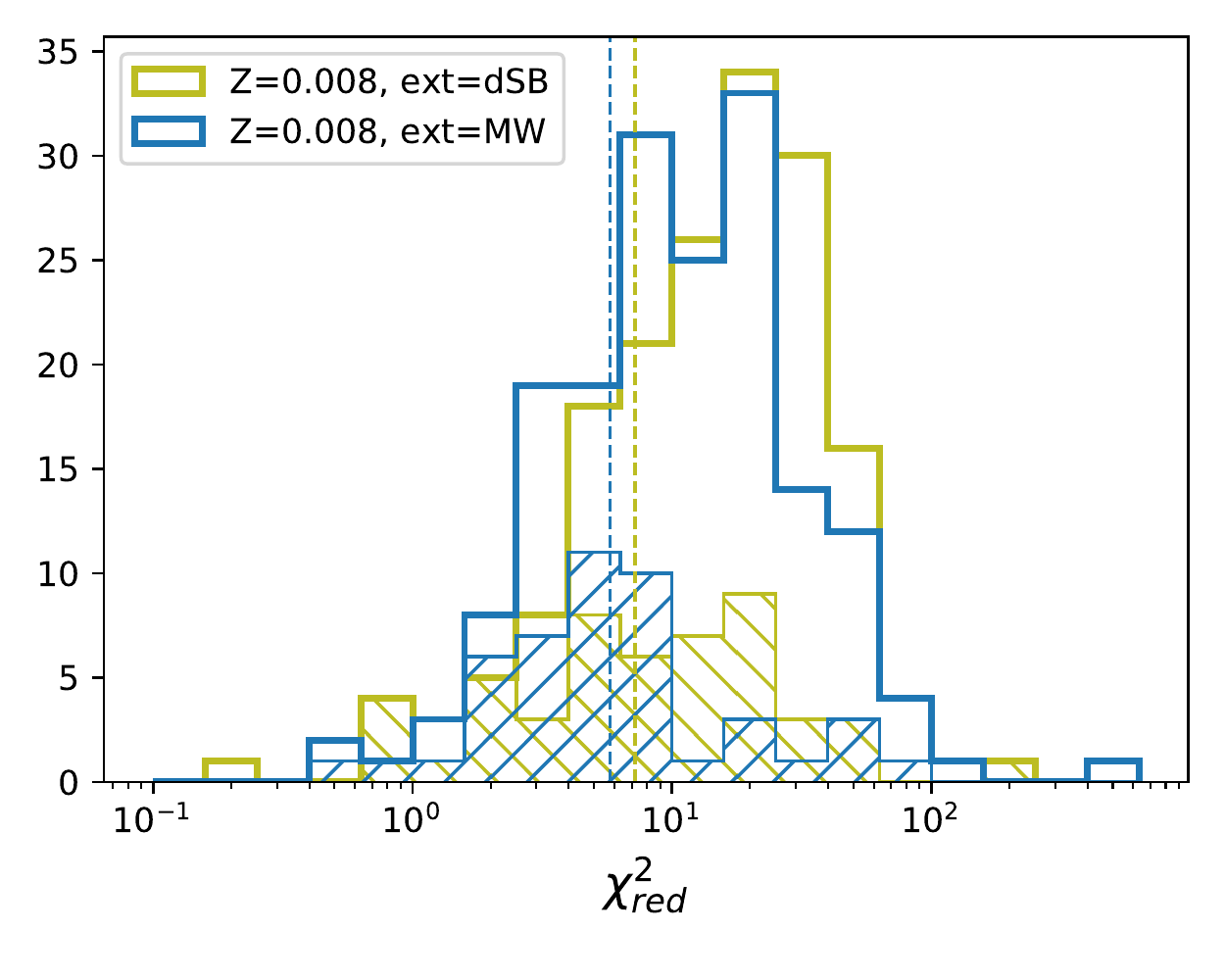}}
\caption{Distribution of the \chisq\ values from the best-fit results. Empty histograms refer to the total \textit{ExtmapCat} sample, hatched histograms to the \textit{ExtmapCat--final} sample. Vertical dashed lines are median \chisq\ values of the final samples, namely 5.8 for MW extinction with $\rm Z=0.008$, 11.4 for $Z=0.020$ and 7.2 for differential starburst extinction.}
\label{fig:chisq_extcat}
\end{figure*}

We consider SSP models with solar metallicity for the fit and select a final sample in the same way as described in the previous chapter. The final sample counts the exact same sources as of \textit{ExtmapCat--final}.
We compare the recovered ages and extinctions using the density plot in Fig.~\ref{fig:age_ebv_extcat} (central panel). 
On average, solar metallicity models recover slightly older and more extincted sources. 
The median $\chi^2_{red}$ recovered with the solar metallicity models is higher than the one of the reference sample (Fig.~\ref{fig:chisq_extcat}, left panel). An inspection of individual sources' SEDs confirms that sub-solar models provide better fit to the photometry; photometric data containing strong emission lines show an abrupt increase of the flux in the filter F555W compared to F547M, despite their pivot wavelength being very similar. This difference is caused by different bandwidths, allowing F555W to include strong emission lines such as $\rm O{[III]}$, in contrast to F547M. Different metallicities for the stellar models imply different strengths for the nebular lines and therefore different predicted fluxes, as shown for some examples in Fig.~\ref{fig:individual_met} in Appendix~\ref{sec:app_testmet}. We conclude that the availability of both F555W and F547M filters allows us to disentangle the model-metallicity that gives a better agreement with the data, in this case sub--solar, $\rm Z=0.008$\footnote{No observable difference was observed if a metallicity $\rm Z=0.004$ was considered.}. 

We repeat all the analyses, included the final selection, using sub--solar metallicity models but implementing a differential starburst attenuation. The final sample selected remains almost identical to the reference \textit{ExtmapCat--final}. The most notable difference is the presence of 2 additional sources as young-extincted cluster candidates, C33 and C175, whose $\chi^2_{red}$ distributions and best--fit fluxes are collected in Fig.~\ref{fig:individual_ext1}. 
The distribution of ages and extinctions shown in the density distribution plots in Fig.\ref{fig:age_ebv_extcat} reveals little difference from the reference case. On average, starburst extinction models recover younger and less extincted sources. The median $\chi^2_{red}$ recovered in this case is slightly larger than in the reference case (Fig.~\ref{fig:chisq_extcat}, right panel).
%

\subsection{Ages, extinctions and masses of PBcompactCat}\label{sec:pbcompact_results}
\begin{figure*}
\centering
\subfigure{\includegraphics[width=0.49\textwidth]{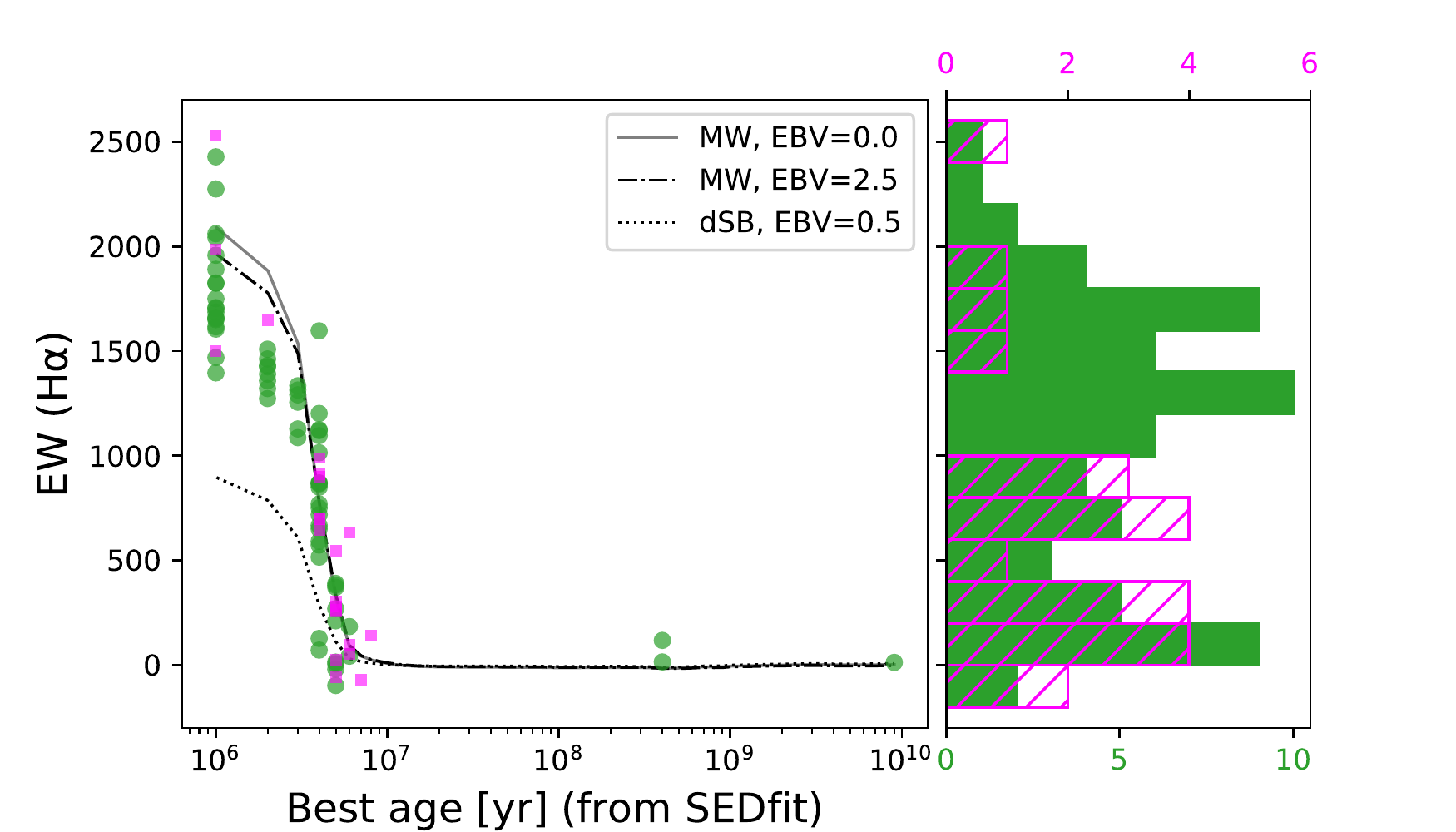}}
\subfigure{\includegraphics[width=0.49\textwidth]{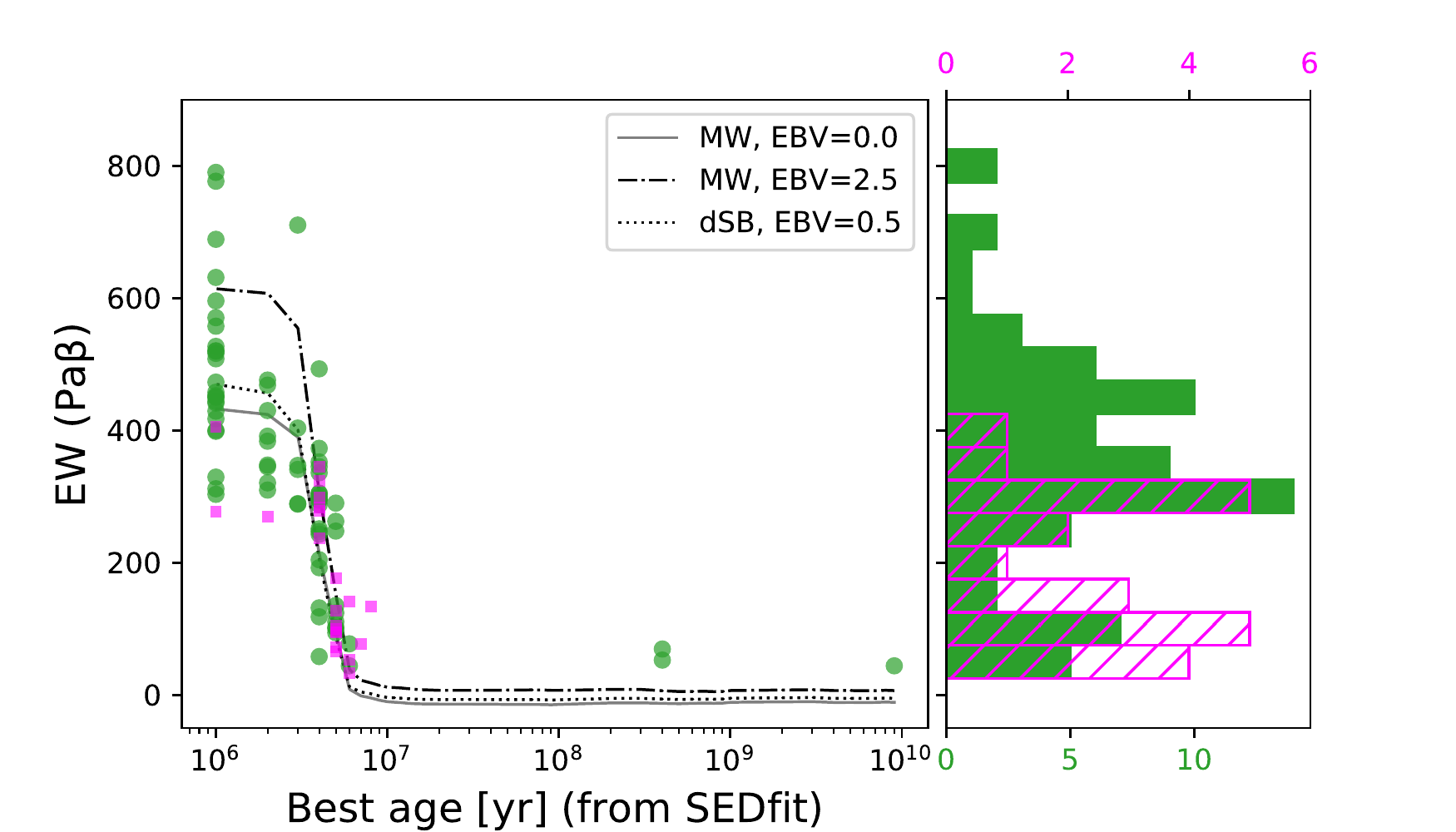}}
\caption{Same as Fig.~\ref{fig:ew_ages_extcat}, but for the \textit{PBcompactCat} sources. Orange color have been changed into magenta, blue into green.}
\label{fig:ew_ages_pbcat}
\end{figure*}

We show in Fig.~\ref{fig:ew_ages_pbcat} the best-fit age values for the sources in the \textit{PBcompactCat}, along with the EW of \ha\ and \pb, comparing them to the EW from the models, again color--coding them according to their light-profile (red for stellar--like, green for larger). 
As in the previous sample, we find good consistency between the ages derived from the SED fit and \ha\ and \pb\ EW values. We note that there are no sources with $\rm EW(Pa\beta)\sim0$, because of sample selection criteria. For the same reason, most of the sources with PSF--like light profiles have large \ha\ and \pb\ equivalent widths. 

\begin{figure*}
\centering
\subfigure{\includegraphics[width=\textwidth]{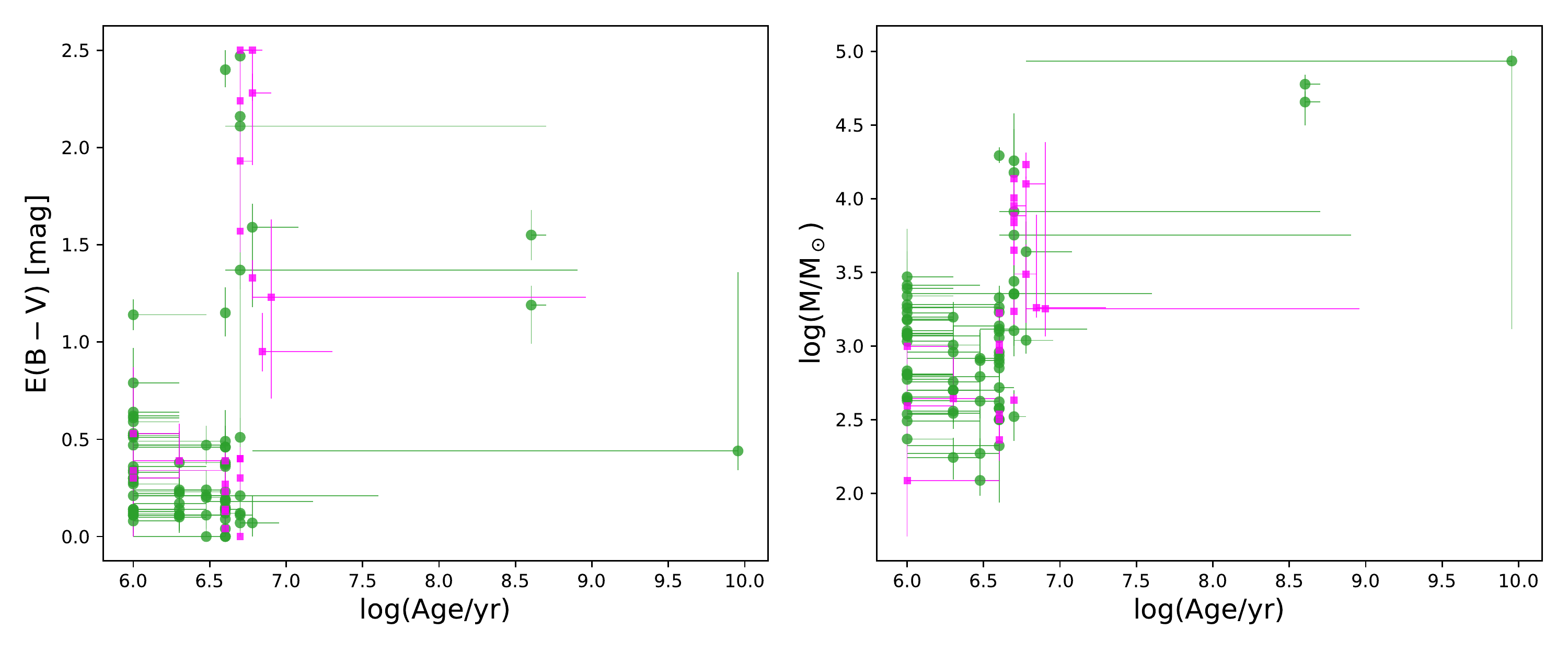}}
\subfigure{\includegraphics[width=\textwidth]{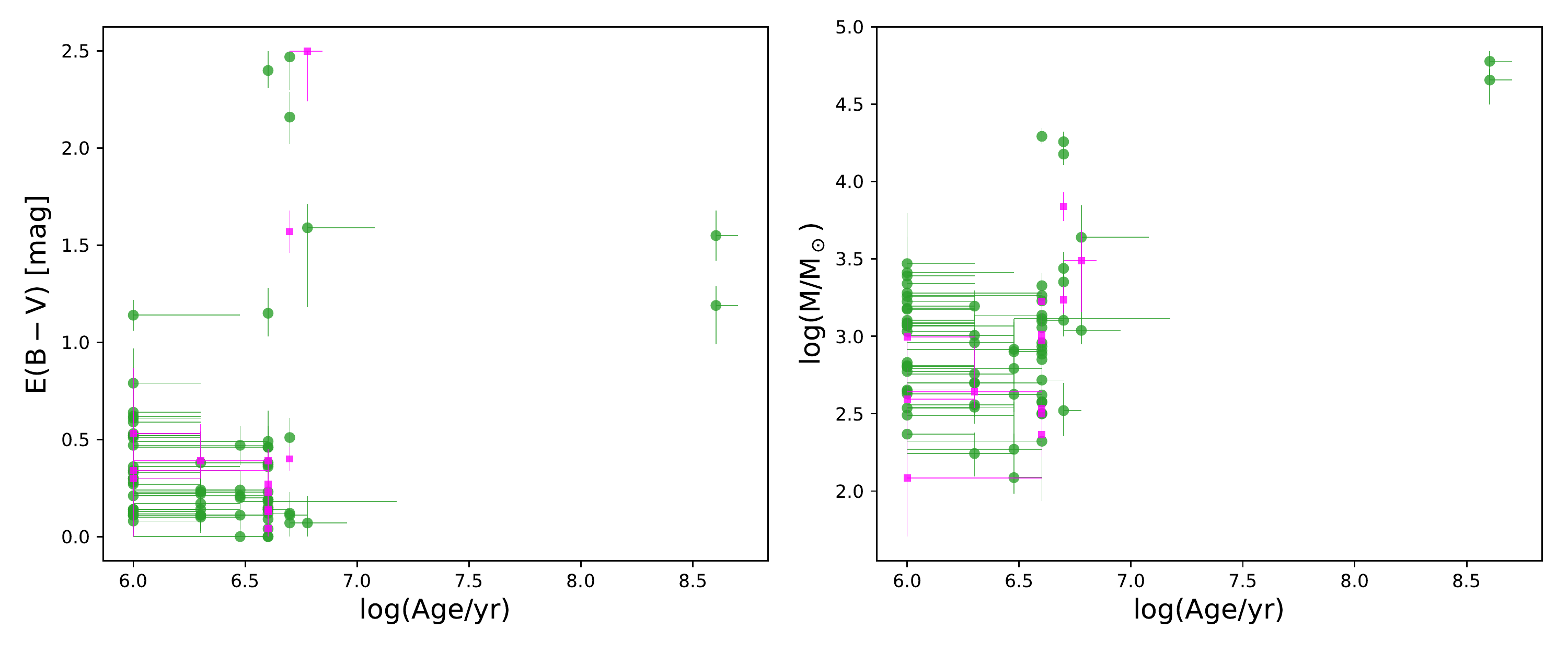}}
\caption{Same as Fig.~\ref{fig:age_ebv_extcat}, but for the \textit{PBcompactCat} sources. Blue color have been changed into green, orange into magenta. The scale of the X--axis in the bottom--row plots was adapted to the narrower range spanned by the data in the final sample. }
\label{fig:age_ebv_mass_pbcomp}
\end{figure*}

We show in Fig.~\ref{fig:age_ebv_mass_pbcomp} the best-fit values of ages, extinctions and masses, along with their uncertainties. The great majority of sources cluster around ages smaller than 10 Myr; we attribute this to the selection method of the sample, based on the observation of compact nebular emission and therefore biased towards young sources.
We implemented the same selection criteria used for cleaning the \textit{ExtmapCat}, reaching a final sample (\textit{PBcompactCat--final}) counting 84 sources, 30 of which are in common with \textit{ExtmapCat--final} (Tab.~\ref{tab:summary_samples}). These make most of the \textit{ExtmapCat}, with some of the excluded sources being the young sources without line emission described in the previous section. We deduce that \textit{PBcompactCat--final}  is a more complete version of \textit{ExtmapCat--final}. The distribution of \textit{PBcompactCat} across the galaxy is shown in Fig.~\ref{fig:coo_dist}. 

We analyze individually all the sources with $\rm E(B-V)>1.0$ mag in Fig.~\ref{fig:individual_ext2}. 
For some of them, the best--fit $\chi^2_{red}$ reaches a high value, above 50 (C203, C228, C233, C240, C274) suggesting inaccurate fit results. 
For two of the sources (C239 and C251), the photometry from F657N filter was not used in the fit due to high uncertainty; nevertheless its value suggest no \ha\ emission, in contrast with the NIR filters suggesting \pb\ emission.
Only three young--and--extincted sources, namely C213 (also included in the \textit{ExtmapCat--final} sample under the ID number C10), C236 and C307, are good candidates for being retained in our final cut.

\begin{table}[]
    \centering
    \begin{tabular}{|l|cc|}
    \hline
    \ & \textit{ExtmapCat} & \textit{PBcompactCat} \\
    \hline
    \hline
    total & 190 & 124 \\
    stellar & 136 (72\%) & 39 (31\%) \\
    resolved & 54 (28\%) & 85 (69\%) \\
    \hline
    \hline
    \ & \textit{ExtmapCat--final} & \textit{PBcompactCat--final} \\
    \hline
    \hline
    total & 46 & 84 \\
    stellar & 5 (11\%) & 13 (15\%) \\
    resolved & 41 (89\%) & 71 (85\%) \\
    \multicolumn{1}{|l|}{in common} & \multicolumn{2}{c|}{30} \\
    \hline
    \end{tabular}
    \caption{Summary of the number of sources in the initial extracted samples (\textit{ExtmapCat} and \textit{PBcompactCat}, Section~\ref{sec:catalogs}) and in the final samples (\textit{ExtmapCat--final} and \textit{PBcompactCat--final}, after the selection described in Section~\ref{sec:extmapcat_results}).}
    \label{tab:summary_samples}
\end{table}

\begin{figure*}
\centering
\subfigure{\includegraphics[width=\textwidth]{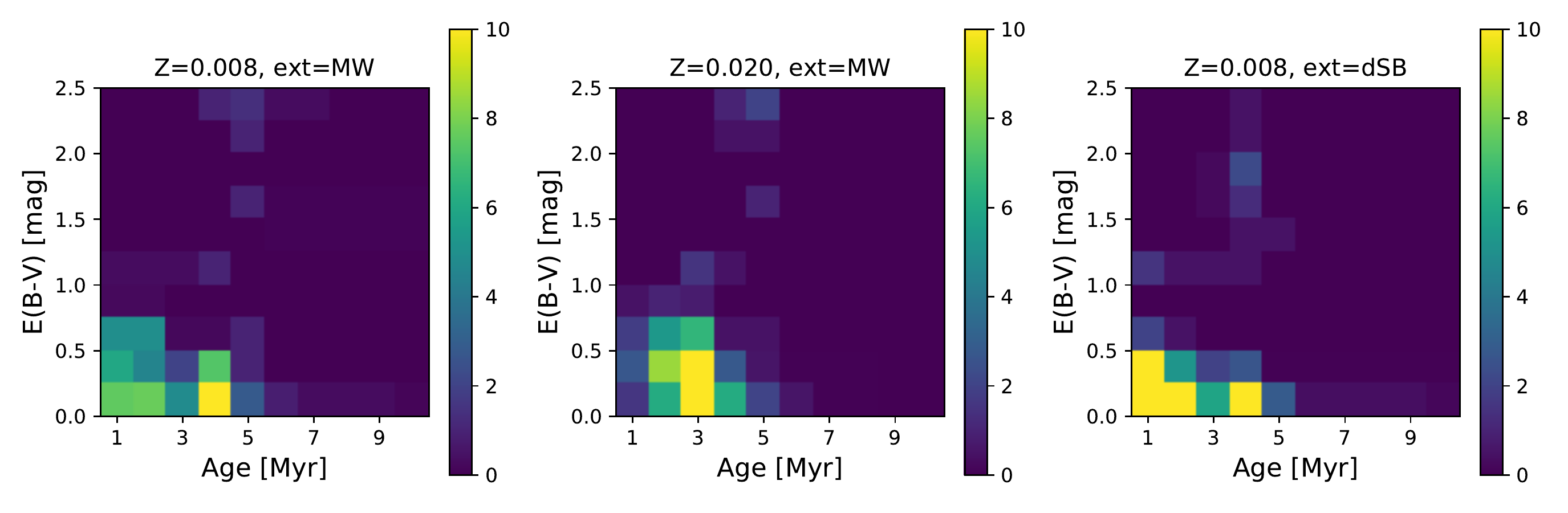}}
\subfigure{\includegraphics[width=0.49\textwidth]{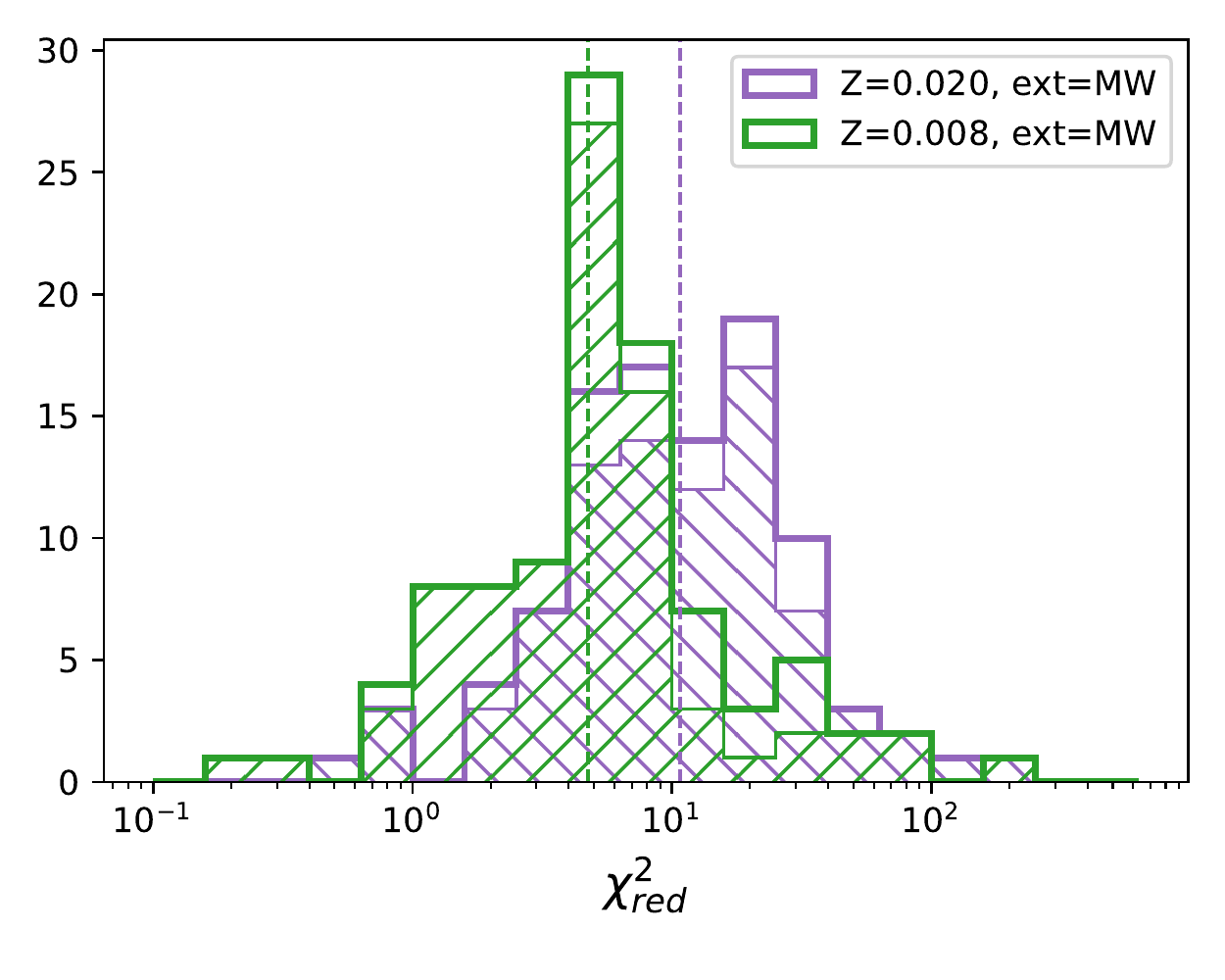}}
\subfigure{\includegraphics[width=0.49\textwidth]{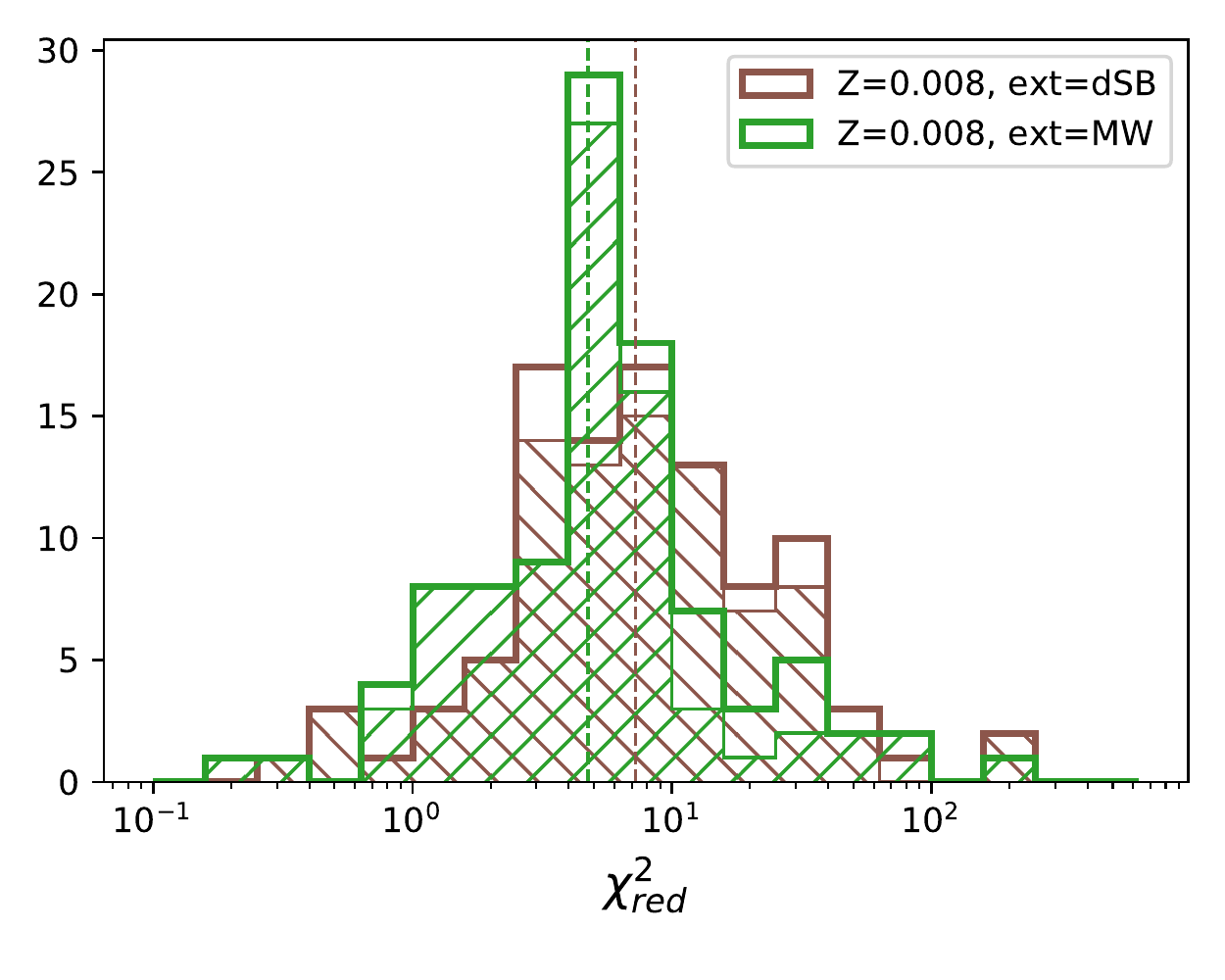}}
\caption{Same as Fig.~\ref{fig:age_ebv_extcat} (top panels) and as Fig.~\ref{fig:chisq_extcat} (bottom panels) for the \textit{PBcompactCat} sample. The \chisq\ values are 4.7 for MW extinction with $\rm Z=0.008$, 10.8 for $Z=0.020$ and 7.2 for differential starburst extinction.}
\label{fig:age_ebv_pbcomp}
\end{figure*}

We show the resulting ages and extinctions from models with solar metallicity and with differential starburst extinction in Fig.~\ref{fig:age_ebv_pbcomp} (top panels). The overall trends discussed in Section~\ref{sec:test_metext} are recovered also in the \textit{PBcompactCat--final} catalog, namely solar metallicity models predict older ages, while differential starburst extinction predicts younger ages and lower $\rm E(B-V)$ values. The reference models remain the most viable also in this case, with the lowest median \chisq\ values (Fig.~\ref{fig:age_ebv_pbcomp}, bottom panels).

\subsection{Sizes}
\begin{figure*}
\centering
\subfigure{\includegraphics[width=0.49\textwidth]{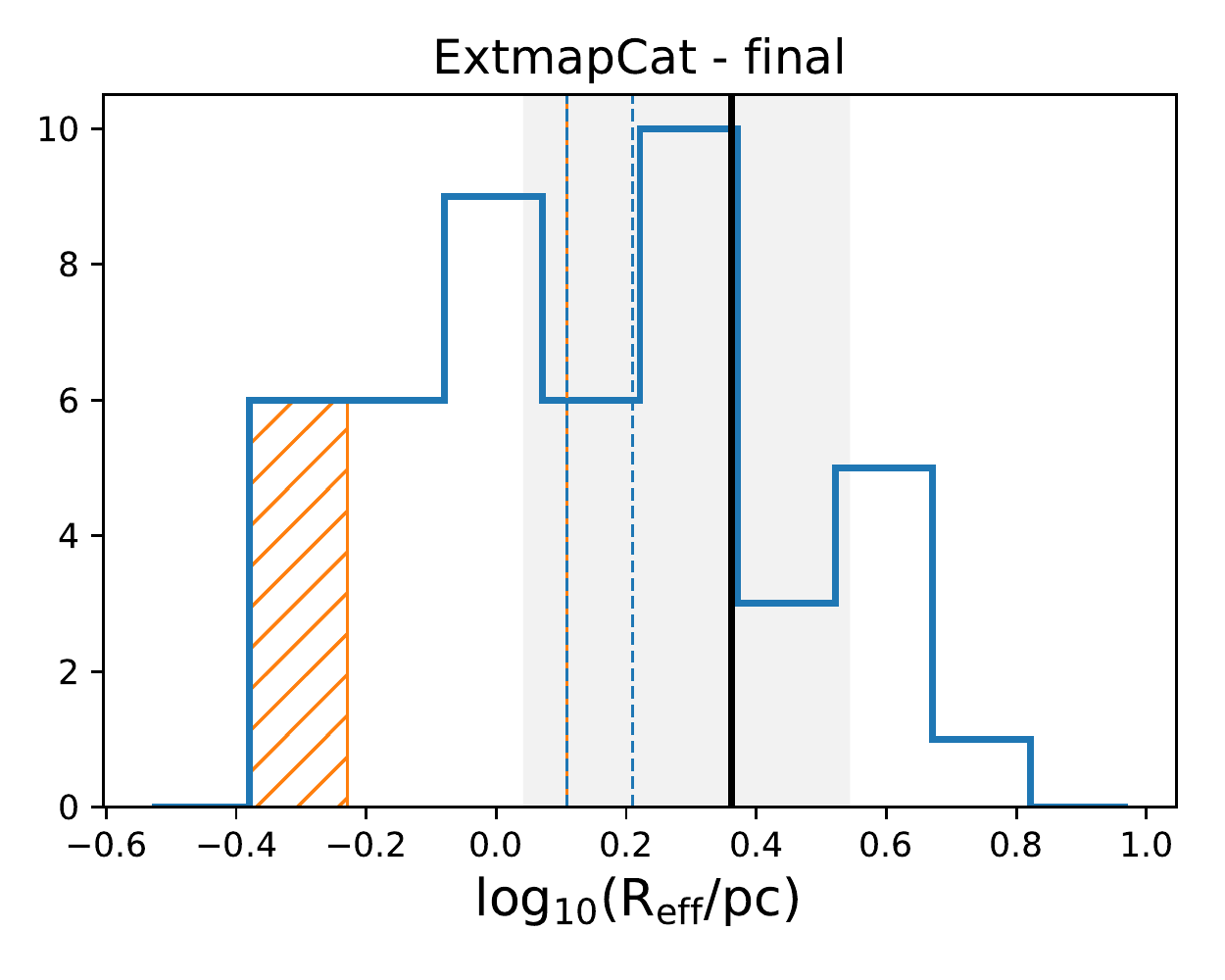}}
\subfigure{\includegraphics[width=0.49\textwidth]{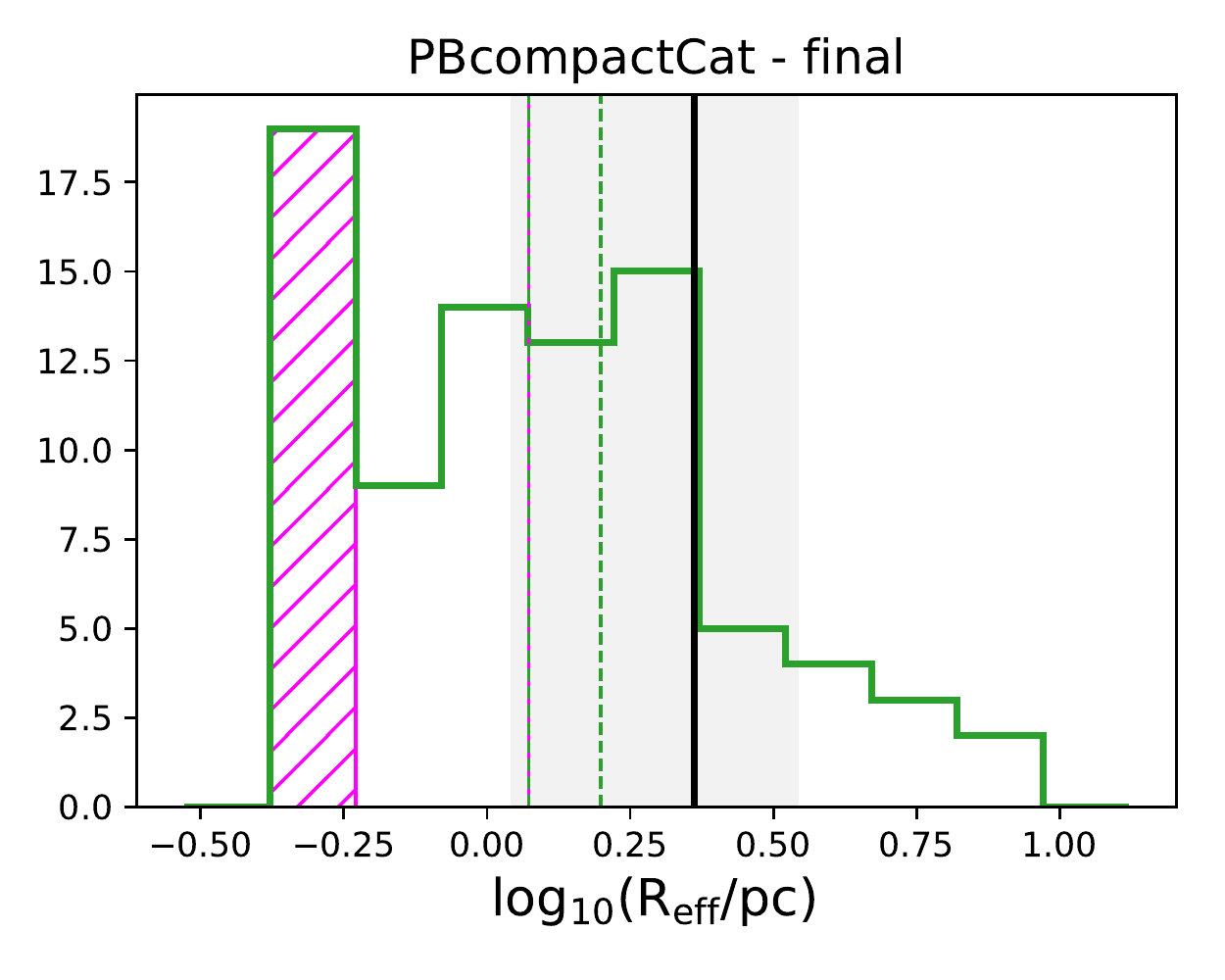}}
\caption{Distribution of sizes of the \textit{ExtmapCat--final} (left) and \textit{PBcompactCat--final} (right) samples. Sources with stellar-like PSF ($\rm R_{eff}<0.59$ pc) have been highlighted using orange and magenta hatched histograms and their sizes have been collapsed into a single bin. Both sources with stellar-like and with resolved profiles contribute to measure of the median sizes, showed as vertical lines: blue-orange and green-magenta for the samples including sources with stellar-like profiles, dashed blue and dashed green excluding them. 
The black solid vertical lines indicate the median $\rm R_{eff}$ of the NGC 1313 clusters in the LEGUS sample, with the gray shaded area marking its uncertainty range, as reported by \citet{ryon2017}.}
\label{fig:reffs}
\end{figure*}

The distributions of cluster sizes, derived in Section~\ref{sec:sizes-phot}, are showed in Fig.~\ref{fig:reffs}.
The distributions have a median at $\rm R_{eff}=1.28$ pc in the case of \textit{ExtmapCat--final} and $\rm R_{eff}=1.18$ pc in the case of \textit{PBcompactCat--final}. 

We can compare our results with those obtained by the study of LEGUS clusters sizes in NGC 1313 \citep{ryon2017}, that found a median effective radius of $2.3$ pc with a dispersion $\pm1.2$ pc. There are four major differences between our analysis and that of \citet{ryon2017}. First, the fitting method used is different (LEGUS used the software \texttt{GALFIT}, \citealp{galfit2002,galfit2010}). Second, the LEGUS analysis was limited to a sample of clusters with masses above $\rm 5000\ M_\odot$ and younger than $200$ Myr. Only 9 sources from our final catalogs reside in that range, while most of our sources have lower masses. Third, we use the F814W filter as the reference for fitting the size, while LEGUS used F555W. 
Finally, the analysis of \citep{ryon2017} is based on the LEGUS catalog and therefore ignores sources with stellar--like PSF (see also \citealp{adamo2017}); for a better comparison we can re-estimate our median values excluding unresolved sources, finding $\rm R_{eff}=1.62$ pc and $\rm R_{eff}=1.58$ pc for \textit{ExtmapCat--final} and \textit{PBcompactCat--final}, respectively.
While the difference in filters should not affect the resulting median sizes (assuming that the size is the same when observed in different bands), the difference in the age--mass range considered makes almost impossible a detailed comparison between the two results. Despite those differences, our values are consistent with \citet{ryon2017} results within their uncertainties.

\section{Discussion}\label{sec:discussion}
\subsection{Mass estimates}\label{sec:masses}
We performed aperture photometry implementing an average aperture correction that assumes all clusters to be well--described by Moffat light profiles with $\rm R_{eff} = 2.5$ pc. While the assumption is justified by studies of cluster sizes in NGC 1313 \citep{ryon2017}, we discuss here whether using the individual size of each cluster in the photometric aperture correction would produce different mass estimates. We recall that different aperture corrections lead to different normalizations of the photometry, and therefore affect only mass estimates, leaving the recovered ages and extinctions unchanged.

\begin{figure*}
\centering
\subfigure{\includegraphics[width=0.49\textwidth]{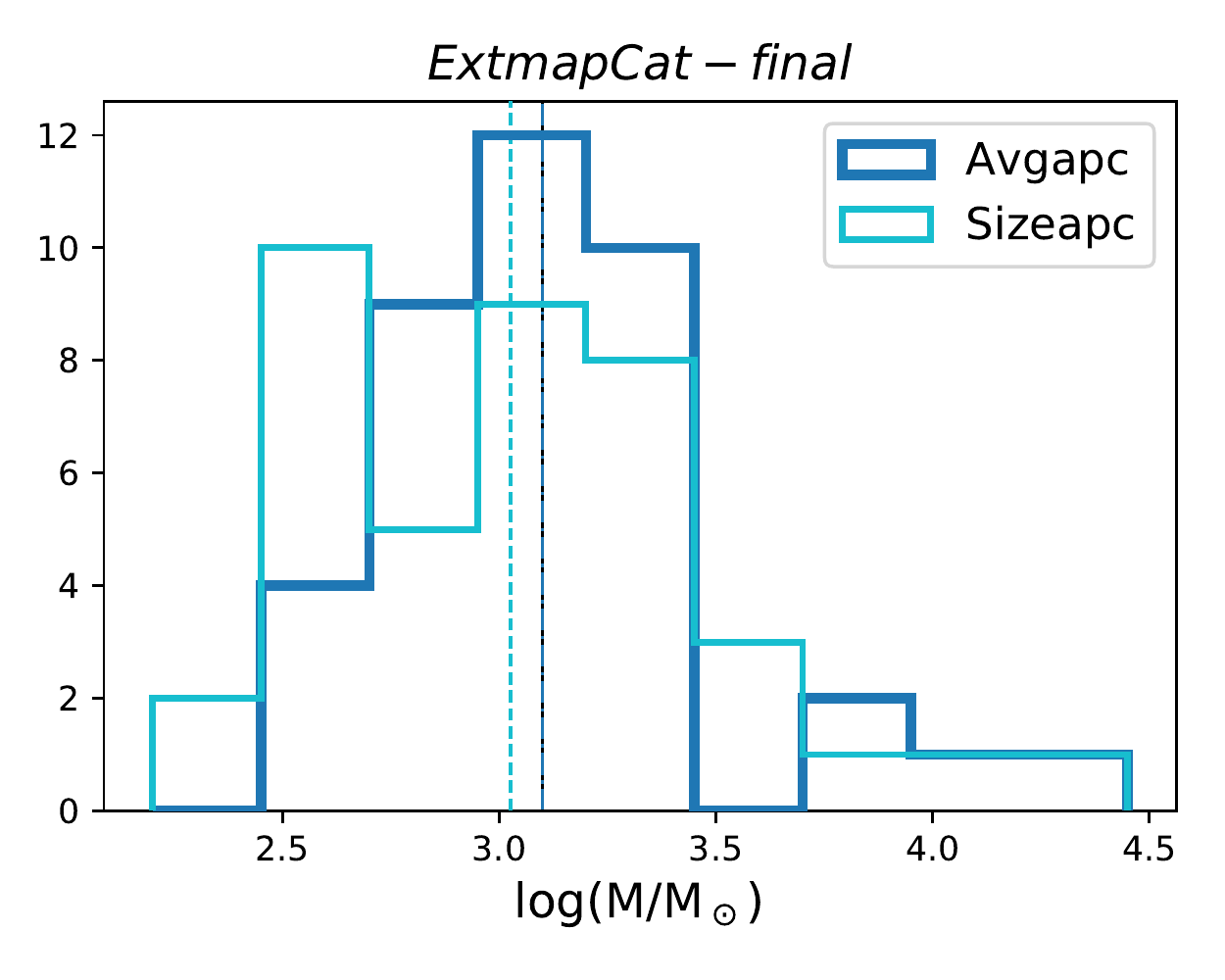}}
\subfigure{\includegraphics[width=0.49\textwidth]{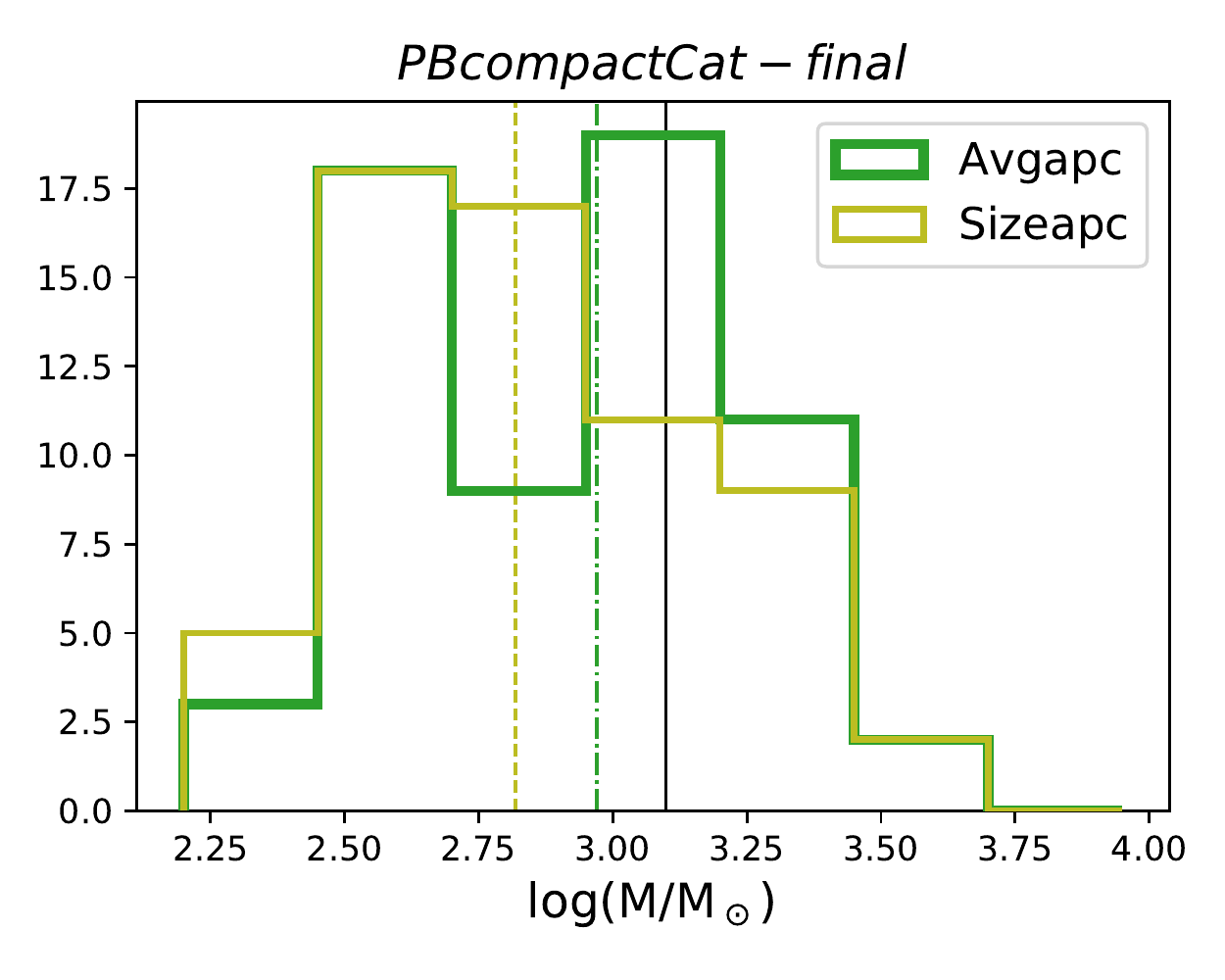}}
\caption{Mass distributions of the \textit{ExtmapCat--final} sample (left panel) and \textit{PBcompactCat--final} (right panel). Two distributions are shown for each catalogs, one referring to photometry with average aperture correction (avgapc), the other referring to the size-photometry analysis (sizeapc). Vertical lines shows their median distributions and the black vertical line is the median mass of the NGC 1313 LEGUS sample in the age range 0-6 Myr.}
\label{fig:mass_comparison}
\end{figure*}

The procedure for deriving photometric results considering the individual sources' sizes was described in Section~\ref{sec:sizes-phot}. We show in Fig.~\ref{fig:mass_comparison} the distributions of masses for \textit{ExtmapCat--final} and \textit{PBcompactCat--final} derived with the two photometric analysis. The mass distribution derived via the aperture photometry analysis is peaked at a slightly larger median value, for both catalogs. This is consistent with having found median effective radii of 1.28 and 1.18 pc for the \textit{ExtmapCat--final} and \textit{PBcompactCat--final} samples, respectively, smaller than $\rm R_{eff} = 2.5$ pc used as reference for the average aperture correction. However, we note that the overall distributions are very similar, and we conclude that the mass distribution is not strongly affected by our choice of the average aperture correction. 

We compare the median masses from our sample to the ones of the NGC1313 LEGUS cluster sample. For this comparison, we select from the LEGUS catalogs only sources with ages $\le6$ Myr (the range covered by our final samples), and with visual class 1, 2 or 3, hence avoiding class 4 (that according to the LEGUS classification contains non--clusters, see \citealp{adamo2017}). The median mass of the LEGUS sample, $\rm M=1256\ M_\odot$ is indicated in Fig.~\ref{fig:mass_comparison} by a black vertical line. This value is identical to the median mass of our \textit{ExtmapCat--final} sample when considering the aperture photometry analysis and is only slightly higher than the median mass of the \textit{PBcompactCat--final} sample, $\rm M=935\ M_\odot$.

As mentioned in Section~\ref{sec:extmapcat_results}, the low cluster masses we are considering raise the problem of stochasticity, i.e. the stellar IMF of some of our clusters may not be fully sampled, as instead assumed by the models used for the SED fit (Section~\ref{sec:fit}). 
On the other hand, we are considering clusters that power ionised gas, that need to host stars more massive than $\rm 8 M_\odot$, therefore mitigating the stochasticity problem.

\subsection{Are we missing sources?}
\subsubsection{Completeness of the samples}\label{sec:completeness}
Only a few young clusters from our final samples have extinctions $\rm E(B-V)>1$ mag. The completeness test of Section~\ref{sec:completeness_set} revealed that we expect less than $10\%$ of sources with extinction up to $\rm E(B-V)=1.5$ mag to be missed by our photometric analysis. We discuss now the possibility that the lack of sources with $\rm E(B-V)>1$ mag can is due to imprecision in recovering properties from the SED fit. 

\begin{figure*}
\centering
\includegraphics[width=0.49\textwidth]{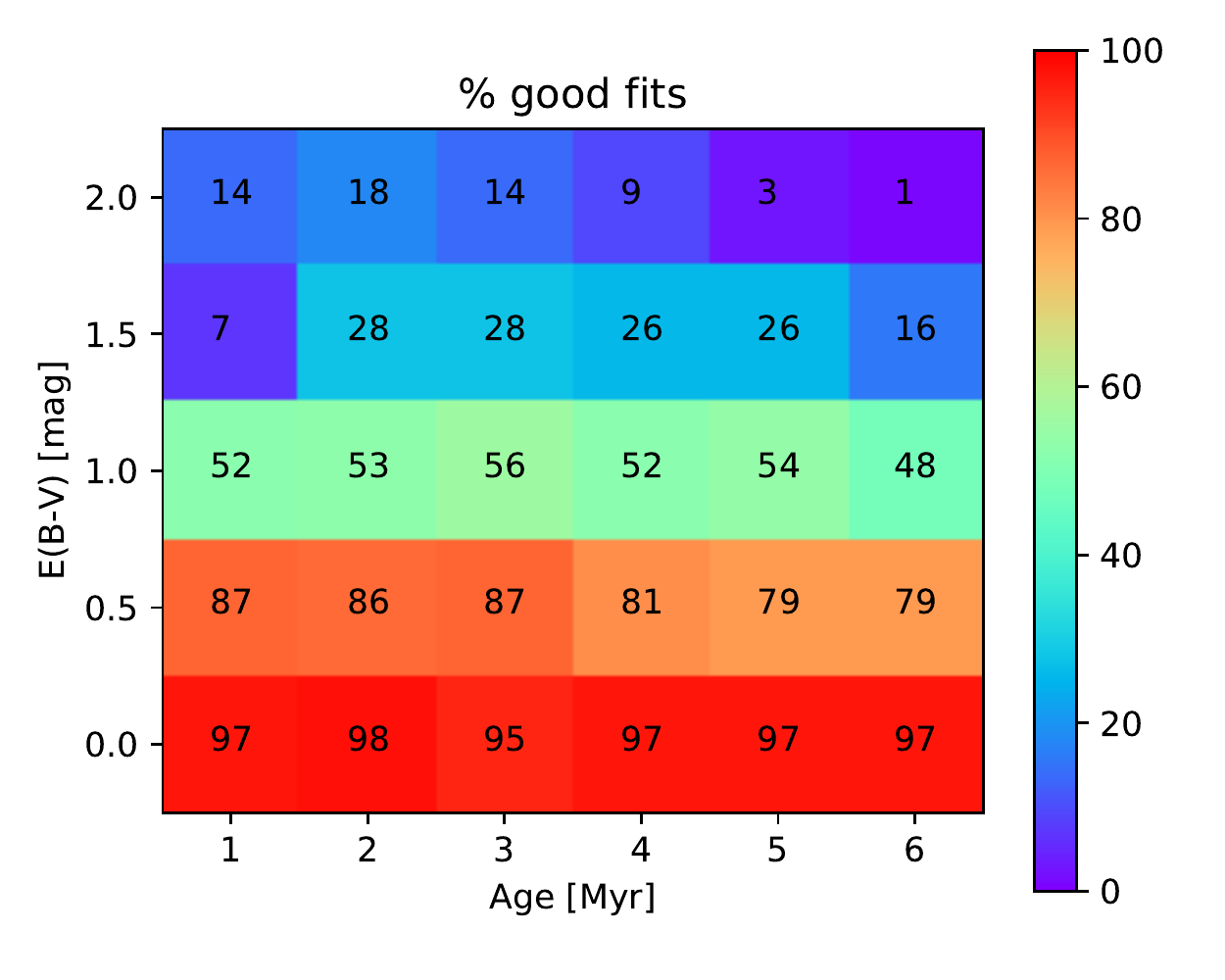}
\caption{Percentage of sources with retrieved good fit, for all the combinations of simulated ages and extinctions. We consider a source to have good fit if its input age and extinction values fall into the uncertainty range given by the SED fitting process.}
\label{fig:completeness_goodness}
\end{figure*}

\begin{figure*}
\centering
\includegraphics[width=0.60\textwidth]{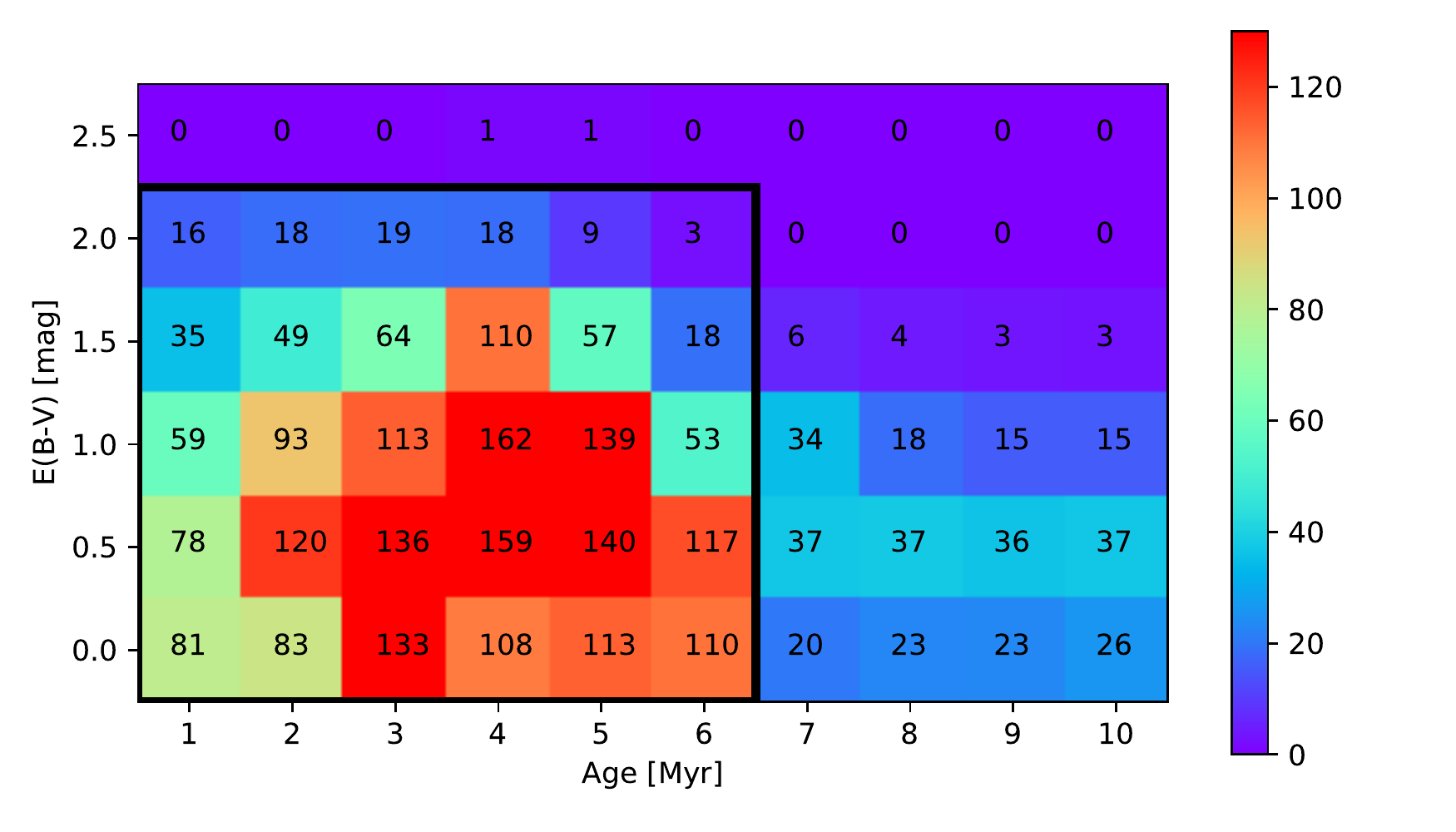}
\caption{Density plots, as in Fig~\ref{fig:age_ebv_extcat} and Fig~\ref{fig:age_ebv_pbcomp}, for the sources created for the completeness test. The input sources were simulated only in the age range [1;6] Myr and $\rm E(B-V)$ range [0.0;2.0] mag, but a larger grid is plotted to account for uncertainties and fit results extending outside the initial grid (marked by a thick solid black contour). This plot indicate what are the ages and extinctions favoured (or disfavoured) by the fitting process.}
\label{fig:completeness_density}
\end{figure*}

We consider the simulated sources from the completeness test described in Section \ref{sec:completeness_set} and we collect the results of their SED fitting; in Fig~\ref{fig:completeness_goodness} we build a matrix showing the fraction of retrieved sources with good fits. We consider good fits the cases where the original age and extinction are consistent with the derived values within the uncertainties. The matrix proves that we have completeness above $80\%$, for sources with low extinction (up to 0.5 mag). The completeness decreases to $\sim50\%$ for $\rm E(B-V)=1.0$ mag and to $\sim25\%$ for $\rm E(B-V)=1.5$ mag. Typically a source can be detected up to an optical depth $\tau\sim$1, which in our case is reached for E(B--V)$\sim1.3$~mag at the  wavelength of \pb. Our findings are consistent with this expectation. In Fig.~\ref{fig:completeness_density} we plot all the derived ages and masses using the method of Fig~\ref{fig:age_ebv_extcat} and Fig.~\ref{fig:age_ebv_pbcomp}, in order to find if some combinations of ages and extinction are favored by the SED-fitting procedure. The figure shows that, overall, ages around 5 Myr are favored by the fitting process compared to younger ages. On the other hand, in our data we don't see many sources with age $\sim4/5$ Myr and extinctions of $\rm E(B-V)\sim0.5/1.0$ mag, therefore we have not over--predicted them.

Fig.~\ref{fig:app_completeness_density} in Appendix~\ref{sec:app_completeness} shows individual age-extinction density plots for each input age; for input ages of 4, 5 and 6 Myr we recover consistent ages, while our fitting process cannot clearly distinguish ages of 1, 2 and 3 Myr. This limitation is related to models having very similar SEDs for such young ages. 

As final note, we remind that the approach used to extract our samples, limit the detection to clusters that host stars massive enough to ionised Hydrogen. As consequence, we are practically blind to clusters that do not have massive stars.


\subsubsection{Extincted sources in our samples} \label{sec:discuss_single_extincted}
From our analysis we recover a total of 13 sources\footnote{this total come from considering both \textit{ExtmapCat--final} and \textit{PBcompact--final} samples and including also the best-fit values coming from the models with differential starburst extinction.} whose best fit properties fall in the ranges of ages$\le 6$ Myr and $\rm E(B-V)>1.0$ mag. Their best fits (distributions of \chisq\ values and photometry) are shown in Fig.~\ref{fig:individual_ext1} and Fig.~\ref{fig:individual_ext2} in Appendix~\ref{sec:ind_fit_extincted}. 
We note that not all of the best-fit values can be considered as robust, as suggested by the \chisq\ values, which in some cases is very high ($>50$).
In particular, C10 and C307 are the two most robust cases, as their best fits follow the overall trend of the broadband SED and recover the line emission. For some other sources the least--\chisq\ values are small and therefore we consider the fit reasonable, but the best--fit SEDs clearly miss to reproduce the flux in one band: F547M in the case of C175--C236 (if starburst extinction is considered), F435W in the case of C251. All the other cases are less robust either because of a limited number of filters with detections (e.g. C93--C272, C187--C239, C233) or because of large least--\chisq\ values (e.g. C142, C228, C274). We note that all those very extincted candidates are clearly displaced from the main distribution of extinctions, that extends up to $\rm E(B-V)\sim1.0$ mag (see Fig.~\ref{fig:age_ebv_mass_extcat} and Fig.~\ref{fig:age_ebv_mass_pbcomp}). In addition they all have ages between 4 and 6 Myr, while naively we would expect the most extincted sources to be the youngest. 
We kept all the discussed candidates in our final samples because the presence of line emission in all of them indicates they are young. However, we discuss in Section~\ref{sec:ebv_distr} how the properties of the sample would change when excluding them.

In both our samples, a fraction of the sources were not fitted due to the lack of enough photometric detections. If those were young and very extincted clusters we would be able to see their line emission, as the extinction has only small impact on \ha\ and \pb\ equivalent widths (see Fig.~\ref{fig:ew_theoretical}). As described in Section~\ref{sec:extmapcat_results}, we take into account in our analyses all sources with high EW values, and we therefore took care to not to leave out of the final samples possible young and extincted candidates.

The completeness analysis of Section~\ref{sec:completeness} suggests that we should find approximately half of the $10^3\ M_\odot$ clusters with $\rm E(B-V)\sim1$ mag, if they were there. We expect proto--clusters to be deeply embedded into their natal cloud of gas and dust; such sources can only be seen through their MIR--FIR emission, and their investigation is beyond the scope of this study. However, we expect them to start forming massive stars and later to clear their cloud. 
Different concurrent factors may cause the absence of young sources with color excess in the range $\sim1.0-2.5$ mag. First of all, it could simply be the case that our data are not sufficiently deep for the observation of such sources. The completeness analysis just discussed, however, rules out this possibility, as even if there is incompleteness for high extinctions, we still should be able to observe $\sim$half of the clusters with $\rm E(B-V)\sim1.0$ mag. We remind that the comparison to the \textit{Spitzer} $\rm 8\ \mu m$ and to the CO emission maps revealed some compact regions missed by our extinction map and therefore by our source selection.
Such regions could be associated to proto-stars/proto-clusters. If this is the case, their study at high spatial resolution will be made possible by the advent of the James Webb Space Telescope.

Another possibility is that GMCs in NGC 1313 have low surface densities, and consequently clusters form in relatively low--density environments. The extinction map derived in Section~\ref{sec:catalogs} (see Fig.~\ref{fig:maps_ext_spitzer}) indeed suggests that the most extincted regions of the nebular emission have $\rm A_V\sim3.0$ mag, consistent with the high--end of the color-excess main distributions at $\rm E(B-V)\sim1.0$ mag derived in Section~\ref{sec:results} (see Fig.~\ref{fig:age_ebv_mass_extcat} and \ref{fig:age_ebv_mass_pbcomp}). 
An observed color excess $\rm E(B-V)=1.0$ mag corresponds to a gas screen with density of $\rm 150\ M_\odot/pc^2$ (following the prescription by \citealp{bohlin1978} adjusting for sub--solar metallicity); this estimate assumes a gas screen only between the source and the observer and constitute therefore only a lower-limit estimate. 
ALMA observations indicate that CO clouds in NGC1313 have surface densities ranging from $\sim10$ to $\rm \sim300\ M_\odot/pc^2$, with median value $\rm \sim50\ M_\odot/pc^2$ (a detailed description of the ALMA observation in NGC1313 and its analysis will be presented in a coming paper, Finn et al., in prep.), consistent with the expectations estimated from the color excess and confirming the hypothesis of clusters form in relatively low--density environments. 

Finally, lack of embedded clusters could be due to (or enhanced by) an extremely short duration of the process of gas clearing due to feedback after the massive stars are formed (as suggested by e.g. \citealp{matthews2018,hannon2019,kruijssen2019,chevance2020a}). We discuss in the next section the typical timescales associated with the clearing of the natal gas cloud.
We point out that in the case of non--uniform clouds and therefore in the presence of holes that give clear lines of sight into the cluster, the fit values will tend to show lower extinction. Such patchy clouds could still be the effect of stellar feedback since early times \citep[see e.g.][]{dale2014}.

\subsection{Typical cluster extinctions at young ages}\label{sec:avg_ext}
\subsubsection{Distribution of extinctions}\label{sec:ebv_distr}
\begin{figure*}
\centering
\subfigure{\includegraphics[width=0.49\textwidth]{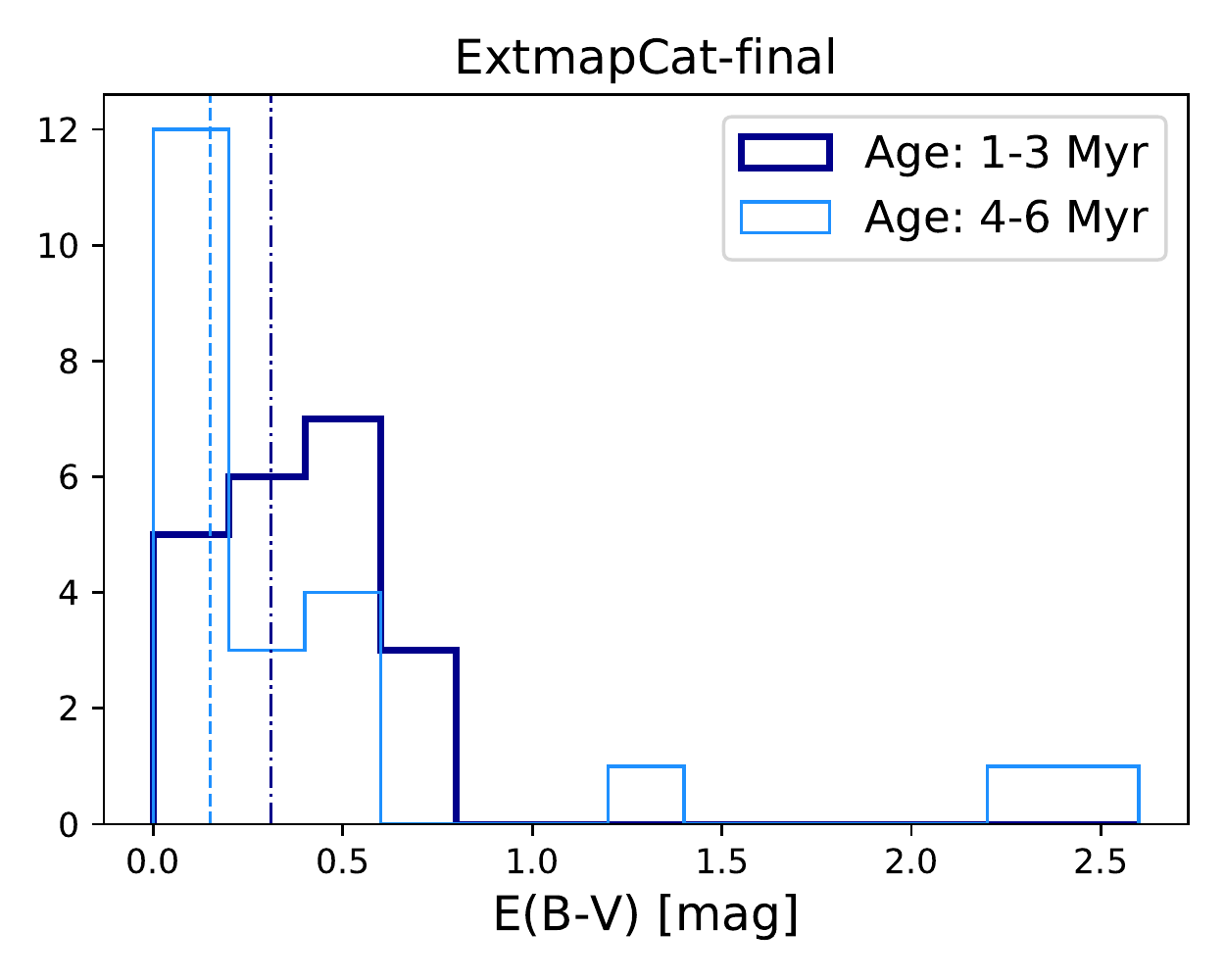}}
\subfigure{\includegraphics[width=0.49\textwidth]{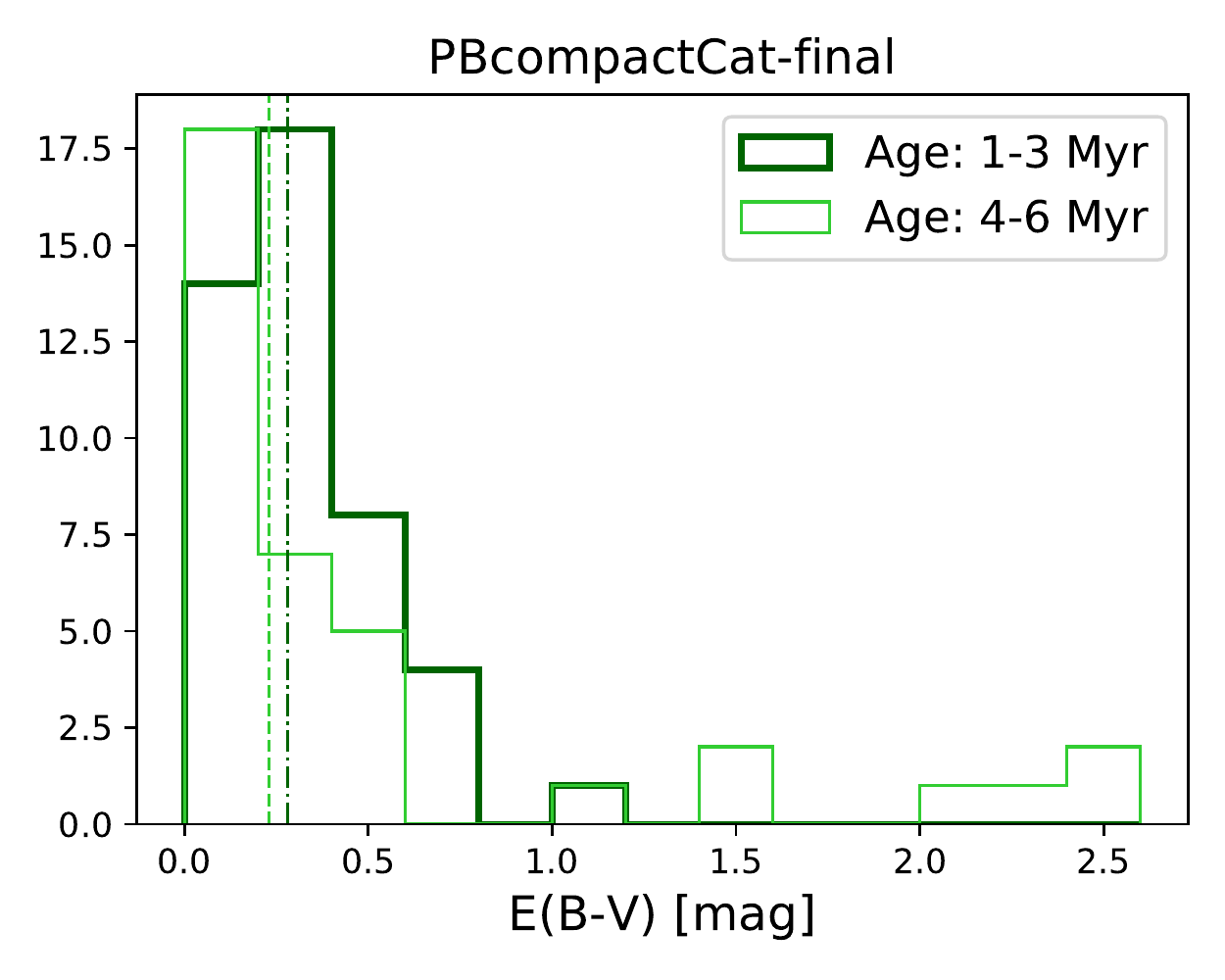}}
\caption{Distributions of extinctions for the \textit{ExtmapCat--final} (left) and the \textit{PBcompactCat--final} (right) samples. Dark colors refer to the age range 1-3 Myr, light colors for 4-6 Myr. The dot--dashed and dashed vertical lines are the median values for the two age samples, respectively.}
\label{fig:ebv_distr}
\end{figure*}

We consider in Fig.~\ref{fig:ebv_distr} the distribution of extinctions, separating the samples in two age ranges 1 to 3 and 4 to 6 Myr, to test for the presence of an age evolution. The median \ebv\ values are 0.31 and 0.15 mag for \textit{ExtmapCat--final} and 0.28 and 0.23 mag for \textit{PBcompactCat--final}, for the age ranges 1-3 and 4-6 Myr, respectively. An age--dependent extinction is suggested for the first sample, with twice the color  excess for the youngest age bin than the older age bin, while the second sample does not show a trend, rather consistent values between the two age bins. We note that, despite having selected the sample to contain extincted clusters, half of our sources are consistent with having relatively low extinctions, $\rm E(B-V)<0.25$ mag, corresponding to $\rm A_V\lesssim0.75$ mag. 

An alternative way of studying the typical extinctions of our samples is to consider the fraction of clusters below a certain limit. In Tab.~\ref{tab:f025} we report the fraction of clusters with \ebv\ values below 0.25 mag, $\rm F_{<0.25}$, i.e. the values of the lowest row in the density plot in Fig.~\ref{fig:age_ebv_extcat} and Fig.~\ref{fig:age_ebv_pbcomp} divided by the total value in each column. These can be considered the fraction of low-extincted clusters. Once again, we separate the samples in two age bins. 

The fractions varies from $41\%$ to $63\%$ for \textit{ExtmapCat--final} and from $45\%$ to $57\%$ for \textit{PBcompactCat--final}. As before, the \textit{ExtmapCat--final} sample shows a slightly stronger age--dependent extinction trend than the other sample. 

\begin{figure}
\centering
\includegraphics[width=0.49\textwidth]{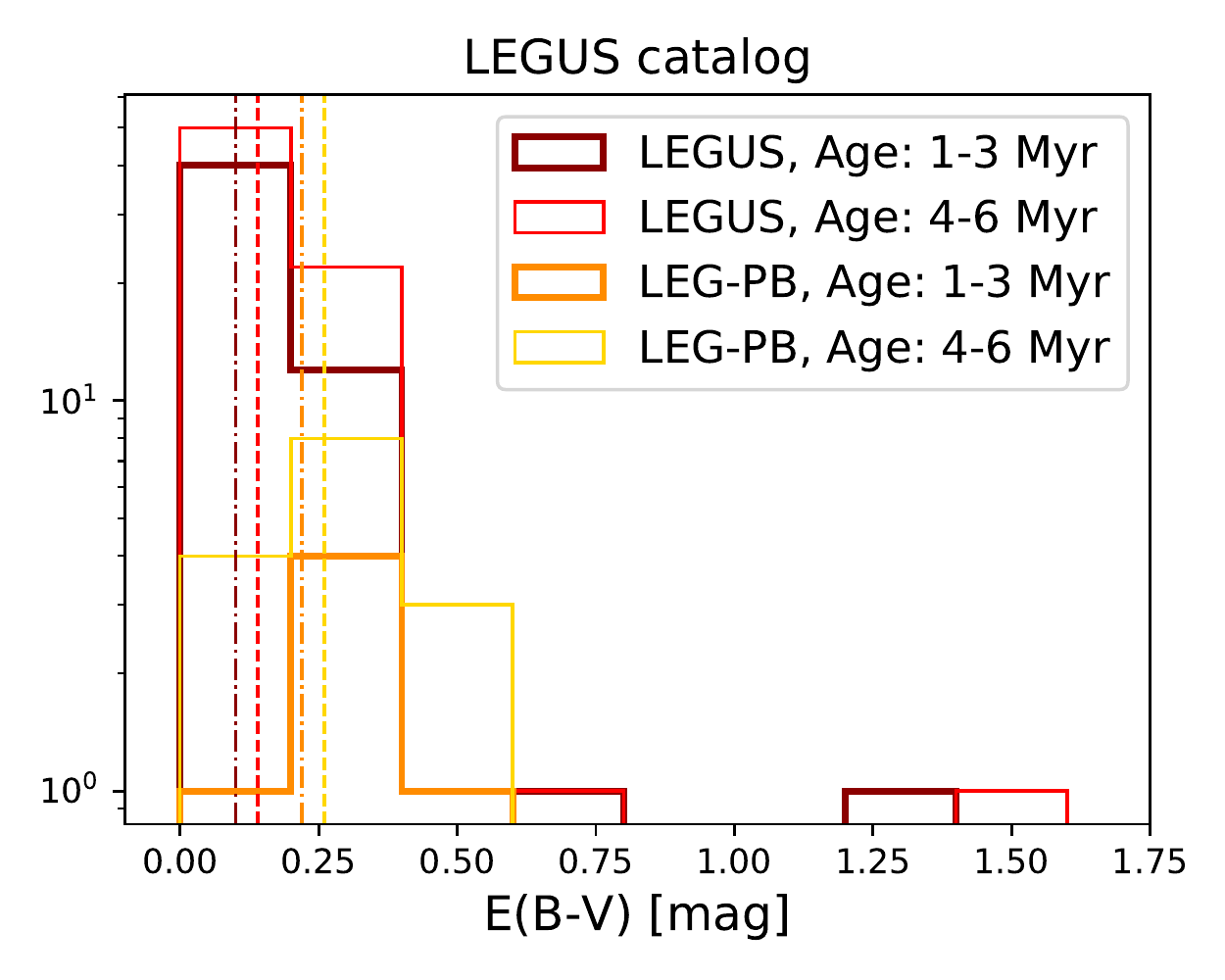}
\caption{Distribution of extinction for the LEGUS sample in two age ranges, 1-3 Myr and 4-6 Myr (dark red and red, respectively). The vertical dot-dashed and dashed lines are the respective median values. The same distributions and medians for the sub--sample of sources in common with our samples (named here LEG-PB) are plotted with orange and yellow colors.}
\label{fig:ebv_distr_legus}
\end{figure}

We can compare our samples to the cluster sample found by the LEGUS collaboration. We recall that the LEGUS sample selection is based on source detection in at least 4 filters in the NUV-optical range. The \ebv\ values of LEGUS clusters with ages $\le6$ Myr are shown in Fig.~\ref{fig:ebv_distr_legus} and the fraction of low-extinction clusters are collected in Tab.~\ref{tab:f025}. Both the median values, 0.10 and 0.14 mag for the 1-3 and 4-6 Myr age ranges, respectively, and the fraction of low-extinction sources, $84\%$ and $75\%$ in the two age bins, respectively, indicate lower extinctions for the LEGUS clusters compared to our sample. This suggests, as expected, that the process of the sample selection in LEGUS is biased against clusters with higher extinctions.

There are 21 sources in common between the LEGUS and our samples. We consider their \ebv\ distribution and fraction of low-extinction sources in Fig~\ref{fig:ebv_distr_legus}; their median \ebv\ values (0.22 and 0.26 mag) are consistent with the values from the overall \textit{PBcompactCat--final} sample. The same is true for the fraction of low-extinction clusters (see Tab.~\ref{tab:f025}). We deduce that our samples are consistent with the higher--extinction portion of the LEGUS sample, i.e. we are biased (by construction) against clusters with low-extinction. This is a direct consequence of the source selection method, that avoided regions with low extinction (in the extinction map derived in Section~\ref{sec:catalogs_highext}) and sources without compact \pb\ emission. 
Similarly, $\sim80$ clusters of our samples are not considered in LEGUS, and therefore the latter sample is biased against sources with high-extinction.

\begin{figure}
\centering
\includegraphics[width=0.49\textwidth]{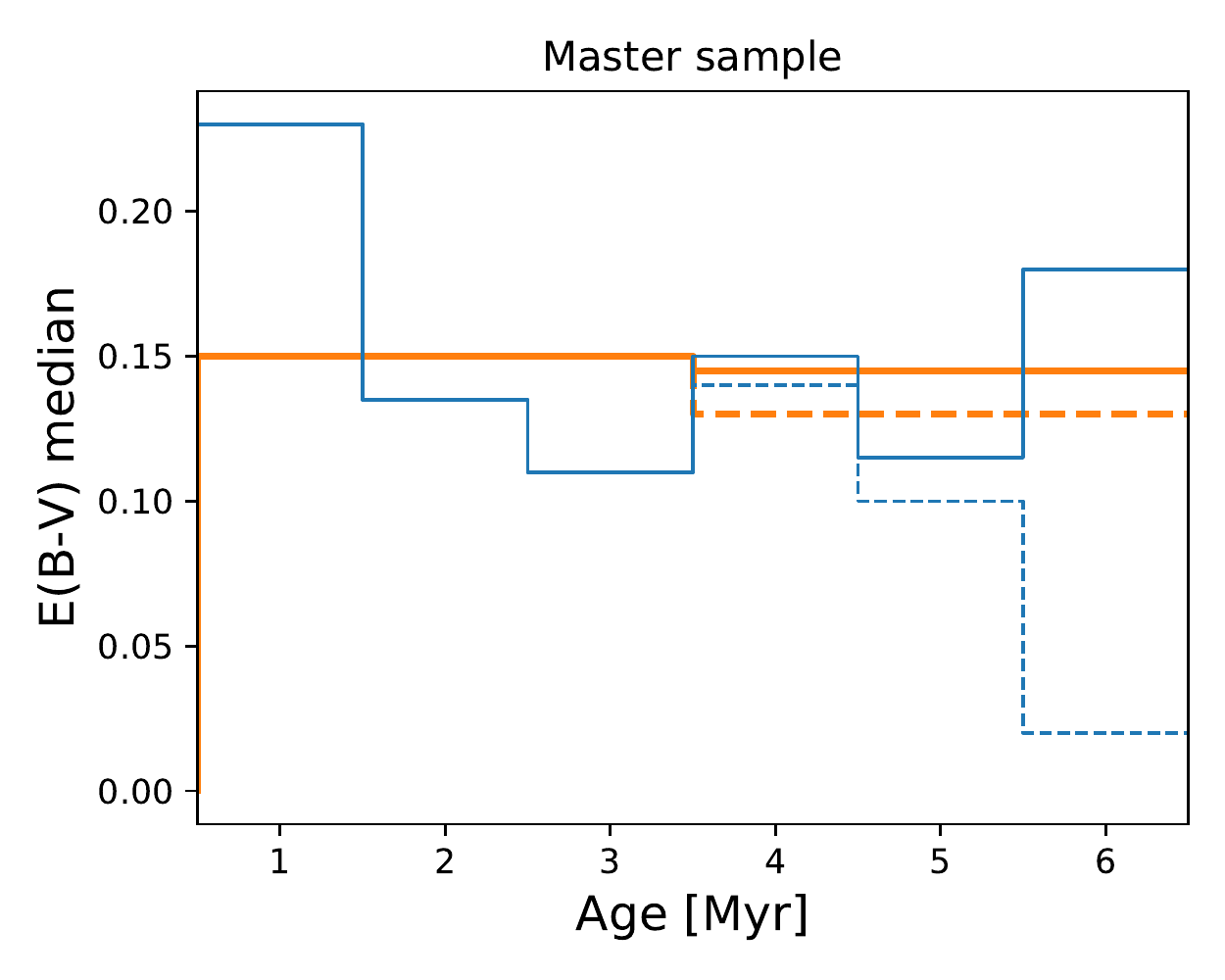}
\caption{Median \ebv\ in function of ages for the master sample obtained by joining our final samples with the LEGUS one. The blue line consider an age division in bins of 1 Myr, the orange histogram divide the sample in two age bins (1-3 and 4-6 Myr). In both cases we consider only sources with age $\le6$ Myr. The dashed histograms exclude from the sample sources with $\rm E(B-V)>1.5$ which, as discussed in the text, are not solid fits.}
\label{fig:ebv_distr_master}
\end{figure}

We create a ``master'' catalog by merging \textit{ExtmapCat--final}, \textit{PBcompactCat--final} and the LEGUS catalog (removing the duplicates, i.e. the sources in common). The master catalog counts 254 sources and we consider it as a more complete version of either our or the LEGUS samples; out of 208 sources with ages between 1 and 6 Myr in the ``master'' catalog, 76 ($\sim37\%$) are found exclusively in this study and were missed by LEGUS, while for 112 ($\sim54\%$) it is true the opposite.
In the case of the ``master'' catalog, we do not find an age evolution of the median \ebv\ values, nor of $\rm F_{<0.25}$ (Fig.~\ref{fig:ebv_distr_master} and Tab.~\ref{tab:f025_master}). 
As discussed in the previous section, most of the sources with $\rm E(B-V)>1.5$ mag are not robust results; repeating the analysis of the master catalog excluding such sources yields an age trend of both the median and $\rm F_{<0.25}$. In any case, we recover overall a low typical value for the cluster extinctions, with median $\rm E(B-V)\approx0.15$ mag or lower.

\begin{table*}[]
    \centering
    \begin{tabular}{|l|r|r|}
    \hline
        \multicolumn{1}{|l}{Sample} & \multicolumn{2}{c|}{$\rm F_{<0.25}$}  \\
        \multicolumn{1}{|l}{} & \multicolumn{1}{l}{1-3 Myr} & \multicolumn{1}{l|}{4-6 Myr}  \\
    \hline
    \hline
        \textit{ExtmapCat--final}    & 41\%    & 63\%  \\
        \textit{PBcompactCat--final} & 45\%    & 57\%  \\
        LEGUS               & 84\%    & 75\%  \\
        LEGUS (in common)   & 25\%    & 48\%  \\
    \hline
    \end{tabular}
    \caption{Percentage of clusters with $\rm E(B-V)<0.25$ mag, $\rm F_{<0.25}$, in two age ranges 1-3 and 4-6 Myr. The last row reports the values of the sample in common between LEGUS and our final samples.}
    \label{tab:f025}
\end{table*}

\begin{table*}[]
    \centering
    \begin{tabular}{|c|c|c|}
    \hline
        Age & $\rm F_{<0.25}$ & w/o high $\rm E(B-V)$
        \\
    \hline
    \hline
        1 Myr   & 64\% & 64\% \\
        2 Myr   & 67\% & 67\% \\
        3 Myr   & 81\% & 81\% \\
        4 Myr   & 71\% & 73\% \\
        5 Myr   & 73\% & 84\% \\
        6 Myr   & 68\% & 87\% \\
    \hline
        1-3 Myr & 70\% & 70\% \\
        4-6 Myr & 71\% & 77\% \\
    \hline
    \end{tabular}
    \caption{Percentage of clusters with $\rm E(B-V)<0.25$ mag, $\rm F_{<0.25}$, for the master catalog, i.e. the merged catalog of our own catalogs with the LEGUS one (see text), both in age bins of 1 Myr and in the two age ranges 1-3 and 4-6 Myr. The second column report $\rm F_{<0.25}$ excluding from the sample the sources with $\rm E(B-V)>1.5$ mag.}
    \label{tab:f025_master}
\end{table*}


\subsubsection{H$\alpha$ and Pa$\beta$ morphology}\label{sec:ha_morphology}
\begin{figure*}
\centering
\subfigure{\includegraphics[width=0.49\textwidth]{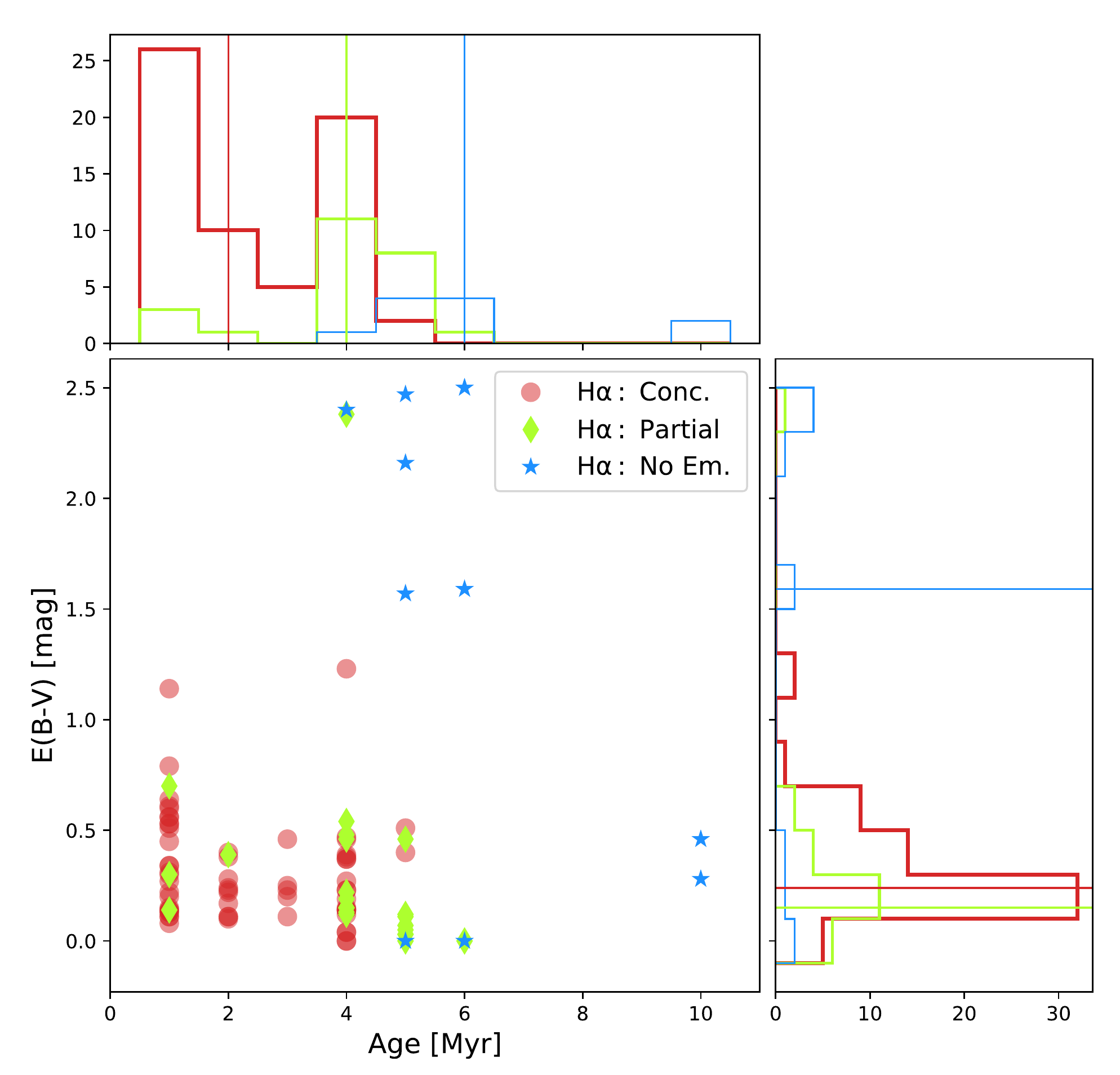}}
\subfigure{\includegraphics[width=0.49\textwidth]{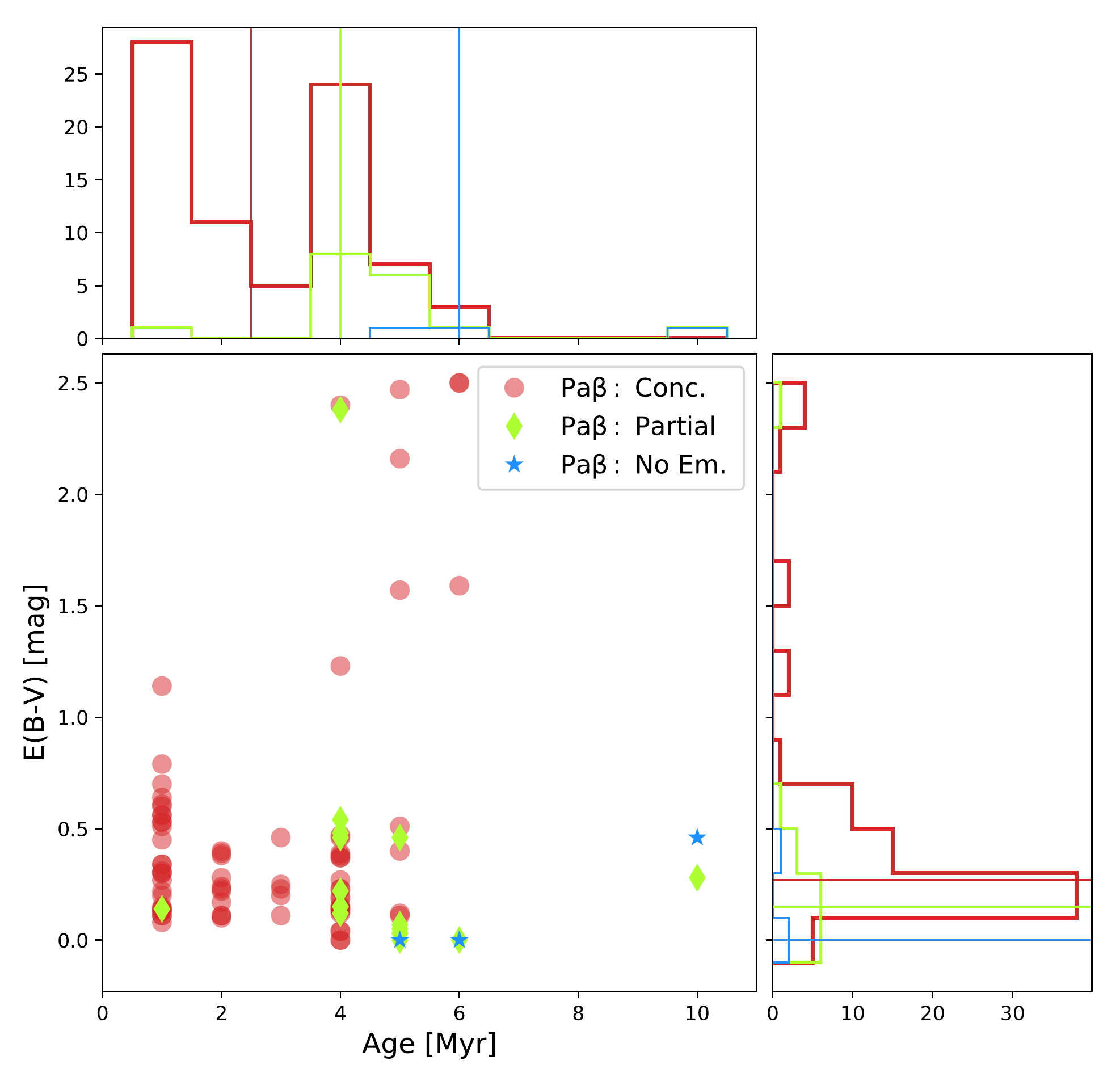}}
\caption{Distributions of ages and extinction for our final samples (merged together), color--coding the clusters based on their \ha\ morphology (left) and \pb\ morphology (right). Horizontal and vertical solid lines in the side histograms refer to the median values of each morphological class. The few sources that lack both \ha\ and \pb\ emission were included in the \textit{ExtmapCat} because  they are located in regions with diffuse nebular emission or very close to regions with elevated nebular emission.}
\label{fig:age_ebv_morphology}
\end{figure*}

Following the analysis of \citet{hannon2019} we visually classify the morphology of the \ha\ and \pb\ emission associated with each source of our samples. We follow their same classification scheme, mediated by \citet{whitmore2011} and \citet{hollyhead2015} who divide the sample in 3 classes: 
\begin{enumerate}
    \item \textbf{concentrated}, i.e. there is a compact HII region on the position of the cluster;
    \item \textbf{partially exposed}, i.e. either the HII region shows bubble--like morphology or only partially cover the cluster or emission is diffuse; 
    \item \textbf{no emission}, i.e. the target cluster is not associated to any nebular emission.
\end{enumerate}
This division in classes is expected to reflect an evolution of the clusters and indeed both \citet{hollyhead2015} and \citet{hannon2019} found an increase of the median age of clusters, when going from concentrated, to partially exposed, to no emission. We show, in Fig.~\ref{fig:age_ebv_morphology}, age and extinction distributions of the clusters in our sample (merging together sources from \textit{ExtmapCat--final} and from \textit{PBcompactCat--final}), separating them in the three classes above. The classification was done independently in \ha\ and in \pb, and therefore we end up with two classifications for each cluster. For both line morphologies, we find a trend with age, that have median values of 2, 4 and 6 Myr for ``concentrated'', ``partially exposed'' and ``no emission'' classes, respectively. We see a possible trend with extinction in the case of \pb\ morphology, as median values decrease going from concentrated to no emission classes, but the same is not true in the case of \ha\ morphology. In the latter case the median extinction of the ``no emission'' class is driven by the group of clusters with high extinctions. 

An interesting feature coming out of the plots in Fig.~\ref{fig:age_ebv_morphology} is that the almost totality of clusters with $\rm E(B-V)>1.5$ have compact \pb\ emission but no \ha. If we assume their extinction values are correct, we expect the absence of \ha\ emission to be driven by the elevated extinction. We test this hypothesis by deriving, from the \pb\ line emission map (Section~\ref{sec:catalogs_highext}), the \pb\ flux of those 7 sources and converting it into an expected \ha\ flux (taking into account the extinction of each source). The derived \ha\ fluxes span the range $\rm 2-25\cdot10^{-20}\ erg\ cm^{-2}$ \AA$^{-1}$. In the assumption of their flux uniformly distributed over a circular region of 1 pc radius ($1.17$ px at the distance of NGC 1313), their surface brightness is between 1 and 14 sigmas above the noise at their coordinates. 4 out of 7 of them have an expected \ha\ emission more than $3\sigma$ above the detection limit. The hypothesis is therefore only partially confirmed.
We also note that all the 7 sources considered are found in regions far from the large star--forming sub--regions (see Fig.~\ref{fig:coo_dist}), and appear in the \pb\ map as isolated sources.

Assuming to have a complete catalog of cluster covering the age range 0-10 Myr, \citet{hannon2019} estimate the timescale associated to each of the \ha\ morphological classes, by simply converting the fraction of sources in the class to a timescale in Myr. We discussed in the previous section how the LEGUS catalog that the authors used for this estimate is incomplete, being biased against extincted sources. 
We repeat their analysis using the master catalog created merging our final samples with the LEGUS catalog. We report in Tab.~\ref{tab:timescales} the timescales for each morphological stage, for the LEGUS sample (of NGC1313 only, while \citealp{hannon2019} reported the value calculated considering together 3 of the LEGUS galaxies), for our final samples alone and for the merged sample. In the case of our samples alone, we assume that they are representative of the age range 0-6 Myr, while for the other two samples we consider an age range 0-10 Myr. The addition of our samples to the LEGUS one prolongs the timescale of the ``concentrated'' stage from $\sim2$ Myr to $\sim3$ Myr, while the intermediate stage of partial emission remains $\sim1.5-2.0$ Myr long. We conclude that already at $\sim3$ Myr, the HII regions typically begin to disperse, and by $\sim5$ Myr there is very little or no nebular emission left around the clusters in NGC 1313.

\begin{table}[]
    \centering
    \begin{tabular}{|lcccccc|}
    \hline
    \multicolumn{1}{|l}{Sample} & \multicolumn{6}{c|}{Age Range} \\
    \multicolumn{1}{|l}{ } & \multicolumn{2}{c}{Concentrated} & \multicolumn{2}{c}{Partial emission} & \multicolumn{2}{c|}{No emission} \\
    \hline
    \hline
    \citet{hannon2019}  & 19.1\% & 1.9 Myr      & 16.3\% & 1.6 Myr      & 64.6\% & 6.5 Myr\\
    This work           & 65.6\% & 3.9 Myr      & 25.0\% & 1.5 Myr      & 9.4\%  & 0.6 Myr\\
    Merged              & 30.6\% & 3.1 Myr      & 18.4\% & 1.8 Myr      & 51.0\% & 5.1 Myr\\
    \hline
    \end{tabular}
    \caption{Relative fractions and inferred timescale of each \ha\ morphological class for the sample of \citet{hannon2019} (considering only the clusters of NGC 1313), our final samples and a master catalog merging them. The timescales for our samples alone has been considered over a total timescale of 6 Myr, and because of that the ``no emission'' stage last only 0.6 Myr.}
    \label{tab:timescales}
\end{table}

\section{Summary and conclusions}\label{sec:summary}
Using NUV--optical--NIR HST broad and narrow--band observations of the nearby galaxy NGC 1313 we looked for  young and embedded star clusters. In particular, new NIR data are used to derive a map of the Paschen$-\beta$ nebular emission and to characterize extincted sources possibly missed by previous analysis focused on the NUV-optical part of the spectrum.

We extract a catalog of possible candidates in two different ways:
\begin{enumerate}
    \item the first selection is made by using a  map of nebular extinction derived from \ha\ and \pb\ observations. Sub--regions of the galaxy where the extinction is higher than $\rm A_V\ge0.8$ mag were searched for sources in the F814W filter with counterparts in the NIR filters;
    \item with a complementary approach, we select sources with compact \pb\ emission within the entire galaxy.
\end{enumerate}
The sources in the two resulting catalogs, named \textit{ExtmapCat} and \textit{PBcompactCat} respectively, are analyzed photometrically and via a least-$\chi^2$ broad and narrow--band SED fit, in order to derive their ages, masses and extinctions. The effective radius of each source is estimated assuming a spherical symmetrical Moffat profile; most of the sources in the initial catalogs have a light profile consistent with stars. Among the star--like sources, only the ones with \ha\ and \pb\ equivalent widths higher than $\rm 317$ \AA and $\rm 95$ \AA respectively, i.e. consistent with being younger than 6 Myr according to our models (see Section~\ref{sec:ew}) are retained in the final samples. In addition, only sources with small uncertainties in the derived ages are considered in the final samples. 

The final samples count 46 and 84 sources for \textit{ExtmapCat--final} and \textit{PBcompactCat--final} respectively. 30 clusters are in common among the two final catalogs. The median size of the sources in the final samples is $\rm R_{eff}\approx1$ pc (Fig.~\ref{fig:reffs}). Overall, they are mainly distributed in the age range $0-6$ Myr, with extinctions $\rm E(B-V)\le1.0$ mag and masses in the range $\rm \log_{10}(M/M_\odot)=2.5-3.5$ (Fig.~\ref{fig:age_ebv_mass_extcat} and Fig.~\ref{fig:age_ebv_mass_pbcomp}). The ages derived via the broad--band SED fit are consistent with the ages that would be estimated from the \ha\ and \pb\ equivalent widths values (Fig.~\ref{fig:ew_ages_extcat} and Fig.~\ref{fig:ew_ages_pbcat}).
A comparison between models with various metallicities and extinction curves reveals that clusters in NGC 1313 are better fit by models with sub--solar metallicity, $\rm Z=0.008$, and Milky Way extinction curve (Fig.~\ref{fig:chisq_extcat}, Fig.~\ref{fig:age_ebv_pbcomp} and Fig.~\ref{fig:individual_met}).

In addition to use the results from photometry and SED fitting, we classify the morphology of the \ha\ and \pb\ emission associated to each of the sources in the final samples; following the example of previous studies \citep[e.g.][]{whitmore2011,hollyhead2015,hannon2019}, we used a 3--classes division, namely ``concentrated'' nebular emission, ``partially exposed'' and sources with ``no (nebular) emission'' associated. The classes are assumed to describe the time evolution of the gas from a dense cloud to the dispersion.
Most of the sources in our final samples fall into the first class, especially for what concerns the \pb\ emission; we remind that the \textit{PBcompactCat} sample is expected to contain sources with concentrated \pb\ emission, by construction.

Despite the majority of the sources in the final catalogs having low extinctions, for 13 sources we derived $\rm E(B-V)>1.0$ mag. Not all of their fits are robust; in some cases the \chisq\ associated to the best fit are high ($>50$), in some other cases we have detections only in a few ($\sim5$) filters. Several of these sources have concentrated \pb\ emission but no \ha\ detected (Fig.~\ref{fig:age_ebv_morphology}); based on their \pb\ flux and their derived $\rm E(B-V)$ values more than half of them should have \ha\ detectable. In addition, they are located far from the main star--forming regions (Fig.~\ref{fig:coo_dist}). Overall, we consider them only as possible candidates for young and extincted regions. Only two of them (with $\rm E(B-V)$ slightly above 1.0 mag) have robust fits.

We expect the young clusters to form in dense clouds of gas and dust, and to be still embedded in their natal cloud at birth. We propose a few hypotheses for the small number of young and extincted sources observed. 
\begin{itemize}
    \item Low surface densities of the GMCs hosting star formation. An extinction of $\rm E(B-V)=1.0$ mag would correspond to a gas screen with a surface density of $\rm 150\ M_\odot pc^{-2}$, consistent with the average values for GMCs in NGC 1313 as revealed by ALMA observations.
    \item An extremely short timescale for the clearing of the natal cloud due to feedback, on the order of $\sim1$ Myr; short timescales could be related to low gas surface density discussed in the previous point.
    \item Via a completeness test, we estimate an incompleteness of $\sim50\%$ at $\rm E(B-V)=1.0$ mag (see Fig.~\ref{fig:completeness_goodness}). According to this test, we would expect to recover $\sim25\%$ of the sources with $\rm E(B-V)$ between $1.0$ and $2.0$ mag but we found only one source in such extinction range and with age $\le3$ Myr (out of 45 total sources with age $\le3$ Myr found).
\end{itemize}

We study the distribution of extinction in function of cluster ages, expecting younger sources to be more extincted. 
\begin{itemize}
\item We recover a weak dependence of $\rm E(B-V)$ with age (Fig.~\ref{fig:ebv_distr}); $\sim40\%$ of the sources with ages in the range $1-3$ Myr have $\rm E(B-V)<0.25$ mag, while the percentage rise to $\sim60\%$ for the sources with ages $4-6$ Myr. 
These data confirm that a significant fraction of sources have low extinction already at ages $\le3$ Myr.
\item A clear trend is observed between the morphology of the \pb\ emission and the age and extinction of the relative cluster; sources with concentrated emission are on average younger and more extincted than sources with partial or no emission (Fig.~\ref{tab:timescales}). 
\end{itemize}

We include the clusters from the LEGUS catalog (focused on NUV-optical sources, while our catalogs are focused on optical-NIR) to create a ``master'' catalog, more representative of the entire young cluster population of NGC 1313. 
\begin{itemize}
\item Repeating the study of the extinctions on the master catalog, we find weak or no evolution with age, depending on whether the sources with $\rm E(B-V)>1.5$ mag discussed above are excluded from consideration or not.  (Fig.~\ref{fig:ebv_distr_master}). 
At best, $70\%$ of the clusters with ages $\le3$ Myr have low extinctions (below $\rm E(B-V)<0.25$)
with the percentage rising to $77\%$ in the age range $4-6$ Myr.
\item Assuming that the ``master'' sample is representative of the NGC 1313 cluster population in the age range $0-10$ Myr we use the fraction of sources in each class to estimate its typical timescale. We recover $3.1$ Myr for the concentrated phase, $1.8$ Myr for the ``partial emission'' and $5.1$ Myr for the ``no emission'' phase. This results prolongs the expected timescale for clearing the cloud by $1$ Myr ($\sim50\%$) compared to a previous estimate based only on NUV--optical data.
\end{itemize}

Clusters in NGC~1313 appear to have cleared the gas cloud around them by the time they reach an age of $5$~Myr and many of them are already almost gas free within the first 3 Myr. 
We conclude that the inclusion of a tracer more transparent to extinction (NIR observations) has allowed to recover the fraction of clusters that are missed in optical--NUV studies and better pin down previous estimates of the duration of the embedded phase to $3$ Myr; including young embedded clusters in NGC~1313 changes previous estimates of the short feedback timescales from clusters by $\sim50\%$ (from $\sim2$ Myr to $\sim3$ Myr), bringing it closer to the timescales 
for the first supernova explosions, which could occur as early as $\sim4$ Myr \citep[e.g.][]{sukhbold2016}. In addition, photoionisation, radiation pressure and winds can open channels before supernovae explode for low mass GMCs ($\sim10^4-10^5\ M_\odot$, e.g. \citealp{dale2014,dale2015}).

The analysis also suggested that the median age of the cluster correlates better with the morphology of the nebular emission (Fig.~\ref{fig:age_ebv_morphology}) than with the cluster extinction (Fig.~\ref{fig:ebv_distr} and Fig.~\ref{fig:ebv_distr_master}); we deduce that the \ha\ (and \pb) morphology is a good tracer of the cluster age evolution. On the other hand, median cluster extinctions also show some correlation with the ionized gas morphology (Fig.~\ref{fig:age_ebv_morphology}) but also reveal that many clusters with 'concentrated' nebular emission associated are not very extincted, with $\rm E(B-V)\sim0.25$ mag, i.e. even when a cluster is still surrounded by a compact gas cloud, its effect on the cluster extinction can be low. We speculate that this could be caused by non-uniform gas--dust clouds, where holes in the line--of--sight direction allow the escape of the stellar radiation.

We point out that the current study consider only clusters in NGC 1313, a single galaxy with a given metallicity (best estimate: $Z=0.008$) and SFR density ($\rm \Sigma_{SFR}\sim0.01\ M_\odot\ yr^{-1}\ kpc^{-2}$). In addition, most of the clusters masses in NGC 1313 are distributed around a mass of $\rm \sim10^3\ M_\odot$. We plan  in the near future to extend this study to galaxies with different properties and clusters with wider range of masses, in order to estimate if and how the interaction between the very young clusters and their cloud is affected by the host galaxy properties and by the properties of cluster themselves.

%

\vspace{5mm}
\facilities{HST}


\acknowledgments{Based on observations made with the NASA/ESA Hubble Space Telescope, obtained  at the Space Telescope Science Institute, which is operated by the 
Association of Universities for Research in Astronomy, Inc., under NASA contract NAS 5--26555. These observations are associated with program \# 15330. 
Support for program \# 15330 was provided by NASA through a grant from the Space Telescope Science Institute.
K.E.J. acknowledge support from  NSF grants 1413231 and 1716335.}



\bibliography{references}{}

\begin{thebibliography}{}
\expandafter\ifx\csname natexlab\endcsname\relax\def\natexlab#1{#1}\fi
\providecommand{\url}[1]{\href{#1}{#1}}
\providecommand{\dodoi}[1]{doi:~\href{http://doi.org/#1}{\nolinkurl{#1}}}
\providecommand{\doeprint}[1]{\href{http://ascl.net/#1}{\nolinkurl{http://ascl.net/#1}}}
\providecommand{\doarXiv}[1]{\href{https://arxiv.org/abs/#1}{\nolinkurl{https://arxiv.org/abs/#1}}}

\bibitem[{{Adamo} {et~al.}(2015){Adamo}, {Kruijssen}, {Bastian}, {Silva-Villa},
  \& {Ryon}}]{adamo2015}
{Adamo}, A., {Kruijssen}, J.~M.~D., {Bastian}, N., {Silva-Villa}, E., \&
  {Ryon}, J. 2015, \mnras, 452, 246, \dodoi{10.1093/mnras/stv1203}

\bibitem[{{Adamo} {et~al.}(2010){Adamo}, {Zackrisson}, {{\"O}stlin}, \&
  {Hayes}}]{adamo2010b}
{Adamo}, A., {Zackrisson}, E., {{\"O}stlin}, G., \& {Hayes}, M. 2010, \apj,
  725, 1620, \dodoi{10.1088/0004-637X/725/2/1620}

\bibitem[{{Adamo} {et~al.}(2017){Adamo}, {Ryon}, {Messa}, {Kim}, {Grasha},
  {Cook}, {Calzetti}, {Lee}, {Whitmore}, {Elmegreen}, {Ubeda}, {Smith},
  {Bright}, {Runnholm}, {Andrews}, {Fumagalli}, {Gouliermis}, {Kahre}, {Nair},
  {Thilker}, {Walterbos}, {Wofford}, {Aloisi}, {Ashworth}, {Brown}, {Chandar},
  {Christian}, {Cignoni}, {Clayton}, {Dale}, {de Mink}, {Dobbs}, {Elmegreen},
  {Evans}, {Gallagher}, {Grebel}, {Herrero}, {Hunter}, {Johnson}, {Kennicutt},
  {Krumholz}, {Lennon}, {Levay}, {Martin}, {Nota}, {{\"O}stlin}, {Pellerin},
  {Prieto}, {Regan}, {Sabbi}, {Sacchi}, {Schaerer}, {Schiminovich}, {Shabani},
  {Tosi}, {Van Dyk}, \& {Zackrisson}}]{adamo2017}
{Adamo}, A., {Ryon}, J.~E., {Messa}, M., {et~al.} 2017, \apj, 841, 131,
  \dodoi{10.3847/1538-4357/aa7132}

\bibitem[{{Aversa} {et~al.}(2011){Aversa}, {Johnson}, {Brogan}, {Goss}, \&
  {Pisano}}]{aversa2011}
{Aversa}, A.~G., {Johnson}, K.~E., {Brogan}, C.~L., {Goss}, W.~M., \& {Pisano},
  D.~J. 2011, \aj, 141, 125, \dodoi{10.1088/0004-6256/141/4/125}

\bibitem[{{Bastian} {et~al.}(2013){Bastian}, {Schweizer}, {Goudfrooij},
  {Larsen}, \& {Kissler-Patig}}]{bastian2013}
{Bastian}, N., {Schweizer}, F., {Goudfrooij}, P., {Larsen}, S.~S., \&
  {Kissler-Patig}, M. 2013, \mnras, 431, 1252, \dodoi{10.1093/mnras/stt253}

\bibitem[{{Bastian} {et~al.}(2012){Bastian}, {Adamo}, {Gieles}, {Silva-Villa},
  {Lamers}, {Larsen}, {Smith}, {Konstantopoulos}, \&
  {Zackrisson}}]{bastian2012}
{Bastian}, N., {Adamo}, A., {Gieles}, M., {et~al.} 2012, \mnras, 419, 2606,
  \dodoi{10.1111/j.1365-2966.2011.19909.x}

\bibitem[{{Bohlin} {et~al.}(1978){Bohlin}, {Savage}, \& {Drake}}]{bohlin1978}
{Bohlin}, R.~C., {Savage}, B.~D., \& {Drake}, J.~F. 1978, \apj, 224, 132,
  \dodoi{10.1086/156357}

\bibitem[{{Calzetti} {et~al.}(2000){Calzetti}, {Armus}, {Bohlin}, {Kinney},
  {Koornneef}, \& {Storchi-Bergmann}}]{calzetti2000}
{Calzetti}, D., {Armus}, L., {Bohlin}, R.~C., {et~al.} 2000, \apj, 533, 682,
  \dodoi{10.1086/308692}

\bibitem[{{Calzetti} {et~al.}(2015{\natexlab{a}}){Calzetti}, {Johnson},
  {Adamo}, {Gallagher}, {Andrews}, {Smith}, {Clayton}, {Lee}, {Sabbi}, {Ubeda},
  {Kim}, {Ryon}, {Thilker}, {Bright}, {Zackrisson}, {Kennicutt}, {de Mink},
  {Whitmore}, {Aloisi}, {Chandar}, {Cignoni}, {Cook}, {Dale}, {Elmegreen},
  {Elmegreen}, {Evans}, {Fumagalli}, {Gouliermis}, {Grasha}, {Grebel},
  {Krumholz}, {Walterbos}, {Wofford}, {Brown}, {Christian}, {Dobbs}, {Herrero},
  {Kahre}, {Messa}, {Nair}, {Nota}, {{\"O}stlin}, {Pellerin}, {Sacchi},
  {Schaerer}, \& {Tosi}}]{calzetti2015b}
{Calzetti}, D., {Johnson}, K.~E., {Adamo}, A., {et~al.} 2015{\natexlab{a}},
  \apj, 811, 75, \dodoi{10.1088/0004-637X/811/2/75}

\bibitem[{{Calzetti} {et~al.}(2015{\natexlab{b}}){Calzetti}, {Lee}, {Sabbi},
  {Adamo}, {Smith}, {Andrews}, {Ubeda}, {Bright}, {Thilker}, {Aloisi}, {Brown},
  {Chandar}, {Christian}, {Cignoni}, {Clayton}, {da Silva}, {de Mink}, {Dobbs},
  {Elmegreen}, {Elmegreen}, {Evans}, {Fumagalli}, {Gallagher}, {Gouliermis},
  {Grebel}, {Herrero}, {Hunter}, {Johnson}, {Kennicutt}, {Kim}, {Krumholz},
  {Lennon}, {Levay}, {Martin}, {Nair}, {Nota}, {{\"O}stlin}, {Pellerin},
  {Prieto}, {Regan}, {Ryon}, {Schaerer}, {Schiminovich}, {Tosi}, {Van Dyk},
  {Walterbos}, {Whitmore}, \& {Wofford}}]{calzetti2015_legus}
{Calzetti}, D., {Lee}, J.~C., {Sabbi}, E., {et~al.} 2015{\natexlab{b}}, \aj,
  149, 51, \dodoi{10.1088/0004-6256/149/2/51}

\bibitem[{{Cardelli} {et~al.}(1989){Cardelli}, {Clayton}, \&
  {Mathis}}]{cardelli1989}
{Cardelli}, J.~A., {Clayton}, G.~C., \& {Mathis}, J.~S. 1989, \apj, 345, 245,
  \dodoi{10.1086/167900}

\bibitem[{{Chandar} {et~al.}(2015){Chandar}, {Fall}, \&
  {Whitmore}}]{chandar2015}
{Chandar}, R., {Fall}, S.~M., \& {Whitmore}, B.~C. 2015, \apj, 810, 1,
  \dodoi{10.1088/0004-637X/810/1/1}

\bibitem[{{Chevance} {et~al.}(2020{\natexlab{a}}){Chevance}, {Kruijssen},
  {Hygate}, {Schruba}, {Longmore}, {Groves}, {Henshaw}, {Herrera}, {Hughes},
  {Jeffreson}, {Lang}, {Leroy}, {Meidt}, {Pety}, {Razza}, {Rosolowsky},
  {Schinnerer}, {Bigiel}, {Blanc}, {Emsellem}, {Faesi}, {Glover}, {Haydon},
  {Ho}, {Kreckel}, {Lee}, {Liu}, {Querejeta}, {Saito}, {Sun}, {Usero}, \&
  {Utomo}}]{chevance2020a}
{Chevance}, M., {Kruijssen}, J.~M.~D., {Hygate}, A. P.~S., {et~al.}
  2020{\natexlab{a}}, \mnras, 493, 2872, \dodoi{10.1093/mnras/stz3525}

\bibitem[{{Chevance} {et~al.}(2020{\natexlab{b}}){Chevance}, {Kruijssen},
  {Vazquez-Semadeni}, {Nakamura}, {Klessen}, {Ballesteros-Paredes}, {Inutsuka},
  {Adamo}, \& {Hennebelle}}]{chevance2020b}
{Chevance}, M., {Kruijssen}, J.~M.~D., {Vazquez-Semadeni}, E., {et~al.}
  2020{\natexlab{b}}, \ssr, 216, 50, \dodoi{10.1007/s11214-020-00674-x}

\bibitem[{{Corbelli} {et~al.}(2017){Corbelli}, {Braine}, {Bandiera},
  {Brouillet}, {Combes}, {Druard}, {Gratier}, {Mata}, {Schuster}, {Xilouris},
  \& {Palla}}]{corbelli2017}
{Corbelli}, E., {Braine}, J., {Bandiera}, R., {et~al.} 2017, \aap, 601, A146,
  \dodoi{10.1051/0004-6361/201630034}

\bibitem[{{Dale}(2015)}]{dale2015}
{Dale}, J.~E. 2015, \nar, 68, 1, \dodoi{10.1016/j.newar.2015.06.001}

\bibitem[{{Dale} {et~al.}(2014){Dale}, {Ngoumou}, {Ercolano}, \&
  {Bonnell}}]{dale2014}
{Dale}, J.~E., {Ngoumou}, J., {Ercolano}, B., \& {Bonnell}, I.~A. 2014, \mnras,
  442, 694, \dodoi{10.1093/mnras/stu816}

\bibitem[{{de Vaucouleurs}(1963)}]{de-vaucouleurs1963}
{de Vaucouleurs}, G. 1963, \apj, 137, 720, \dodoi{10.1086/147550}

\bibitem[{{Elmegreen} \& {Elmegreen}(2019)}]{elmegreen2019}
{Elmegreen}, B.~G., \& {Elmegreen}, D.~M. 2019, \apjs, 245, 14,
  \dodoi{10.3847/1538-4365/ab4903}

\bibitem[{{Elmegreen} \& {Elmegreen}(2020)}]{elmegreen2020}
---. 2020, \apj, 895, 71, \dodoi{10.3847/1538-4357/ab8d20}

\bibitem[{{Elmegreen} {et~al.}(2018){Elmegreen}, {Elmegreen}, \&
  {Efremov}}]{elmegreen2018}
{Elmegreen}, B.~G., {Elmegreen}, D.~M., \& {Efremov}, Y.~N. 2018, \apj, 863,
  59, \dodoi{10.3847/1538-4357/aacf9a}

\bibitem[{{Elson} {et~al.}(1987){Elson}, {Fall}, \& {Freeman}}]{elson1987}
{Elson}, R. A.~W., {Fall}, S.~M., \& {Freeman}, K.~C. 1987, \apj, 323, 54,
  \dodoi{10.1086/165807}

\bibitem[{{Ferland} {et~al.}(2013){Ferland}, {Porter}, {van Hoof}, {Williams},
  {Abel}, {Lykins}, {Shaw}, {Henney}, \& {Stancil}}]{ferland2013}
{Ferland}, G.~J., {Porter}, R.~L., {van Hoof}, P.~A.~M., {et~al.} 2013, \rmxaa,
  49, 137.
\newblock \doarXiv{1302.4485}

\bibitem[{{Finn} {et~al.}(2019){Finn}, {Johnson}, {Brogan}, {Wilson},
  {Indebetouw}, {Harris}, {Kamenetzky}, \& {Bemis}}]{finn2019}
{Finn}, M.~K., {Johnson}, K.~E., {Brogan}, C.~L., {et~al.} 2019, \apj, 874,
  120, \dodoi{10.3847/1538-4357/ab0d1e}

\bibitem[{{Fitzpatrick}(1999)}]{fitzpatrick1999}
{Fitzpatrick}, E.~L. 1999, \pasp, 111, 63, \dodoi{10.1086/316293}

\bibitem[{{Gaia Collaboration} {et~al.}(2018){Gaia Collaboration}, {Brown},
  {Vallenari}, {Prusti}, {de Bruijne}, {Babusiaux}, {Bailer-Jones}, {Biermann},
  {Evans}, {Eyer}, {Jansen}, {Jordi}, {Klioner}, {Lammers}, {Lindegren},
  {Luri}, {Mignard}, {Panem}, {Pourbaix}, {Randich}, {Sartoretti}, {Siddiqui},
  {Soubiran}, {van Leeuwen}, {Walton}, {Arenou}, {Bastian}, {Cropper},
  {Drimmel}, {Katz}, {Lattanzi}, {Bakker}, {Cacciari}, {Casta{\~n}eda},
  {Chaoul}, {Cheek}, {De Angeli}, {Fabricius}, {Guerra}, {Holl}, {Masana},
  {Messineo}, {Mowlavi}, {Nienartowicz}, {Panuzzo}, {Portell}, {Riello},
  {Seabroke}, {Tanga}, {Th{\'e}venin}, {Gracia-Abril}, {Comoretto},
  {Garcia-Reinaldos}, {Teyssier}, {Altmann}, {Andrae}, {Audard},
  {Bellas-Velidis}, {Benson}, {Berthier}, {Blomme}, {Burgess}, {Busso},
  {Carry}, {Cellino}, {Clementini}, {Clotet}, {Creevey}, {Davidson}, {De
  Ridder}, {Delchambre}, {Dell'Oro}, {Ducourant},
  {Fern{\'a}ndez-Hern{\'a}ndez}, {Fouesneau}, {Fr{\'e}mat}, {Galluccio},
  {Garc{\'\i}a-Torres}, {Gonz{\'a}lez-N{\'u}{\~n}ez}, {Gonz{\'a}lez-Vidal},
  {Gosset}, {Guy}, {Halbwachs}, {Hambly}, {Harrison}, {Hern{\'a}ndez},
  {Hestroffer}, {Hodgkin}, {Hutton}, {Jasniewicz}, {Jean-Antoine-Piccolo},
  {Jordan}, {Korn}, {Krone-Martins}, {Lanzafame}, {Lebzelter}, {L{\"o}ffler},
  {Manteiga}, {Marrese}, {Mart{\'\i}n-Fleitas}, {Moitinho}, {Mora}, {Muinonen},
  {Osinde}, {Pancino}, {Pauwels}, {Petit}, {Recio-Blanco}, {Richards},
  {Rimoldini}, {Robin}, {Sarro}, {Siopis}, {Smith}, {Sozzetti}, {S{\"u}veges},
  {Torra}, {van Reeven}, {Abbas}, {Abreu Aramburu}, {Accart}, {Aerts},
  {Altavilla}, {{\'A}lvarez}, {Alvarez}, {Alves}, {Anderson}, {Andrei},
  {Anglada Varela}, {Antiche}, {Antoja}, {Arcay}, {Astraatmadja}, {Bach},
  {Baker}, {Balaguer-N{\'u}{\~n}ez}, {Balm}, {Barache}, {Barata}, {Barbato},
  {Barblan}, {Barklem}, {Barrado}, {Barros}, {Barstow}, {Bartholom{\'e}
  Mu{\~n}oz}, {Bassilana}, {Becciani}, {Bellazzini}, {Berihuete}, {Bertone},
  {Bianchi}, {Bienaym{\'e}}, {Blanco-Cuaresma}, {Boch}, {Boeche}, {Bombrun},
  {Borrachero}, {Bossini}, {Bouquillon}, {Bourda}, {Bragaglia}, {Bramante},
  {Breddels}, {Bressan}, {Brouillet}, {Br{\"u}semeister}, {Brugaletta},
  {Bucciarelli}, {Burlacu}, {Busonero}, {Butkevich}, {Buzzi}, {Caffau},
  {Cancelliere}, {Cannizzaro}, {Cantat-Gaudin}, {Carballo}, {Carlucci},
  {Carrasco}, {Casamiquela}, {Castellani}, {Castro-Ginard}, {Charlot},
  {Chemin}, {Chiavassa}, {Cocozza}, {Costigan}, {Cowell}, {Crifo}, {Crosta},
  {Crowley}, {Cuypers}, {Dafonte}, {Damerdji}, {Dapergolas}, {David}, {David},
  {de Laverny}, {De Luise}, {De March}, {de Martino}, {de Souza}, {de Torres},
  {Debosscher}, {del Pozo}, {Delbo}, {Delgado}, {Delgado}, {Di Matteo},
  {Diakite}, {Diener}, {Distefano}, {Dolding}, {Drazinos}, {Dur{\'a}n},
  {Edvardsson}, {Enke}, {Eriksson}, {Esquej}, {Eynard Bontemps}, {Fabre},
  {Fabrizio}, {Faigler}, {Falc{\~a}o}, {Farr{\`a}s Casas}, {Federici},
  {Fedorets}, {Fernique}, {Figueras}, {Filippi}, {Findeisen}, {Fonti},
  {Fraile}, {Fraser}, {Fr{\'e}zouls}, {Gai}, {Galleti}, {Garabato},
  {Garc{\'\i}a-Sedano}, {Garofalo}, {Garralda}, {Gavel}, {Gavras}, {Gerssen},
  {Geyer}, {Giacobbe}, {Gilmore}, {Girona}, {Giuffrida}, {Glass}, {Gomes},
  {Granvik}, {Gueguen}, {Guerrier}, {Guiraud}, {Guti{\'e}rrez-S{\'a}nchez},
  {Haigron}, {Hatzidimitriou}, {Hauser}, {Haywood}, {Heiter}, {Helmi}, {Heu},
  {Hilger}, {Hobbs}, {Hofmann}, {Holland}, {Huckle}, {Hypki}, {Icardi},
  {Jan{\ss}en}, {Jevardat de Fombelle}, {Jonker}, {Juh{\'a}sz}, {Julbe},
  {Karampelas}, {Kewley}, {Klar}, {Kochoska}, {Kohley}, {Kolenberg},
  {Kontizas}, {Kontizas}, {Koposov}, {Kordopatis}, {Kostrzewa-Rutkowska},
  {Koubsky}, {Lambert}, {Lanza}, {Lasne}, {Lavigne}, {Le Fustec}, {Le
  Poncin-Lafitte}, {Lebreton}, {Leccia}, {Leclerc}, {Lecoeur-Taibi},
  {Lenhardt}, {Leroux}, {Liao}, {Licata}, {Lindstr{\o}m}, {Lister}, {Livanou},
  {Lobel}, {L{\'o}pez}, {Managau}, {Mann}, {Mantelet}, {Marchal}, {Marchant},
  {Marconi}, {Marinoni}, {Marschalk{\'o}}, {Marshall}, {Martino}, {Marton},
  {Mary}, {Massari}, {Matijevi{\v{c}}}, {Mazeh}, {McMillan}, {Messina},
  {Michalik}, {Millar}, {Molina}, {Molinaro}, {Moln{\'a}r}, {Montegriffo},
  {Mor}, {Morbidelli}, {Morel}, {Morris}, {Mulone}, {Muraveva}, {Musella},
  {Nelemans}, {Nicastro}, {Noval}, {O'Mullane}, {Ord{\'e}novic},
  {Ord{\'o}{\~n}ez-Blanco}, {Osborne}, {Pagani}, {Pagano}, {Pailler},
  {Palacin}, {Palaversa}, {Panahi}, {Pawlak}, {Piersimoni}, {Pineau}, {Plachy},
  {Plum}, {Poggio}, {Poujoulet}, {Pr{\v{s}}a}, {Pulone}, {Racero}, {Ragaini},
  {Rambaux}, {Ramos-Lerate}, {Regibo}, {Reyl{\'e}}, {Riclet}, {Ripepi}, {Riva},
  {Rivard}, {Rixon}, {Roegiers}, {Roelens}, {Romero-G{\'o}mez}, {Rowell},
  {Royer}, {Ruiz-Dern}, {Sadowski}, {Sagrist{\`a} Sell{\'e}s}, {Sahlmann},
  {Salgado}, {Salguero}, {Sanna}, {Santana-Ros}, {Sarasso}, {Savietto},
  {Schultheis}, {Sciacca}, {Segol}, {Segovia}, {S{\'e}gransan}, {Shih},
  {Siltala}, {Silva}, {Smart}, {Smith}, {Solano}, {Solitro}, {Sordo}, {Soria
  Nieto}, {Souchay}, {Spagna}, {Spoto}, {Stampa}, {Steele},
  {Steidelm{\"u}ller}, {Stephenson}, {Stoev}, {Suess}, {Surdej}, {Szabados},
  {Szegedi-Elek}, {Tapiador}, {Taris}, {Tauran}, {Taylor}, {Teixeira},
  {Terrett}, {Teyssand ier}, {Thuillot}, {Titarenko}, {Torra Clotet}, {Turon},
  {Ulla}, {Utrilla}, {Uzzi}, {Vaillant}, {Valentini}, {Valette}, {van Elteren},
  {Van Hemelryck}, {van Leeuwen}, {Vaschetto}, {Vecchiato}, {Veljanoski},
  {Viala}, {Vicente}, {Vogt}, {von Essen}, {Voss}, {Votruba}, {Voutsinas},
  {Walmsley}, {Weiler}, {Wertz}, {Wevers}, {Wyrzykowski}, {Yoldas},
  {{\v{Z}}erjal}, {Ziaeepour}, {Zorec}, {Zschocke}, {Zucker}, {Zurbach}, \&
  {Zwitter}}]{gaia2018}
{Gaia Collaboration}, {Brown}, A.~G.~A., {Vallenari}, A., {et~al.} 2018, \aap,
  616, A1, \dodoi{10.1051/0004-6361/201833051}

\bibitem[{{Grasha} {et~al.}(2017{\natexlab{a}}){Grasha}, {Elmegreen},
  {Calzetti}, {Adamo}, {Aloisi}, {Bright}, {Cook}, {Dale}, {Fumagalli},
  {Gallagher}, {Gouliermis}, {Grebel}, {Kahre}, {Kim}, {Krumholz}, {Lee},
  {Messa}, {Ryon}, \& {Ubeda}}]{grasha2017b}
{Grasha}, K., {Elmegreen}, B.~G., {Calzetti}, D., {et~al.} 2017{\natexlab{a}},
  \apj, 842, 25, \dodoi{10.3847/1538-4357/aa740b}

\bibitem[{{Grasha} {et~al.}(2017{\natexlab{b}}){Grasha}, {Calzetti}, {Adamo},
  {Kim}, {Elmegreen}, {Gouliermis}, {Dale}, {Fumagalli}, {Grebel}, {Johnson},
  {Kahre}, {Kennicutt}, {Messa}, {Pellerin}, {Ryon}, {Smith}, {Shabani},
  {Thilker}, \& {Ubeda}}]{grasha2017a}
{Grasha}, K., {Calzetti}, D., {Adamo}, A., {et~al.} 2017{\natexlab{b}}, \apj,
  840, 113, \dodoi{10.3847/1538-4357/aa6f15}

\bibitem[{{Grasha} {et~al.}(2018){Grasha}, {Calzetti}, {Bittle}, {Johnson},
  {Donovan Meyer}, {Kennicutt}, {Elmegreen}, {Adamo}, {Krumholz}, {Fumagalli},
  {Grebel}, {Gouliermis}, {Cook}, {Gallagher}, {Aloisi}, {Dale}, {Linden},
  {Sacchi}, {Thilker}, {Walterbos}, {Messa}, {Wofford}, \&
  {Smith}}]{grasha2018}
{Grasha}, K., {Calzetti}, D., {Bittle}, L., {et~al.} 2018, \mnras, 481, 1016,
  \dodoi{10.1093/mnras/sty2154}

\bibitem[{{Grasha} {et~al.}(2019){Grasha}, {Calzetti}, {Adamo}, {Kennicutt},
  {Elmegreen}, {Messa}, {Dale}, {Fedorenko}, {Mahadevan}, {Grebel},
  {Fumagalli}, {Kim}, {Dobbs}, {Gouliermis}, {Ashworth}, {Gallagher}, {Smith},
  {Tosi}, {Whitmore}, {Schinnerer}, {Colombo}, {Hughes}, {Leroy}, \&
  {Meidt}}]{grasha2019}
{Grasha}, K., {Calzetti}, D., {Adamo}, A., {et~al.} 2019, \mnras, 483, 4707,
  \dodoi{10.1093/mnras/sty3424}

\bibitem[{{Hannon} {et~al.}(2019){Hannon}, {Lee}, {Whitmore}, {Chand ar},
  {Adamo}, {Mobasher}, {Aloisi}, {Calzetti}, {Cignoni}, {Cook}, {Dale},
  {Deger}, {Della Bruna}, {Elmegreen}, {Gouliermis}, {Grasha}, {Grebel},
  {Herrero}, {Hunter}, {Johnson}, {Kennicutt}, {Kim}, {Sacchi}, {Smith},
  {Thilker}, {Turner}, {Walterbos}, \& {Wofford}}]{hannon2019}
{Hannon}, S., {Lee}, J.~C., {Whitmore}, B.~C., {et~al.} 2019, \mnras, 490,
  4648, \dodoi{10.1093/mnras/stz2820}

\bibitem[{{Hollyhead} {et~al.}(2015){Hollyhead}, {Bastian}, {Adamo},
  {Silva-Villa}, {Dale}, {Ryon}, \& {Gazak}}]{hollyhead2015}
{Hollyhead}, K., {Bastian}, N., {Adamo}, A., {et~al.} 2015, \mnras, 449, 1106,
  \dodoi{10.1093/mnras/stv331}

\bibitem[{{Jacobs} {et~al.}(2009){Jacobs}, {Rizzi}, {Tully}, {Shaya},
  {Makarov}, \& {Makarova}}]{jacobs2009}
{Jacobs}, B.~A., {Rizzi}, L., {Tully}, R.~B., {et~al.} 2009, \aj, 138, 332,
  \dodoi{10.1088/0004-6256/138/2/332}

\bibitem[{{Johnson} {et~al.}(2018){Johnson}, {Brogan}, {Indebetouw}, {Testi},
  {Wilner}, {Reines}, {Chen}, \& {Vanzi}}]{johnson2018}
{Johnson}, K.~E., {Brogan}, C.~L., {Indebetouw}, R., {et~al.} 2018, \apj, 853,
  125, \dodoi{10.3847/1538-4357/aa9ff8}

\bibitem[{{Johnson} {et~al.}(2009){Johnson}, {Hunt}, \& {Reines}}]{johnson2009}
{Johnson}, K.~E., {Hunt}, L.~K., \& {Reines}, A.~E. 2009, \aj, 137, 3788,
  \dodoi{10.1088/0004-6256/137/4/3788}

\bibitem[{{Johnson} {et~al.}(2004){Johnson}, {Indebetouw}, {Watson}, \&
  {Kobulnicky}}]{johnson2004}
{Johnson}, K.~E., {Indebetouw}, R., {Watson}, C., \& {Kobulnicky}, H.~A. 2004,
  \aj, 128, 610, \dodoi{10.1086/422017}

\bibitem[{{Johnson} \& {Kobulnicky}(2003)}]{johnson2003}
{Johnson}, K.~E., \& {Kobulnicky}, H.~A. 2003, \apj, 597, 923,
  \dodoi{10.1086/378585}

\bibitem[{{Johnson} {et~al.}(2001){Johnson}, {Kobulnicky}, {Massey}, \&
  {Conti}}]{johnson2001}
{Johnson}, K.~E., {Kobulnicky}, H.~A., {Massey}, P., \& {Conti}, P.~S. 2001,
  \apj, 559, 864, \dodoi{10.1086/322335}

\bibitem[{{Johnson} {et~al.}(2015){Johnson}, {Leroy}, {Indebetouw}, {Brogan},
  {Whitmore}, {Hibbard}, {Sheth}, \& {Evans}}]{johnson2015}
{Johnson}, K.~E., {Leroy}, A.~K., {Indebetouw}, R., {et~al.} 2015, \apj, 806,
  35, \dodoi{10.1088/0004-637X/806/1/35}

\bibitem[{{Johnson} {et~al.}(2016){Johnson}, {Seth}, {Dalcanton}, {Beerman},
  {Fouesneau}, {Lewis}, {Weisz}, {Williams}, {Bell}, {Dolphin}, {Larsen},
  {Sandstrom}, \& {Skillman}}]{johnson2016}
{Johnson}, L.~C., {Seth}, A.~C., {Dalcanton}, J.~J., {et~al.} 2016, \apj, 827,
  33, \dodoi{10.3847/0004-637X/827/1/33}

\bibitem[{{Kennicutt} {et~al.}(2008){Kennicutt}, {Lee}, {Funes}, {J.}, {Sakai},
  \& {Akiyama}}]{kennicutt2008}
{Kennicutt}, Robert~C., J., {Lee}, J.~C., {Funes}, J.~G., {et~al.} 2008, \apjs,
  178, 247, \dodoi{10.1086/590058}

\bibitem[{{Kepley} {et~al.}(2014){Kepley}, {Reines}, {Johnson}, \&
  {Walker}}]{kepley2014}
{Kepley}, A.~A., {Reines}, A.~E., {Johnson}, K.~E., \& {Walker}, L.~M. 2014,
  \aj, 147, 43, \dodoi{10.1088/0004-6256/147/2/43}

\bibitem[{{Kewley} \& {Dopita}(2002)}]{kewley2002}
{Kewley}, L.~J., \& {Dopita}, M.~A. 2002, \apjs, 142, 35,
  \dodoi{10.1086/341326}

\bibitem[{{Kobulnicky} \& {Johnson}(1999)}]{kobulnicky1999}
{Kobulnicky}, H.~A., \& {Johnson}, K.~E. 1999, \apj, 527, 154,
  \dodoi{10.1086/308075}

\bibitem[{{Kroupa}(2001)}]{kroupa2001}
{Kroupa}, P. 2001, \mnras, 322, 231, \dodoi{10.1046/j.1365-8711.2001.04022.x}

\bibitem[{{Kruijssen} {et~al.}(2019){Kruijssen}, {Schruba}, {Chevance},
  {Longmore}, {Hygate}, {Haydon}, {McLeod}, {Dalcanton}, {Tacconi}, \& {van
  Dishoeck}}]{kruijssen2019}
{Kruijssen}, J.~M.~D., {Schruba}, A., {Chevance}, M., {et~al.} 2019, \nat, 569,
  519, \dodoi{10.1038/s41586-019-1194-3}

\bibitem[{{Krumholz} {et~al.}(2019){Krumholz}, {McKee}, \& {Bland
  -Hawthorn}}]{krumholz2019}
{Krumholz}, M.~R., {McKee}, C.~F., \& {Bland -Hawthorn}, J. 2019, \araa, 57,
  227, \dodoi{10.1146/annurev-astro-091918-104430}

\bibitem[{{Lada} \& {Lada}(2003)}]{lada2003}
{Lada}, C.~J., \& {Lada}, E.~A. 2003, \araa, 41, 57,
  \dodoi{10.1146/annurev.astro.41.011802.094844}

\bibitem[{{Leitherer} {et~al.}(1999){Leitherer}, {Schaerer}, {Goldader},
  {Delgado}, {Robert}, {Kune}, {de Mello}, {Devost}, \&
  {Heckman}}]{leitherer1999}
{Leitherer}, C., {Schaerer}, D., {Goldader}, J.~D., {et~al.} 1999, \apjs, 123,
  3, \dodoi{10.1086/313233}

\bibitem[{{Lin} {et~al.}(2020){Lin}, {Calzetti}, {Kong}, {Adamo}, {Cignoni},
  {Cook}, {Dale}, {Grasha}, {Grebel}, {Messa}, {Sacchi}, \& {Smith}}]{lin2020}
{Lin}, Z., {Calzetti}, D., {Kong}, X., {et~al.} 2020, \apj, 896, 16,
  \dodoi{10.3847/1538-4357/ab9106}

\bibitem[{{Matthews} {et~al.}(2018){Matthews}, {Johnson}, {Whitmore}, {Brogan},
  {Leroy}, \& {Indebetouw}}]{matthews2018}
{Matthews}, A.~M., {Johnson}, K.~E., {Whitmore}, B.~C., {et~al.} 2018, \apj,
  862, 147, \dodoi{10.3847/1538-4357/aac958}

\bibitem[{{Messa} {et~al.}(2019){Messa}, {Adamo}, {{\"O}stlin}, {Melinder},
  {Hayes}, {Bridge}, \& {Cannon}}]{messa2019}
{Messa}, M., {Adamo}, A., {{\"O}stlin}, G., {et~al.} 2019, \mnras, 487, 4238,
  \dodoi{10.1093/mnras/stz1337}

\bibitem[{{Messa} {et~al.}(2018{\natexlab{a}}){Messa}, {Adamo}, {{\"O}stlin},
  {Calzetti}, {Grasha}, {Grebel}, {Shabani}, {Chandar}, {Dale}, {Dobbs},
  {Elmegreen}, {Fumagalli}, {Gouliermis}, {Kim}, {Smith}, {Thilker}, {Tosi},
  {Ubeda}, {Walterbos}, {Whitmore}, {Fedorenko}, {Mahadevan}, {Andrews},
  {Bright}, {Cook}, {Kahre}, {Nair}, {Pellerin}, {Ryon}, {Ahmad}, {Beale},
  {Brown}, {Clarkson}, {Guidarelli}, {Parziale}, {Turner}, \&
  {Weber}}]{messa2018a}
---. 2018{\natexlab{a}}, \mnras, 473, 996, \dodoi{10.1093/mnras/stx2403}

\bibitem[{{Messa} {et~al.}(2018{\natexlab{b}}){Messa}, {Adamo}, {Calzetti},
  {Reina-Campos}, {Colombo}, {Schinnerer}, {Chand ar}, {Dale}, {Gouliermis},
  {Grasha}, {Grebel}, {Elmegreen}, {Fumagalli}, {Johnson}, {Kruijssen},
  {{\"O}stlin}, {Shabani}, {Smith}, \& {Whitmore}}]{messa2018b}
{Messa}, M., {Adamo}, A., {Calzetti}, D., {et~al.} 2018{\natexlab{b}}, \mnras,
  477, 1683, \dodoi{10.1093/mnras/sty577}

\bibitem[{{Peng} {et~al.}(2002){Peng}, {Ho}, {Impey}, \& {Rix}}]{galfit2002}
{Peng}, C.~Y., {Ho}, L.~C., {Impey}, C.~D., \& {Rix}, H.-W. 2002, \aj, 124,
  266, \dodoi{10.1086/340952}

\bibitem[{{Peng} {et~al.}(2010){Peng}, {Ho}, {Impey}, \& {Rix}}]{galfit2010}
---. 2010, \aj, 139, 2097, \dodoi{10.1088/0004-6256/139/6/2097}

\bibitem[{{Peters} {et~al.}(1994){Peters}, {Freeman}, {Forster}, {Manchester},
  \& {Ables}}]{peters1994}
{Peters}, W.~L., {Freeman}, K.~C., {Forster}, J.~R., {Manchester}, R.~N., \&
  {Ables}, J.~G. 1994, \mnras, 269, 1025, \dodoi{10.1093/mnras/269.4.1025}

\bibitem[{{Reines} {et~al.}(2010){Reines}, {Nidever}, {Whelan}, \&
  {Johnson}}]{reines2010}
{Reines}, A.~E., {Nidever}, D.~L., {Whelan}, D.~G., \& {Johnson}, K.~E. 2010,
  \apj, 708, 26, \dodoi{10.1088/0004-637X/708/1/26}

\bibitem[{{Ryon} {et~al.}(2015){Ryon}, {Bastian}, {Adamo}, {Konstantopoulos},
  {Gallagher}, {Larsen}, {Hollyhead}, {Silva-Villa}, \& {Smith}}]{ryon2015}
{Ryon}, J.~E., {Bastian}, N., {Adamo}, A., {et~al.} 2015, \mnras, 452, 525,
  \dodoi{10.1093/mnras/stv1282}

\bibitem[{{Ryon} {et~al.}(2017){Ryon}, {Gallagher}, {Smith}, {Adamo},
  {Calzetti}, {Bright}, {Cignoni}, {Cook}, {Dale}, {Elmegreen}, {Fumagalli},
  {Gouliermis}, {Grasha}, {Grebel}, {Kim}, {Messa}, {Thilker}, \&
  {Ubeda}}]{ryon2017}
{Ryon}, J.~E., {Gallagher}, J.~S., {Smith}, L.~J., {et~al.} 2017, \apj, 841,
  92, \dodoi{10.3847/1538-4357/aa719e}

\bibitem[{{Silva-Villa} \& {Larsen}(2012)}]{silva-villa2012}
{Silva-Villa}, E., \& {Larsen}, S.~S. 2012, \mnras, 423, 213,
  \dodoi{10.1111/j.1365-2966.2012.20797.x}

\bibitem[{{Sukhbold} {et~al.}(2016){Sukhbold}, {Ertl}, {Woosley}, {Brown}, \&
  {Janka}}]{sukhbold2016}
{Sukhbold}, T., {Ertl}, T., {Woosley}, S.~E., {Brown}, J.~M., \& {Janka}, H.~T.
  2016, \apj, 821, 38, \dodoi{10.3847/0004-637X/821/1/38}

\bibitem[{{Tsai} {et~al.}(2009){Tsai}, {Turner}, {Beck}, {Meier}, \&
  {Ho}}]{tsai2009}
{Tsai}, C.-W., {Turner}, J.~L., {Beck}, S.~C., {Meier}, D.~S., \& {Ho}, P.
  T.~P. 2009, \aj, 137, 4655, \dodoi{10.1088/0004-6256/137/6/4655}

\bibitem[{{Turner} \& {Beck}(2004)}]{turner2004}
{Turner}, J.~L., \& {Beck}, S.~C. 2004, \apjl, 602, L85, \dodoi{10.1086/382699}

\bibitem[{{Vacca} {et~al.}(2002){Vacca}, {Johnson}, \& {Conti}}]{vacca2002}
{Vacca}, W.~D., {Johnson}, K.~E., \& {Conti}, P.~S. 2002, \aj, 123, 772,
  \dodoi{10.1086/338644}

\bibitem[{{V{\'a}zquez} \& {Leitherer}(2005)}]{vazquez2005}
{V{\'a}zquez}, G.~A., \& {Leitherer}, C. 2005, \apj, 621, 695,
  \dodoi{10.1086/427866}

\bibitem[{{Whelan} {et~al.}(2011){Whelan}, {Johnson}, {Whitney}, {Indebetouw},
  \& {Wood}}]{whelan2011}
{Whelan}, D.~G., {Johnson}, K.~E., {Whitney}, B.~A., {Indebetouw}, R., \&
  {Wood}, K. 2011, \apj, 729, 111, \dodoi{10.1088/0004-637X/729/2/111}

\bibitem[{{Whitmore} {et~al.}(2011){Whitmore}, {Chandar}, {Kim}, {Kaleida},
  {Mutchler}, {Stankiewicz}, {Calzetti}, {Saha}, {O'Connell}, {Balick}, {Bond},
  {Carollo}, {Disney}, {Dopita}, {Frogel}, {Hall}, {Holtzman}, {Kimble},
  {McCarthy}, {Paresce}, {Silk}, {Trauger}, {Walker}, {Windhorst}, \&
  {Young}}]{whitmore2011}
{Whitmore}, B.~C., {Chandar}, R., {Kim}, H., {et~al.} 2011, \apj, 729, 78,
  \dodoi{10.1088/0004-637X/729/2/78}

\bibitem[{{Zackrisson} {et~al.}(2001){Zackrisson}, {Bergvall}, {Olofsson}, \&
  {Siebert}}]{zackrisson2001}
{Zackrisson}, E., {Bergvall}, N., {Olofsson}, K., \& {Siebert}, A. 2001, \aap,
  375, 814, \dodoi{10.1051/0004-6361:20010912}

\bibitem[{{Zackrisson} {et~al.}(2011){Zackrisson}, {Rydberg}, {Schaerer},
  {{\"O}stlin}, \& {Tuli}}]{zackrisson2011}
{Zackrisson}, E., {Rydberg}, C.-E., {Schaerer}, D., {{\"O}stlin}, G., \&
  {Tuli}, M. 2011, \apj, 740, 13, \dodoi{10.1088/0004-637X/740/1/13}

\end{thebibliography}
\bibliographystyle{aasjournal}

\appendix

\section{Further details of the size analysis}\label{sec:app_sizeflux}
\begin{figure*}
\centering
\includegraphics[width=0.99\textwidth]{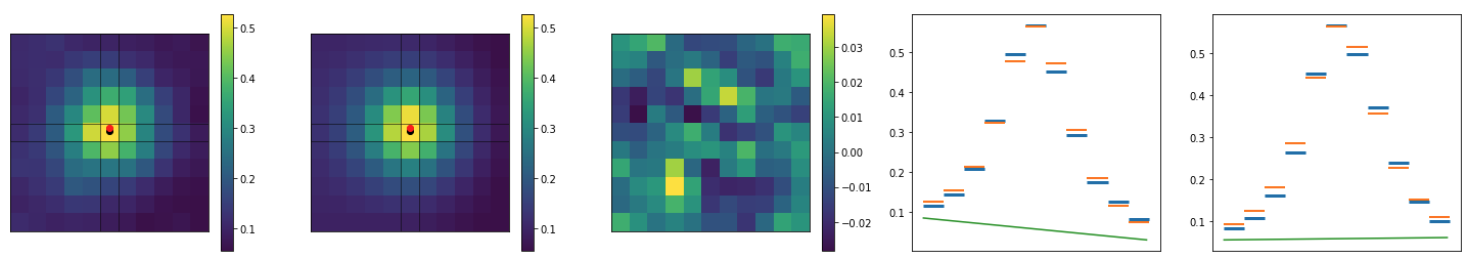}
\caption{Example of a source whose size and photometry are fitted with the size-photometry approach. from left to right are shown: a 2D cutout of the data in F814W (in count/s), its 2D best estimated model, the 2D residuals and the 1D profiles (central row and column) showing data in blue, and best model of the source in orange and best model of the background in green.}
\label{fig:sizeflux}
\end{figure*}

In Section~\ref{sec:sizes-phot} we measure the size of clusters using a size-photometry routine; at the same time this routine allows to measure the cluster fluxes taking into account the width of their light profile, without relying on an aperture correction.
In each filter the source is modelled as a convolution between the instrumental PSF $K_f$, where $f$ denote the current filter, and a Moffat profile \citep{elson1987}. In order to take a non--uniform sky background into  account, the source model was added to a $1^{st}$--degree polynomial (described by 3 parameters, $c_0$, $c_x$ and $c_y$). The observable model (M) is therefore parametrized as:
\begin{equation}
    M_f(x,y|x_0, y_0, F, r_c, c_0, c_x, c_y) = \left[K_f\ast \left(F/F_0\cdot\left(1+(r/r_c)^2\right)^{-1.5}\right)\right] +c_0+c_xx+c_yy
\end{equation}
where $r_c$ is related to the effective radius by $R_{eff}=\sqrt{3}r_c$ and the radial distance $r$ is defined as $r=\sqrt{(x-x_0)^2+(y-y_0)^2}$, where $x,y$ are the pixel coordinates and $x_0,y_0$ are the source coordinates. $F$ and $r_c$ parametrize the flux and size of the source, respectively. We point out that the PSF profile $K_f$ is normalized and that the Moffat profile is also normalized via the $F_0$. 
Since nearby sources can cause contamination especially in the bluer filters, we use F814W to estimate the coordinates and the size of each of the sources. Fig.~\ref{fig:sizeflux} shows an example from the fit of one of the clusters. 
The flux uncertainties are correlated with uncertainties on the other parameters, especially with the size. In order to take it into account, we repeat the fit in the F814W filter twice, the second time keeping fixed coordinates and size. We consider for each source the relative uncertainty given by the size and position uncertainties:
\begin{equation}
    R_{r_cxy}= \frac{\sqrt{F_{F814W,err,1}^2-F_{F814W,err,2}^2}}{F}
\end{equation}
where $F_{F814W,err,1}$ and $F_{F814W,err,1}$ are the total uncertainties on the flux uncertainties in the two repetitions of the fit. Then for every filter $f$ we use this relative value to correct the uncertainty derived from the fit with fixed size, $F_{f,err}$:
\begin{equation}
F_{f,err,tot} = \sqrt{(R_{r_cxy}\cdot F_f)^2+(F_{f,err})^2} 
\end{equation}
where $F_f$ is the flux measured for the current source in the filter $f$.

\section{Additional plots for the completeness test}\label{sec:app_completeness}
\begin{figure*}
\centering
\includegraphics[width=0.49\textwidth]{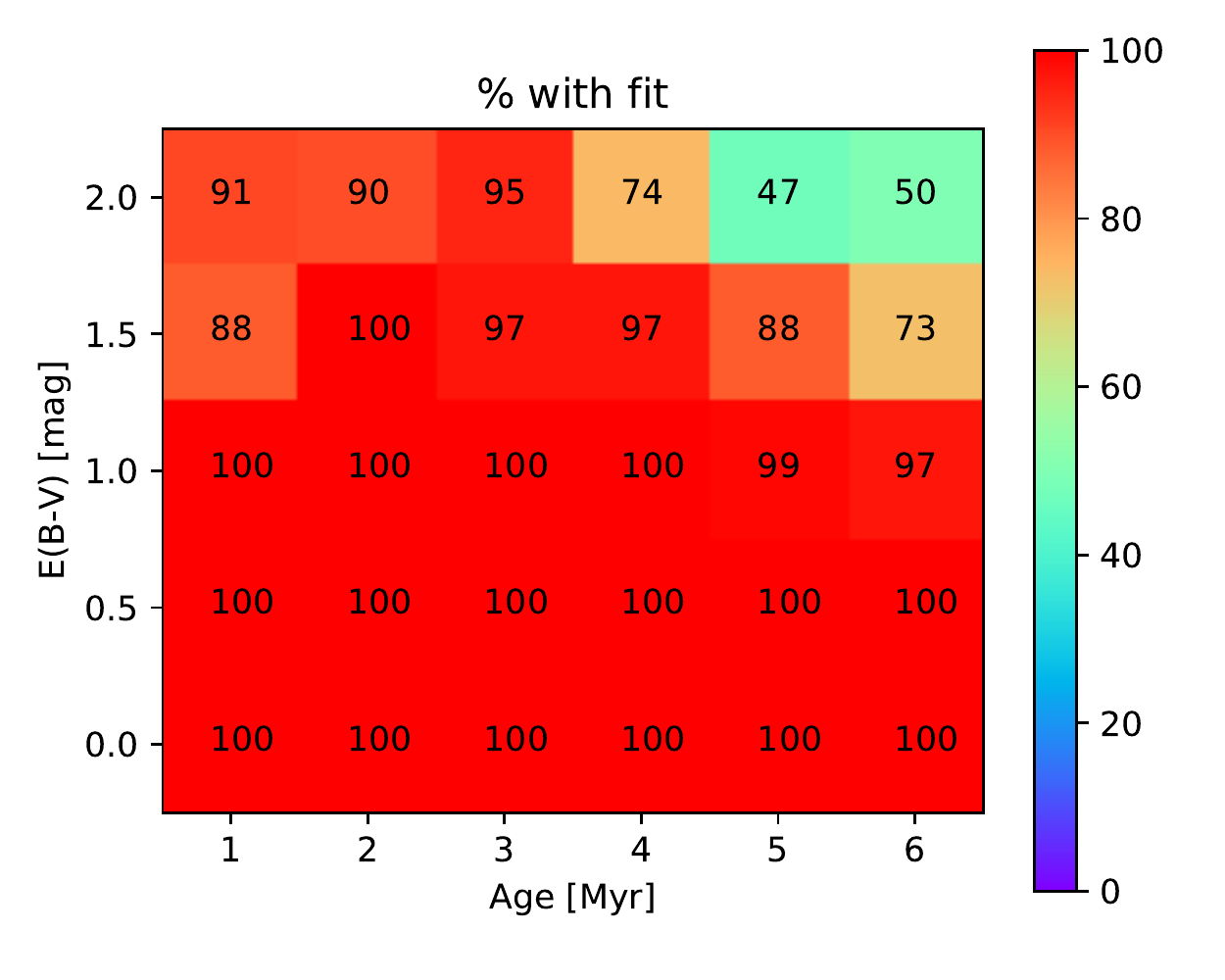}
\caption{Percentage of simulated clusters with detections in at least 5 filters (counting both broad and narrow ones). This is the minimum number of filters required for the fit.}
\label{fig:completeness_percentage}
\end{figure*}

\begin{figure*}
\centering
\subfigure{\includegraphics[width=0.49\textwidth]{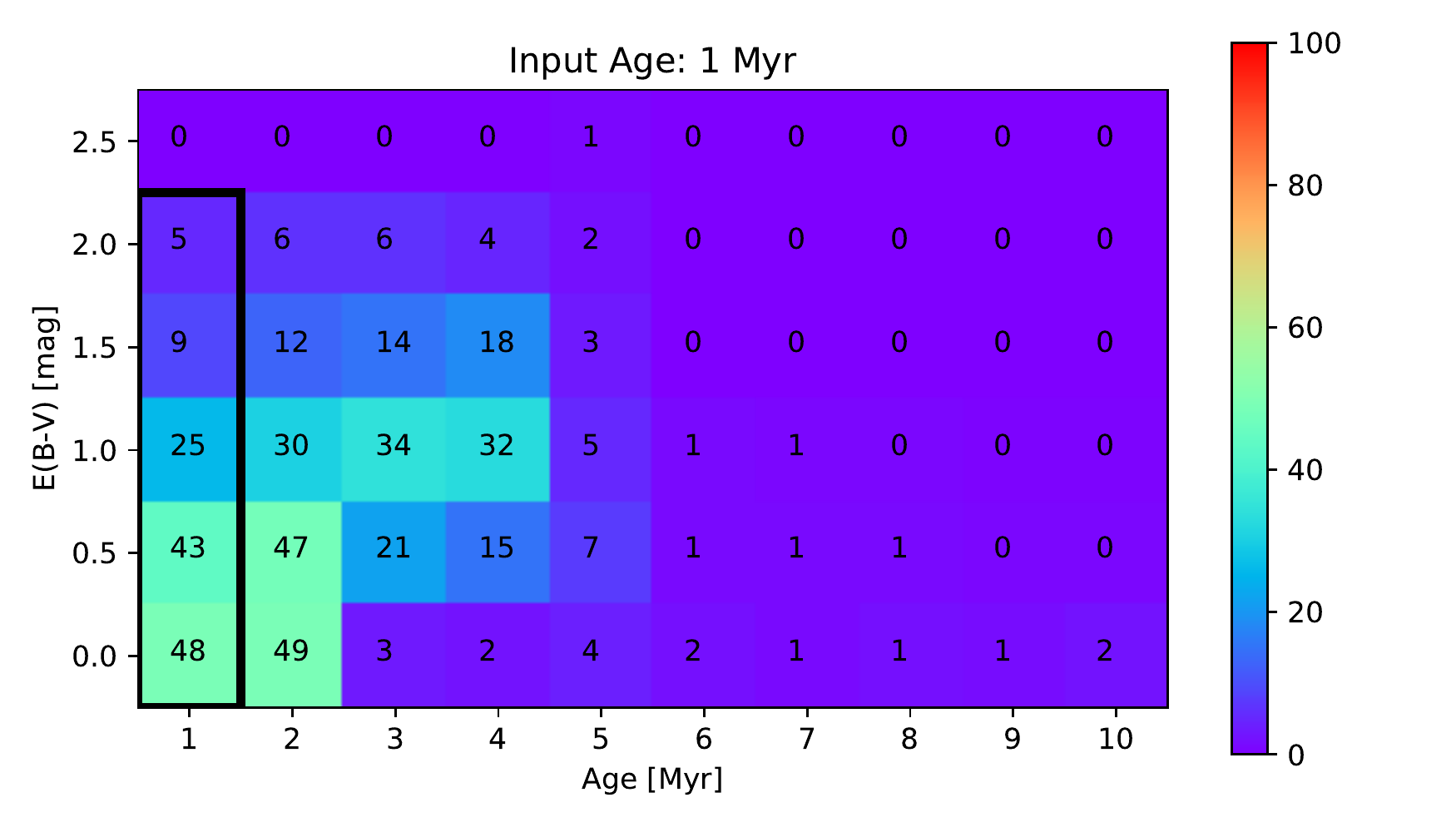}}
\subfigure{\includegraphics[width=0.49\textwidth]{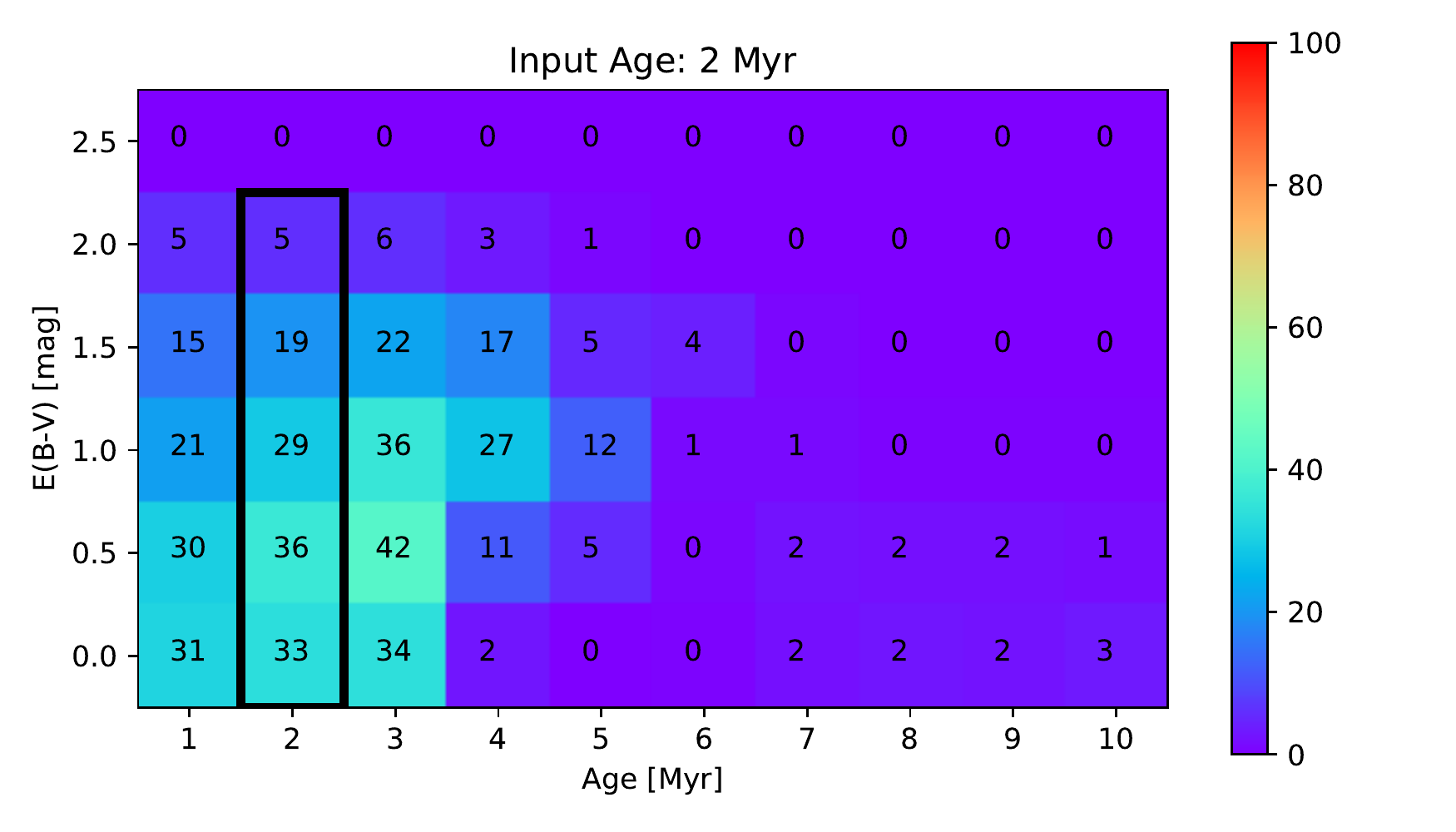}}
\subfigure{\includegraphics[width=0.49\textwidth]{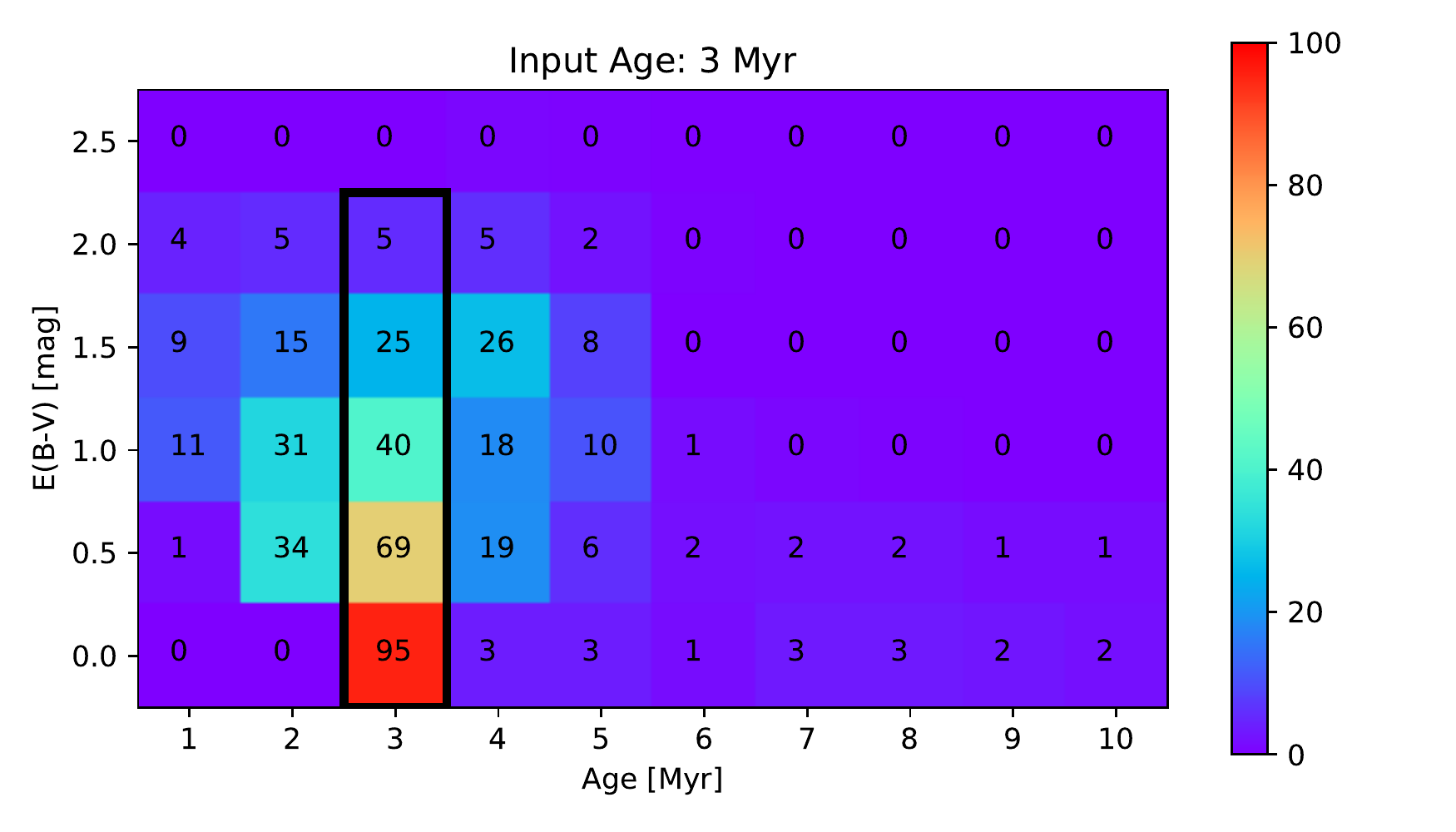}}
\subfigure{\includegraphics[width=0.49\textwidth]{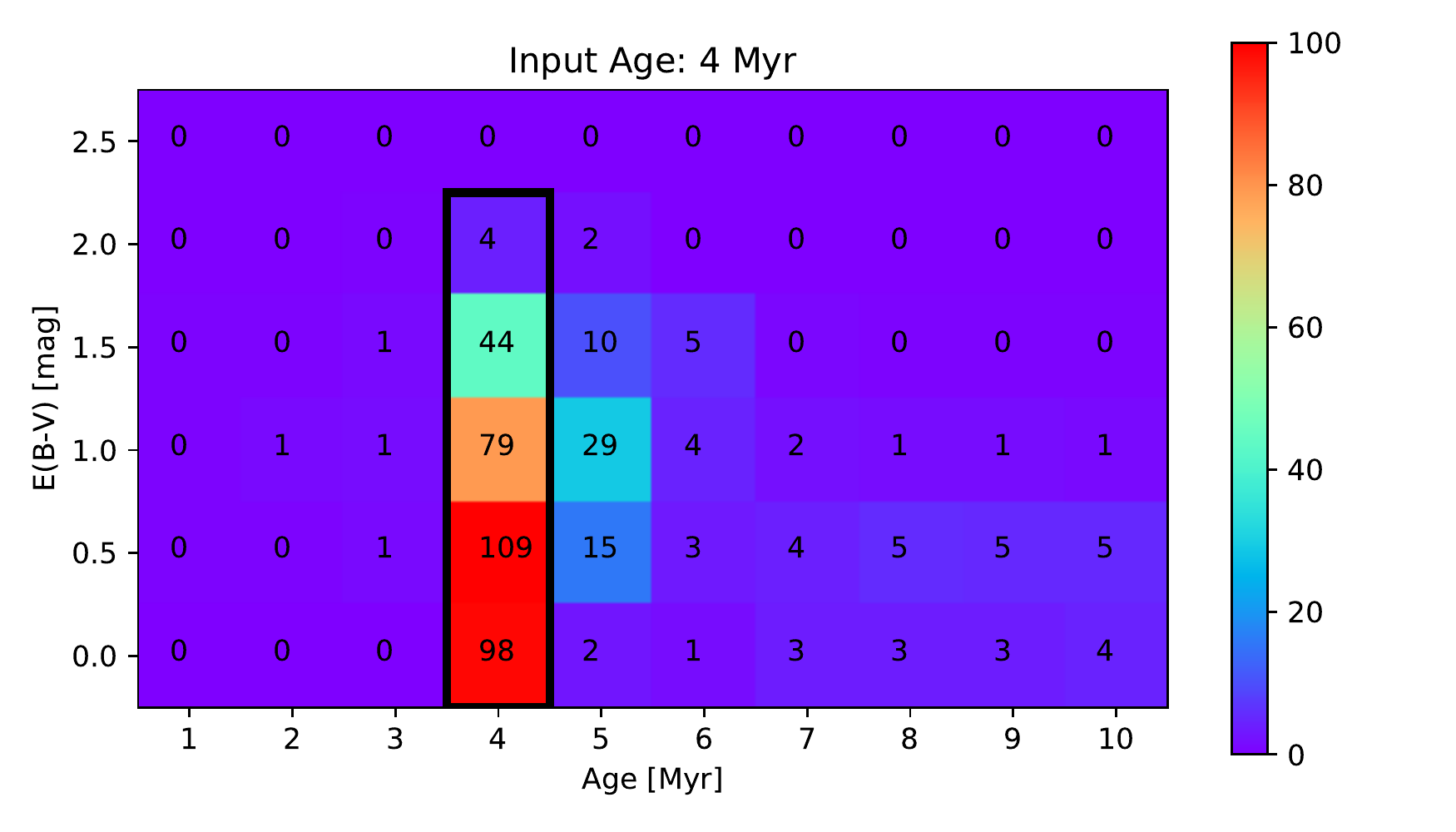}}
\subfigure{\includegraphics[width=0.49\textwidth]{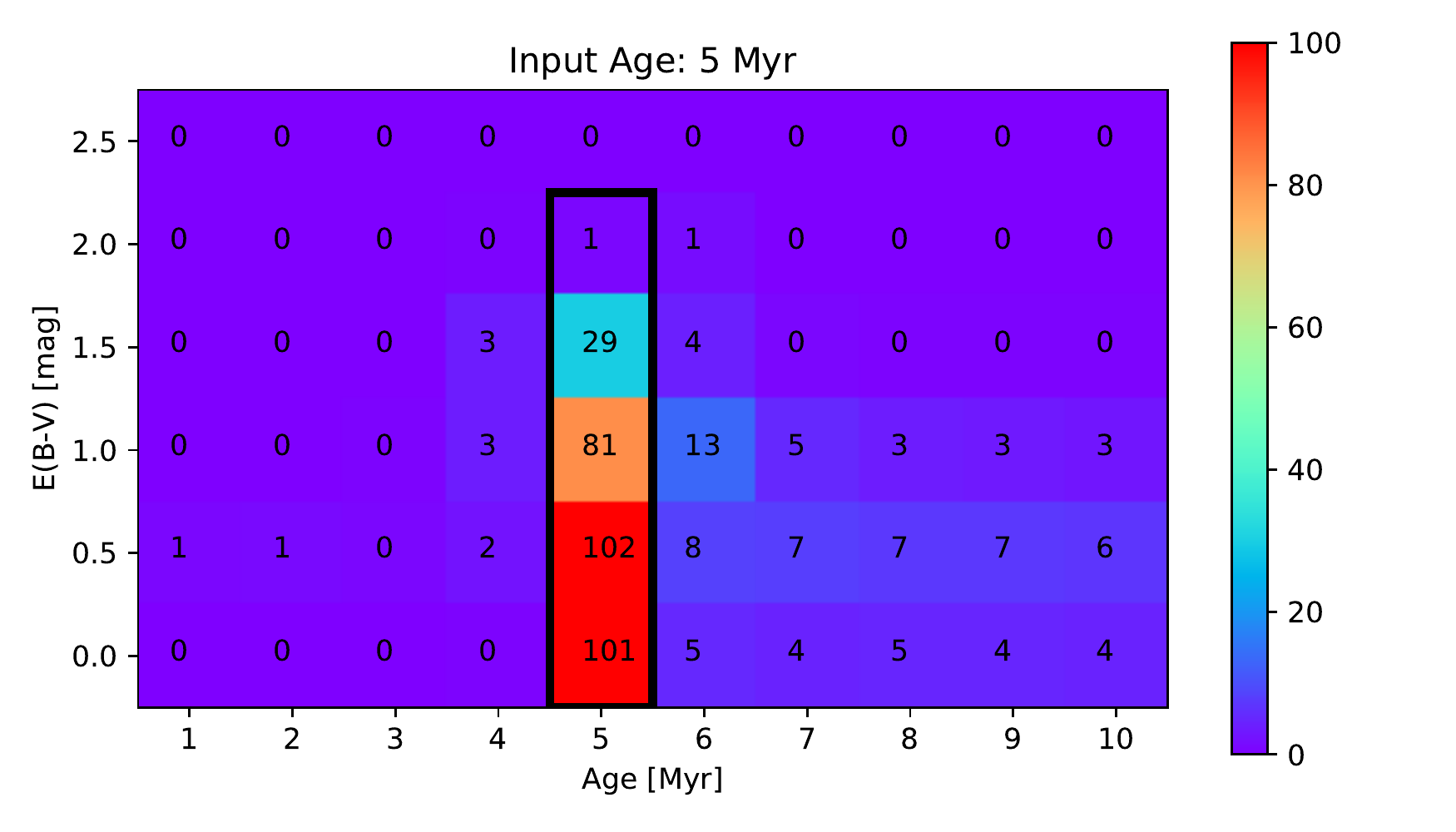}}
\subfigure{\includegraphics[width=0.49\textwidth]{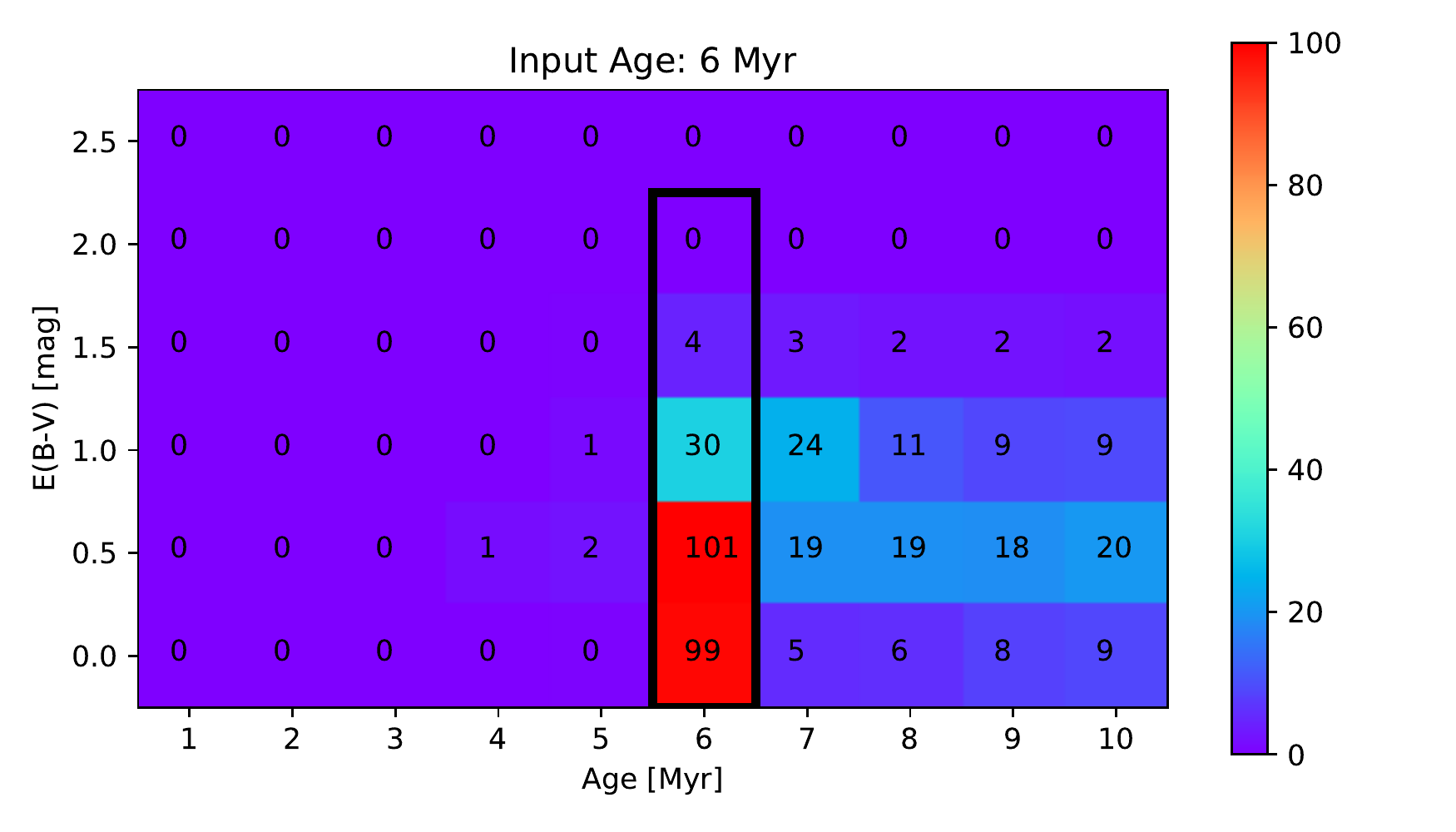}}
\caption{Density plot of the recovered ages and extinctions, as in Fig.~\ref{fig:completeness_density}, but separating the input sources per age bins (as indicated by the black thick contours). }
\label{fig:app_completeness_density}
\end{figure*}

We report in Tab.~\ref{tab:completeness_extraction} and Fig.~\ref{fig:completeness_percentage} a summary of the completeness in detecting photometrically the input sources in the context of the completeness test described in the main text in Section~\ref{sec:completeness_set} and Section~\ref{sec:completeness}. In Fig.~\ref{fig:app_completeness_density} we report the density plot showing how the recovered ages and extinctions are distributed, for each of the input ages used. The main conclusion that can be drawn form Fig.~\ref{fig:app_completeness_density} is that our age determination via broad-- (and narrow--) band SED fitting is degenerate for very young ages; in fact, for the simulated clusters with age of 1 Myr, we retrieved estimated ages equally split between 1 and 2 Myr. Similarly the estimated ages for the simulated clusters of 2 Myr span the entire range up to 3 Myr with almost constant frequency. Only for simulated clusters with ages of 4 Myr (and older) the SED fitting returns consistent ages, with little age spread. This effect is mainly caused by the stellar models used being very similar with each other for young ages, causing degeneracy in the SED fitting process.
For this reason, in the main text we consider as degenerate the SEDs with ages of 3 Myr and below. 

\begin{table*}
    \centering
    \begin{tabular}{|c|ccccc|ccccc|}
    \hline
    \multicolumn{1}{|c|}{Age} & \multicolumn{5}{c|}{E(B-V)\  F814W} & \multicolumn{5}{c|}{E(B-V)\  F128N} \\
    \ & 0.0 & 0.5 & 1.0 & 1.5 & 2.0 & 0.0 & 0.5 & 1.0 & 1.5 & 2.0\\
    \hline
    1 Myr & 100 & 100 & 99  & 95 & 86 & 99  & 98  & 99 & 99  & 98 \\  
    2 Myr & 100 & 100 & 99  & 97 & 80 & 100 & 100 & 99 & 100 & 96 \\  
    3 Myr & 100 & 99  & 100 & 92 & 85 & 100 & 99  & 99 & 93  & 93 \\  
    4 Myr & 100 & 99  & 100 & 94 & 84 & 99  & 97  & 98 & 90  & 86 \\  
    5 Myr & 100 & 99  & 100 & 95 & 75 & 98  & 97  & 90 & 92  & 73 \\  
    6 Myr & 100 & 100 & 100 & 97 & 90 & 97  & 97  & 96 & 85  & 82 \\  
    \hline
   \end{tabular}
    \caption{Number of simulated sources with photometry recovered with an uncertainty below 0.3 mag (i.e. $\rm S/N>3$) for each combination of age and extinction (out of an initial value of 100). Values are reported for two pivotal filters in the catalog extraction, F814W (for \textit{ExtmapCat}) and F128N (for \textit{PBcompactCat}).}
    \label{tab:completeness_extraction}
\end{table*}

\section{Individual fit results}
We report in this section the plots of the SED fitting process for some of the sources discussed in the main text.

\subsection{Comparing solar and sub-solar metallicities}\label{sec:app_testmet}
\begin{figure*}
\centering
\subfigure{\includegraphics[width=0.49\textwidth]{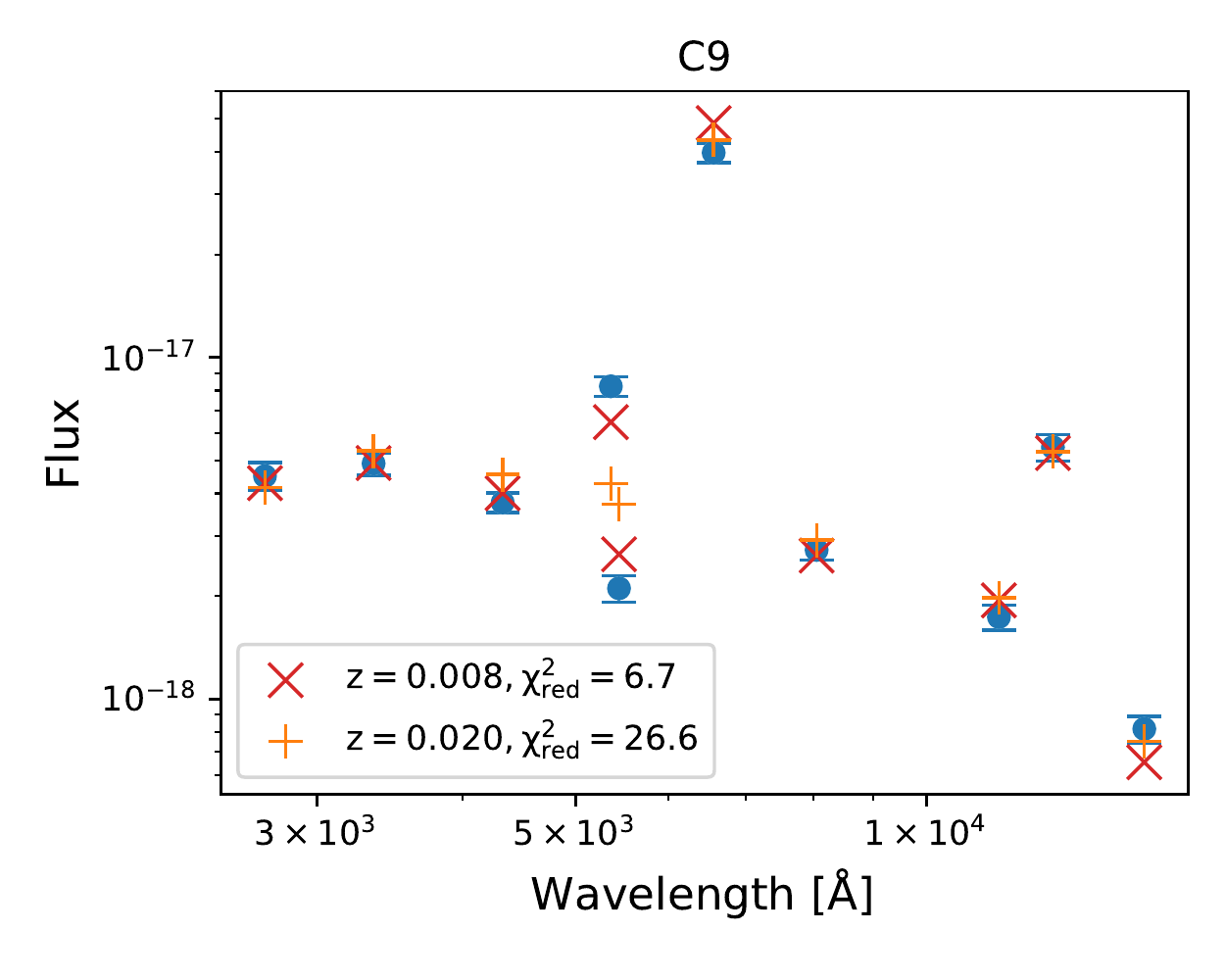}}
\subfigure{\includegraphics[width=0.49\textwidth]{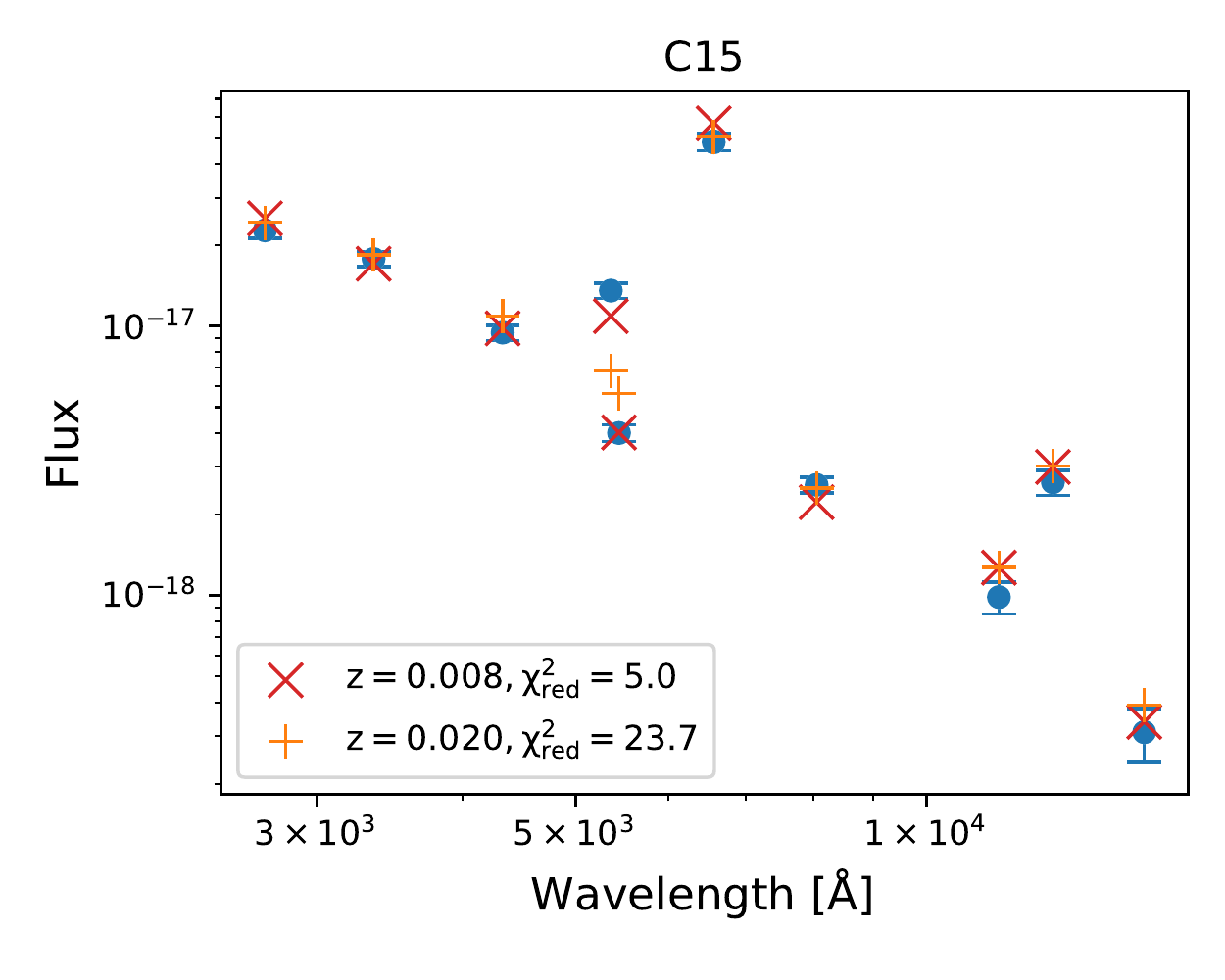}}
\subfigure{\includegraphics[width=0.49\textwidth]{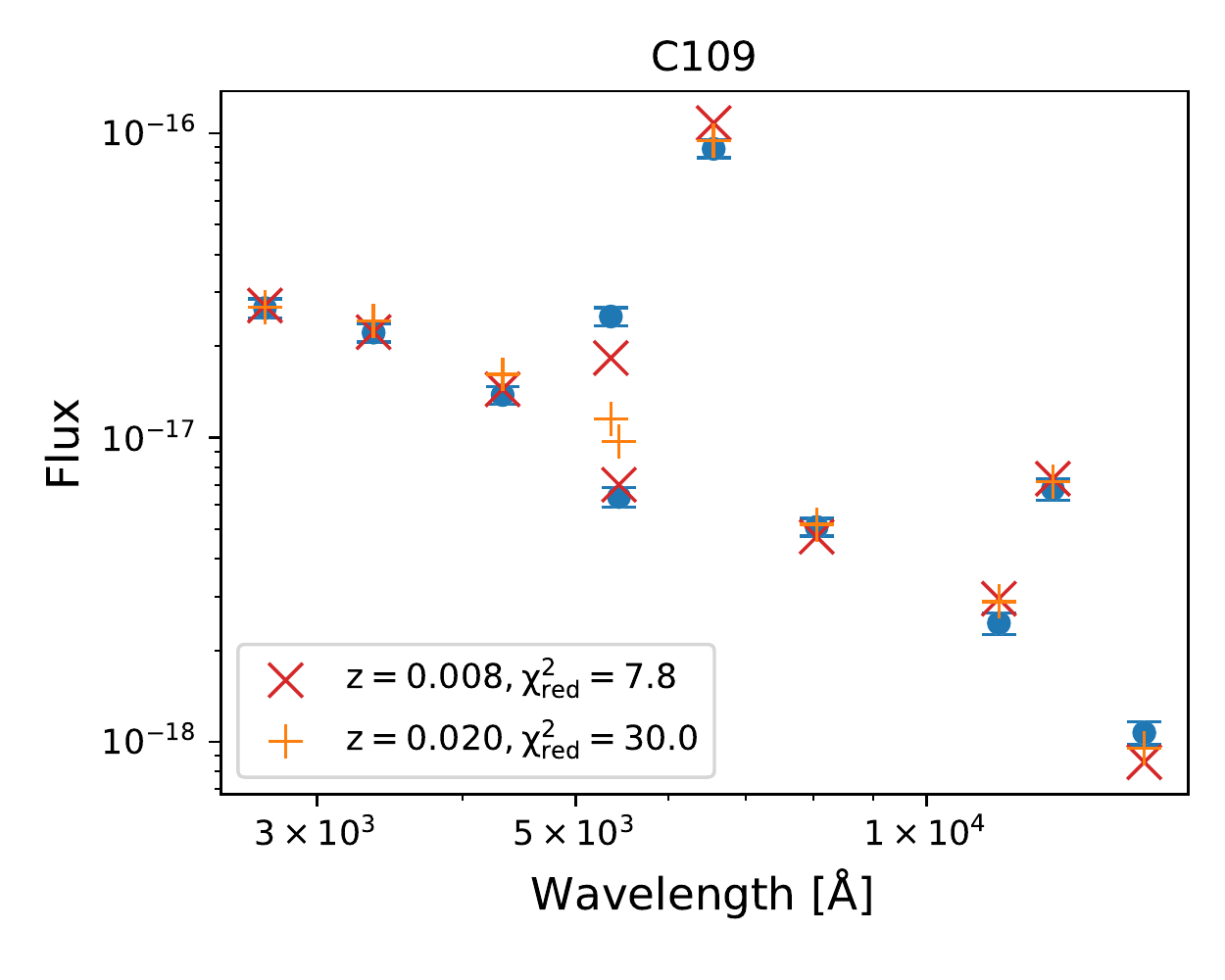}}
\subfigure{\includegraphics[width=0.49\textwidth]{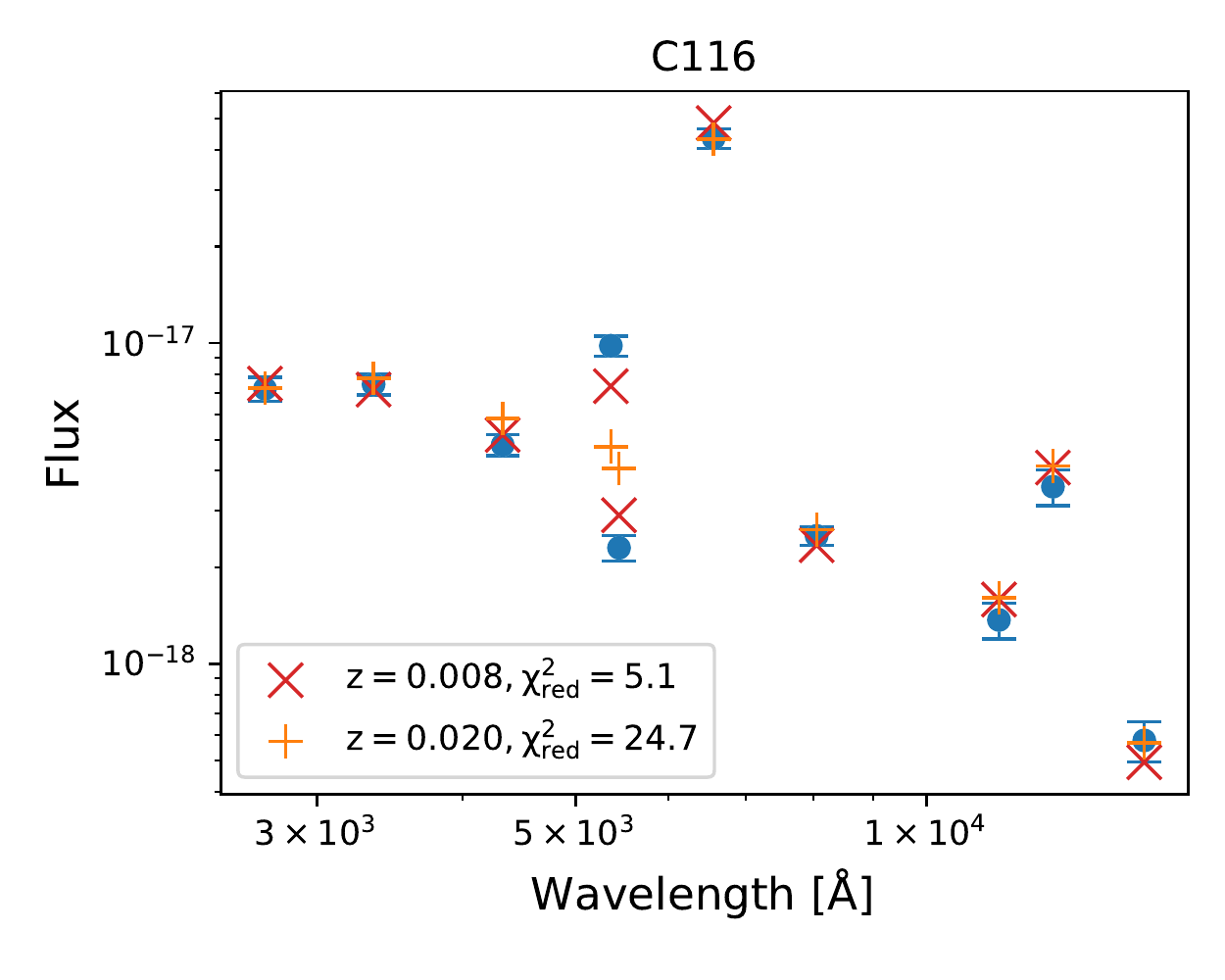}}
\caption{Examples of fits of young clusters using models with different metallicities, Z=0.008 (red crosses) and Z=0.020 (orange plus). Filters F547M and F555W determine the main difference in the recovered \chisq\ values, suggesting that models with sub-solar metallicites are a better assumption for the young clusters in NGC 1313.}
\label{fig:individual_met}
\end{figure*}

We show in Fig.~\ref{fig:individual_met} some examples of the observed broad-band SED of young sources with line emission and the best-fit SED values using solar and sub-solar metallicites. While in both cases the best-fit values for the ages denote young sources ($\rm age\le5\ Myr$), the models with sub-solar metallicity are more capable of capturing the flux difference observed between the F555W and the F547M filters. Also the values of the best $\chi^2_{red}$ in the two cases confirm this trend.

\subsection{young and extincted candidates}\label{sec:ind_fit_extincted}
We show in Fig.~\ref{fig:individual_ext1} and Fig.~\ref{fig:individual_ext2} the sources whose best fit results suggest them being young and extincted. The single cases are discussed in Section~\ref{sec:discuss_single_extincted}. 

\begin{figure*}
\centering
\subfigure{\includegraphics[width=0.26\textwidth]{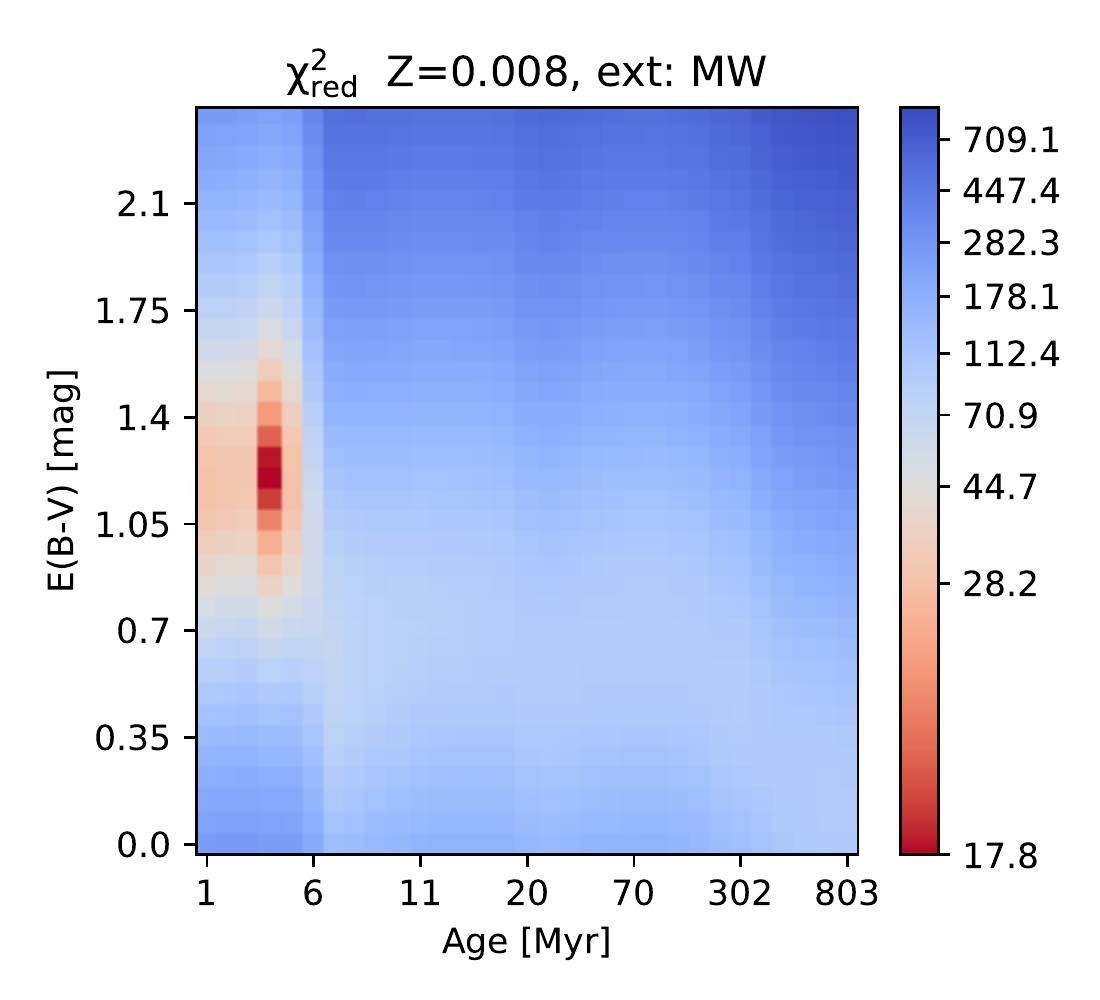}}
\subfigure{\includegraphics[width=0.23\textwidth]{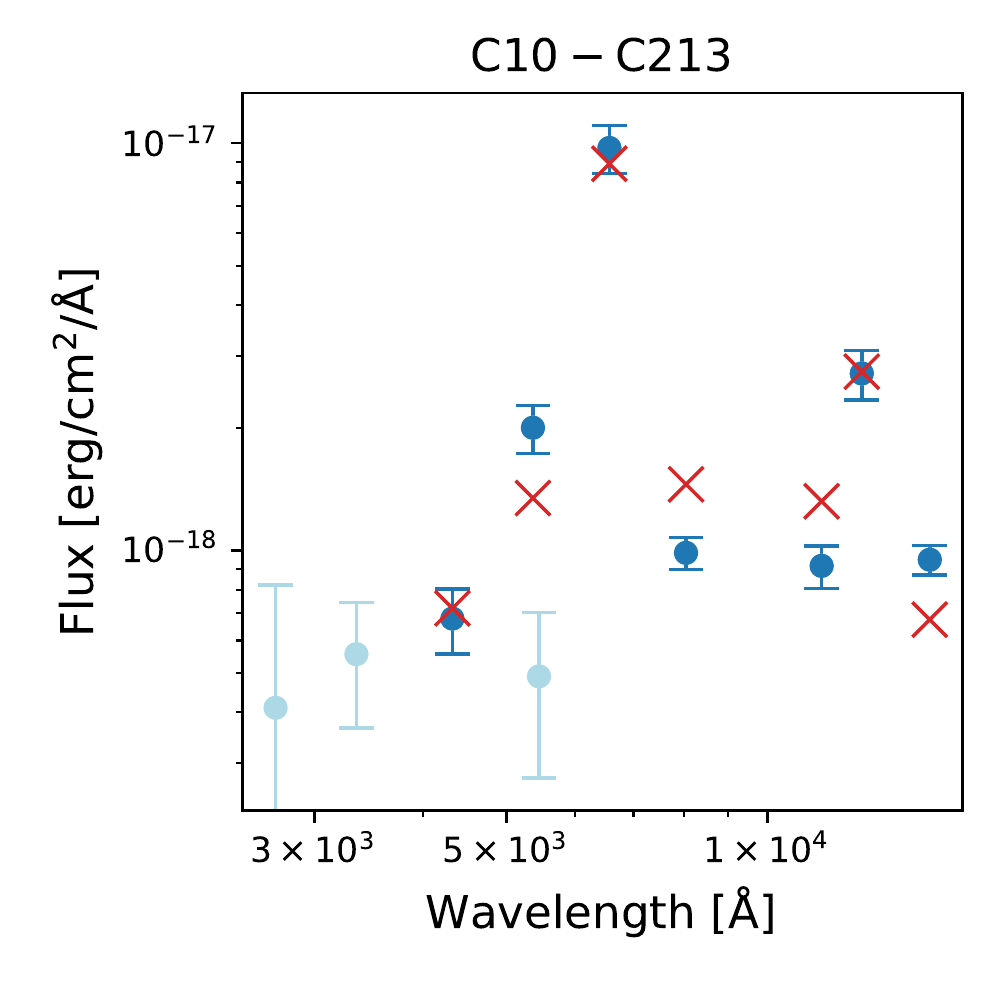}}
\subfigure{\includegraphics[width=0.26\textwidth]{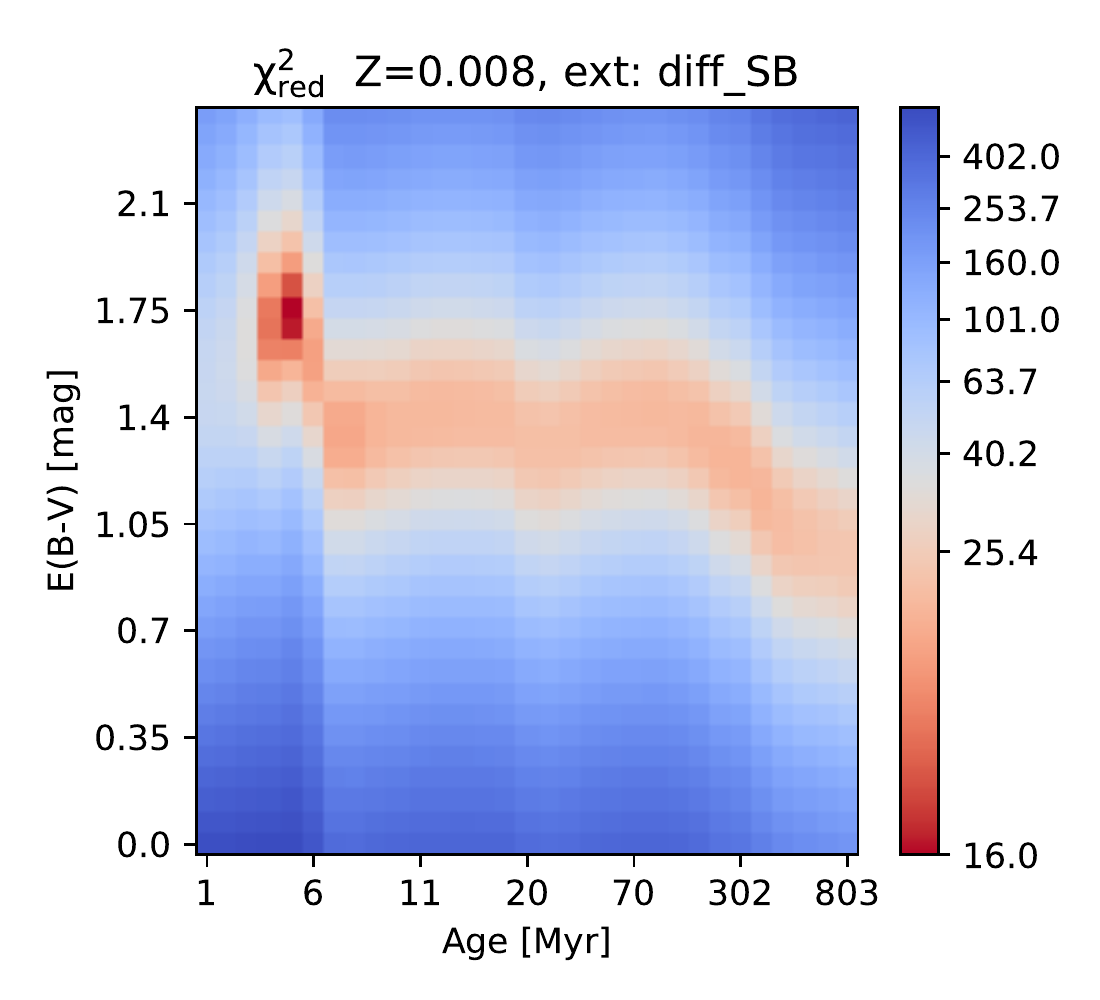}}
\subfigure{\includegraphics[width=0.23\textwidth]{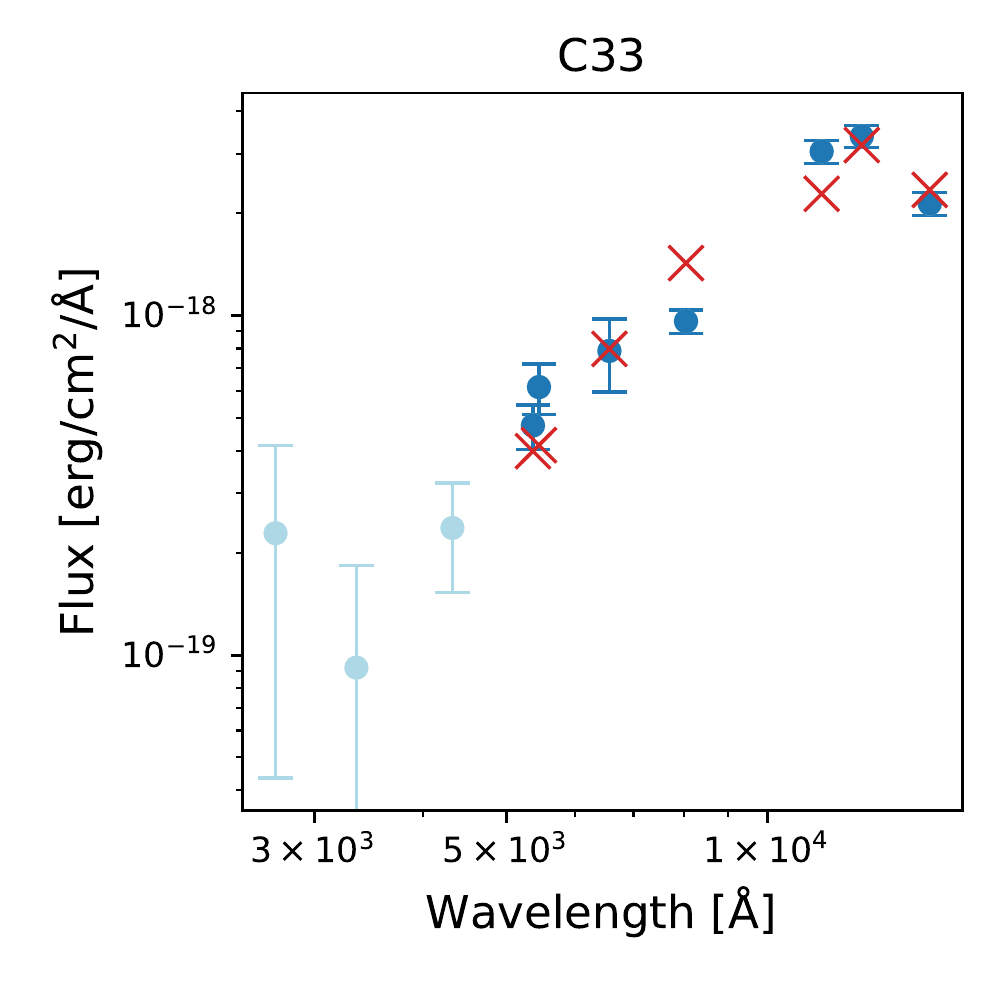}}
\subfigure{\includegraphics[width=0.26\textwidth]{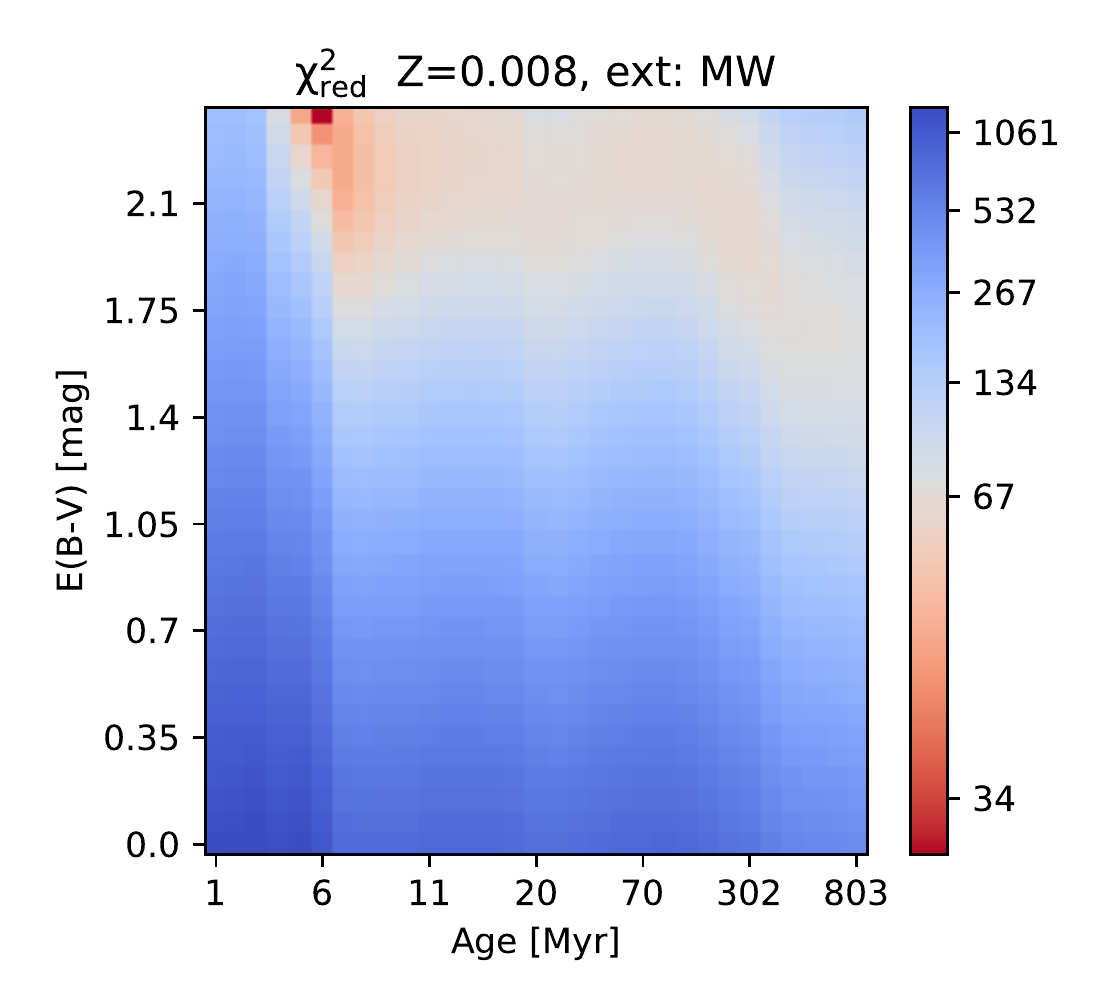}}
\subfigure{\includegraphics[width=0.23\textwidth]{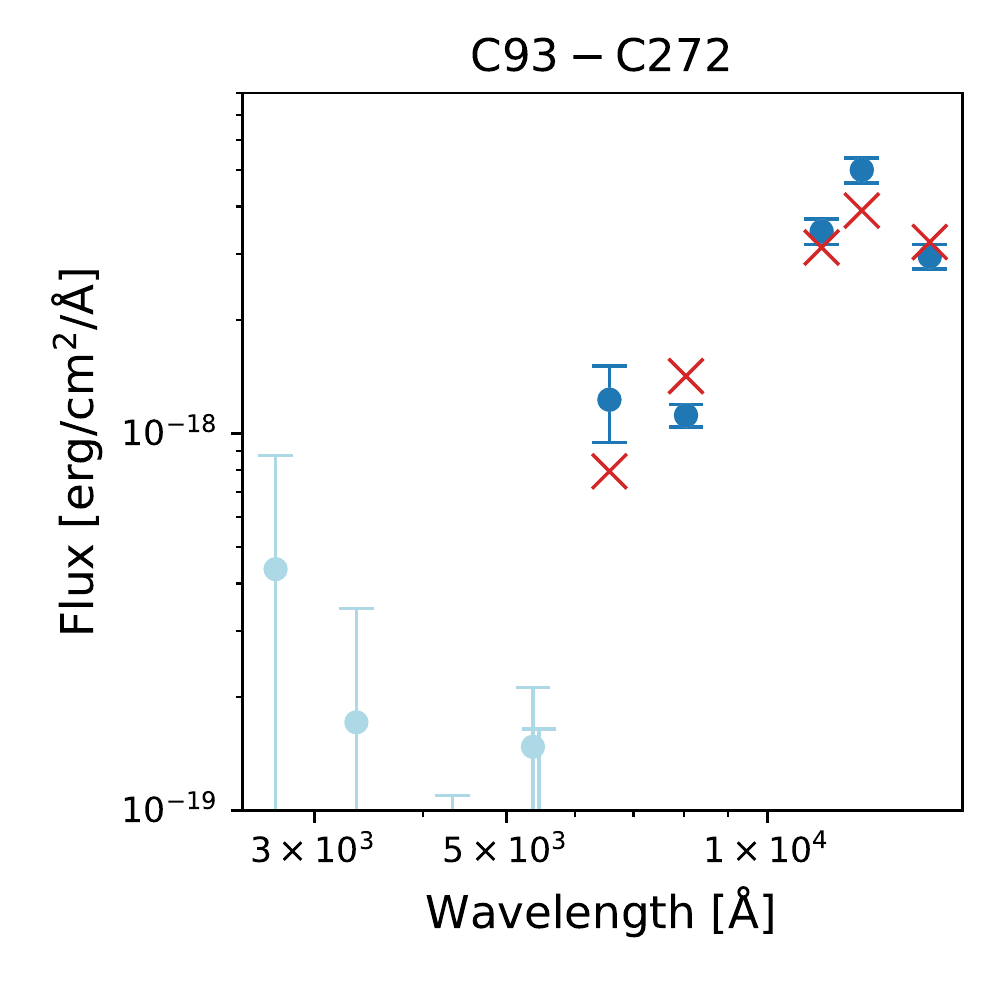}}
\subfigure{\includegraphics[width=0.26\textwidth]{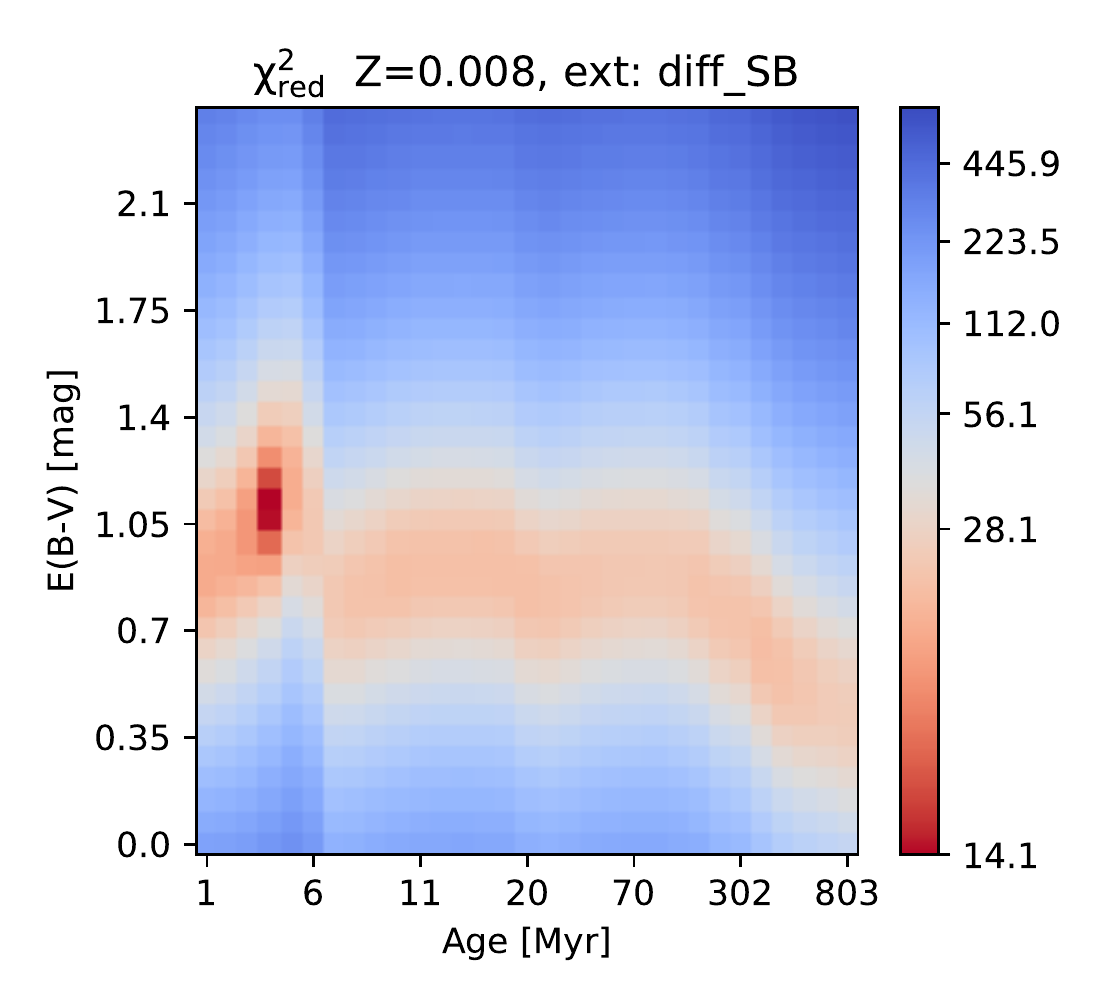}}
\subfigure{\includegraphics[width=0.23\textwidth]{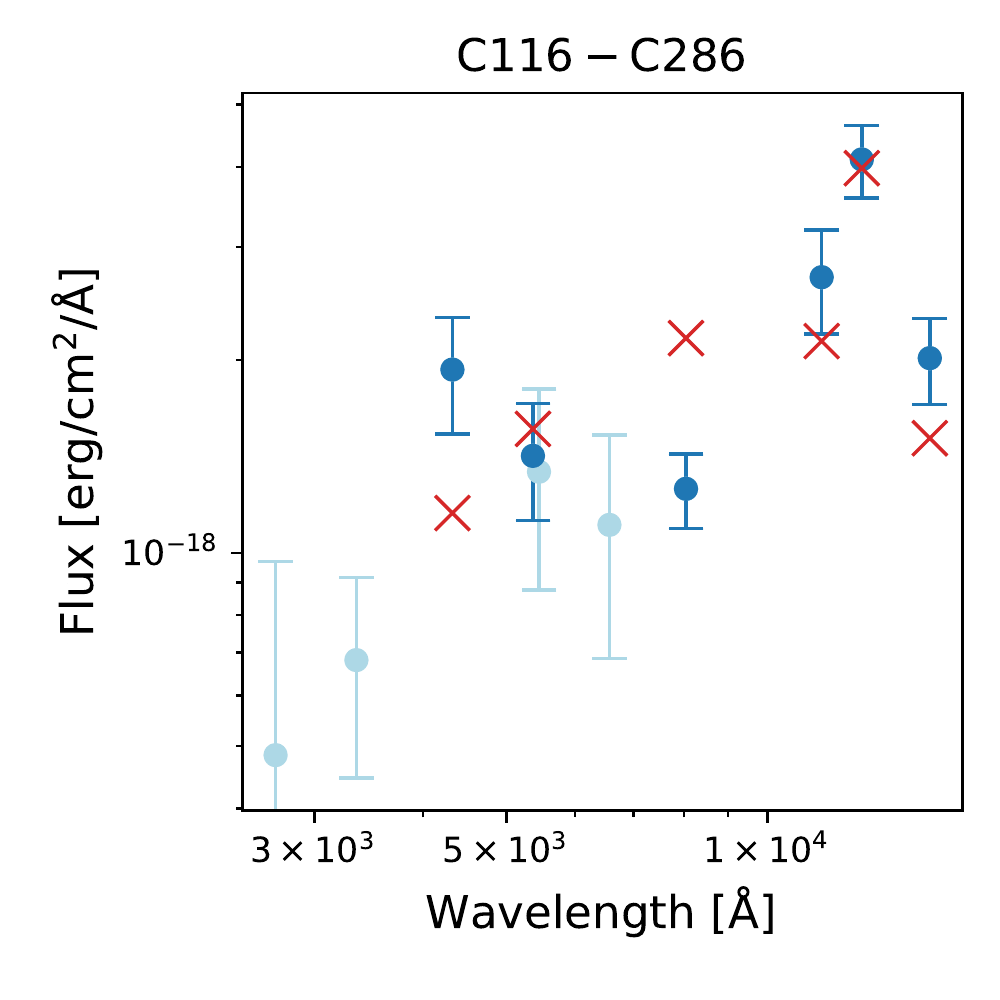}}
\subfigure{\includegraphics[width=0.26\textwidth]{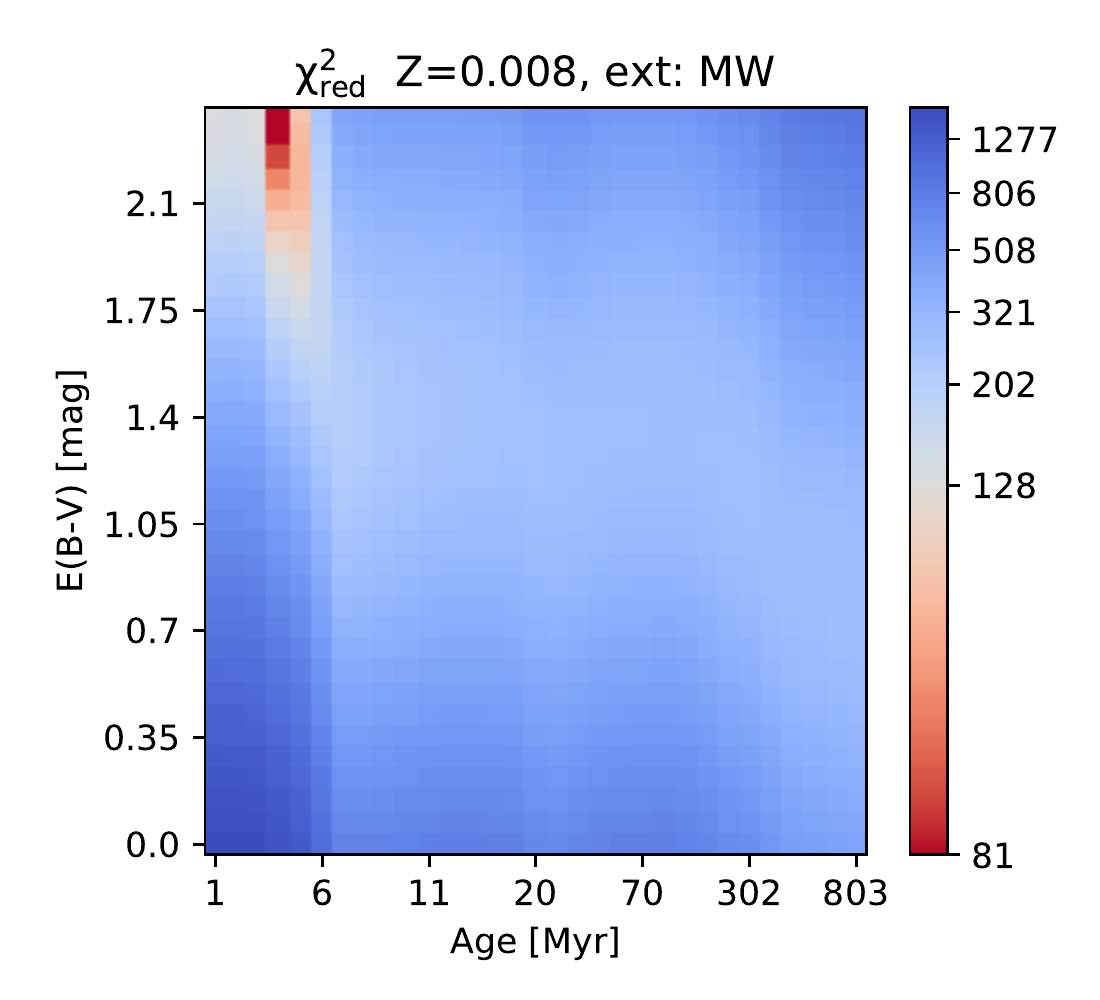}}
\subfigure{\includegraphics[width=0.23\textwidth]{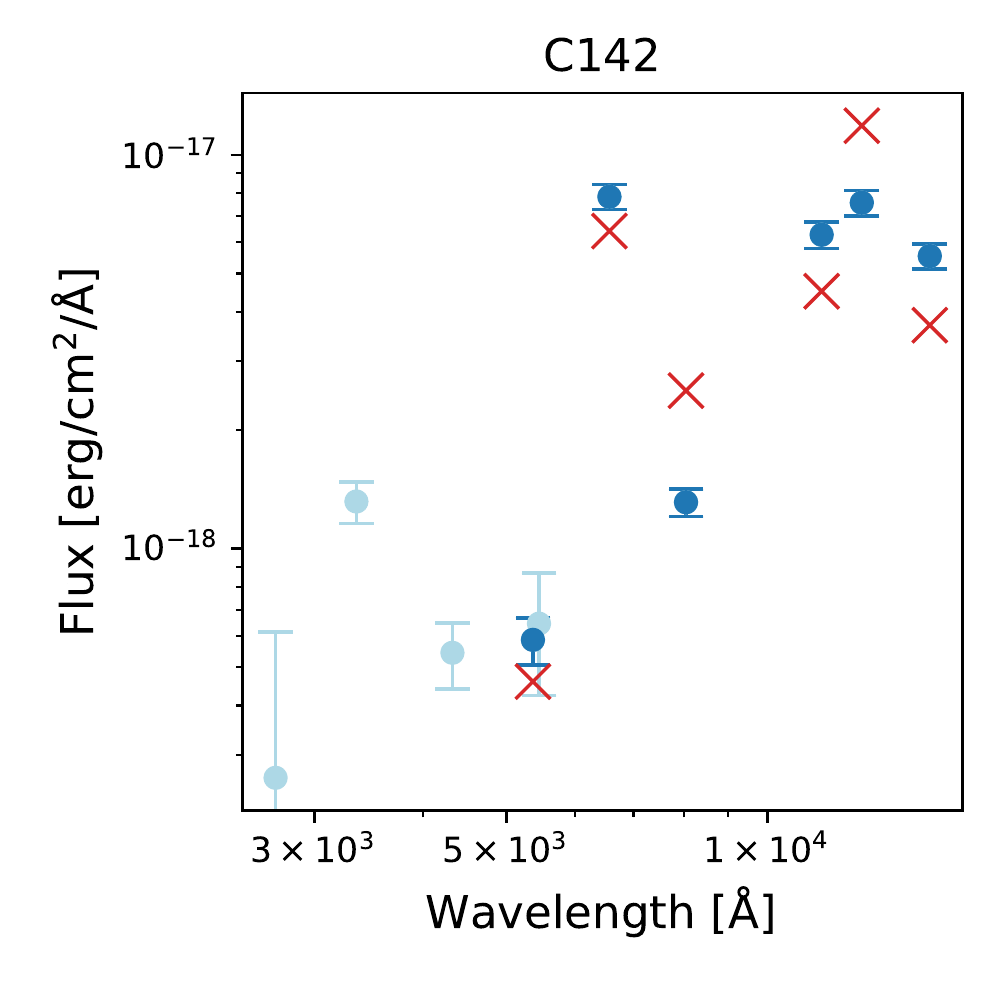}}
\subfigure{\includegraphics[width=0.26\textwidth]{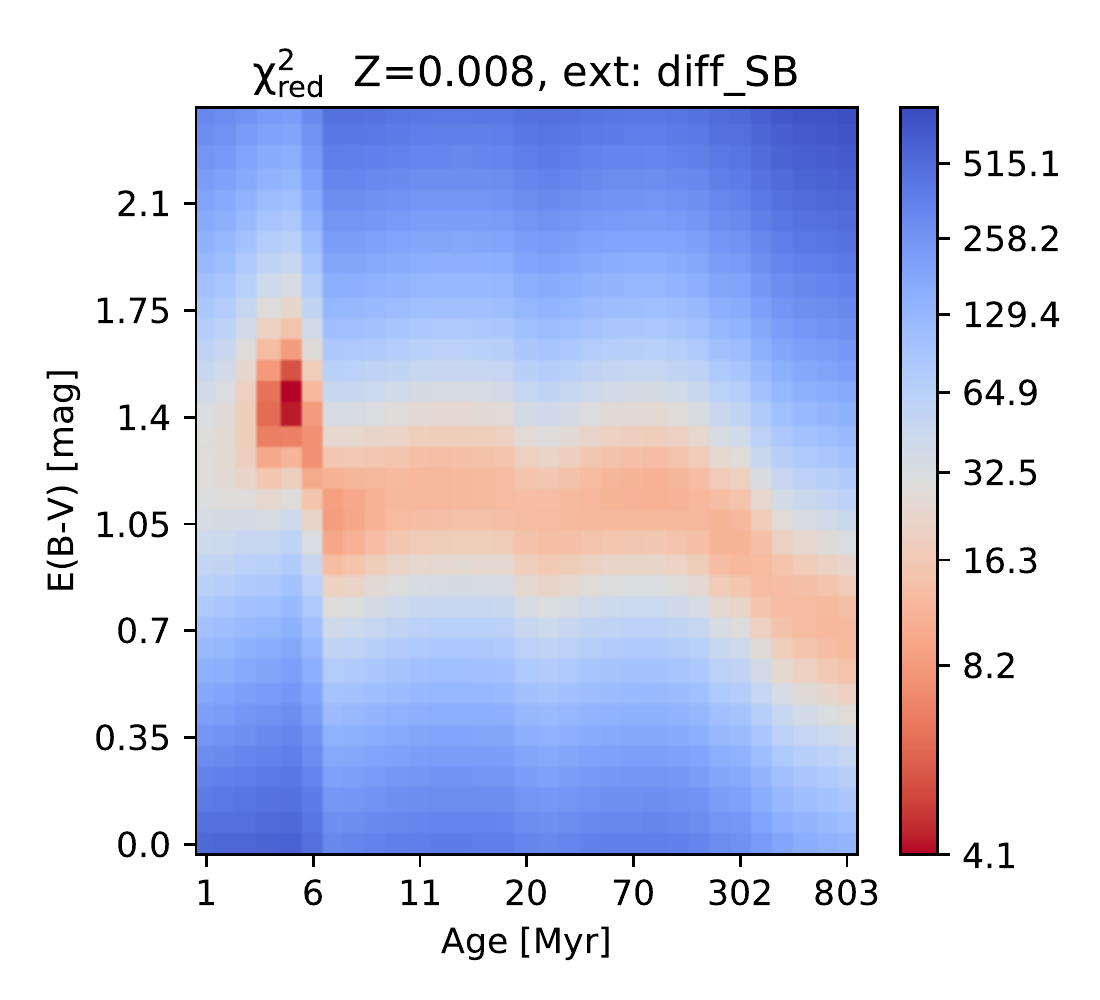}}
\subfigure{\includegraphics[width=0.23\textwidth]{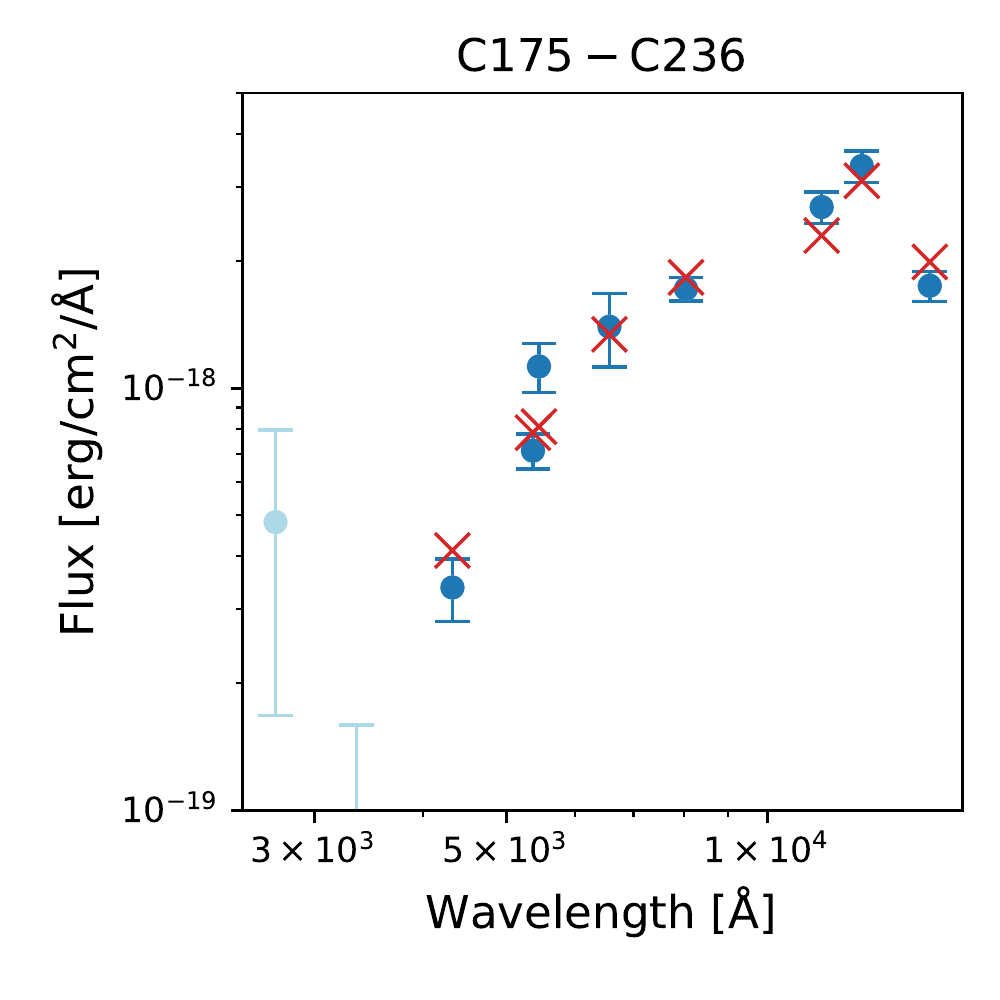}}
\subfigure{\includegraphics[width=0.26\textwidth]{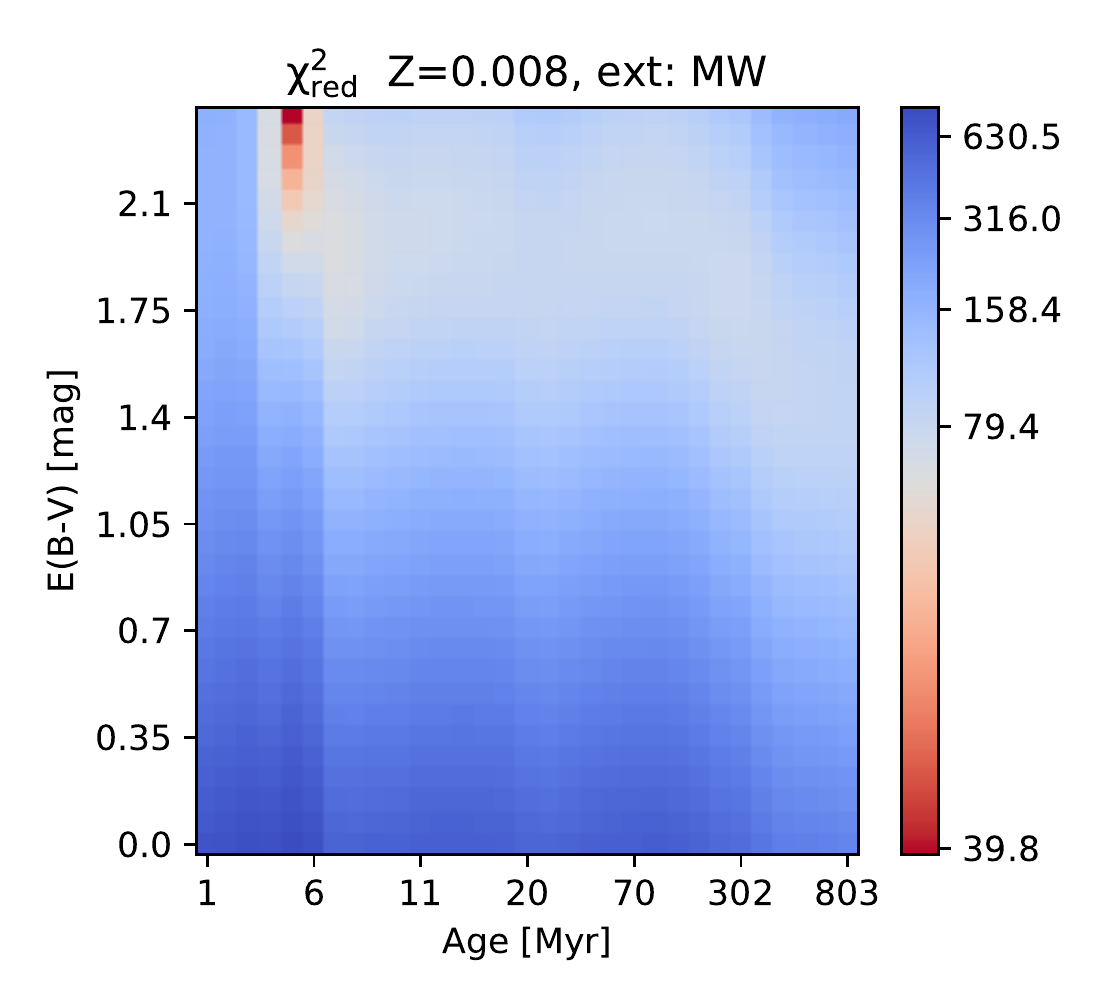}}
\subfigure{\includegraphics[width=0.23\textwidth]{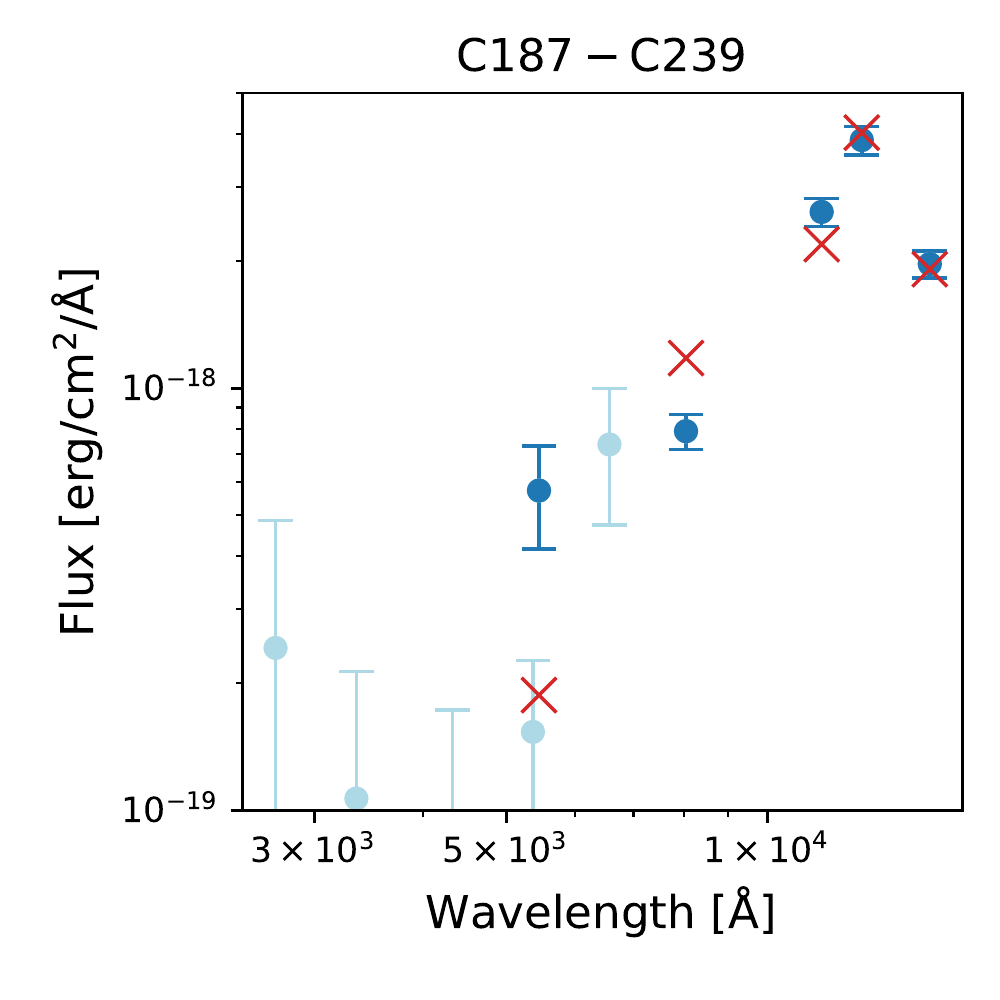}}
\subfigure{\includegraphics[width=0.26\textwidth]{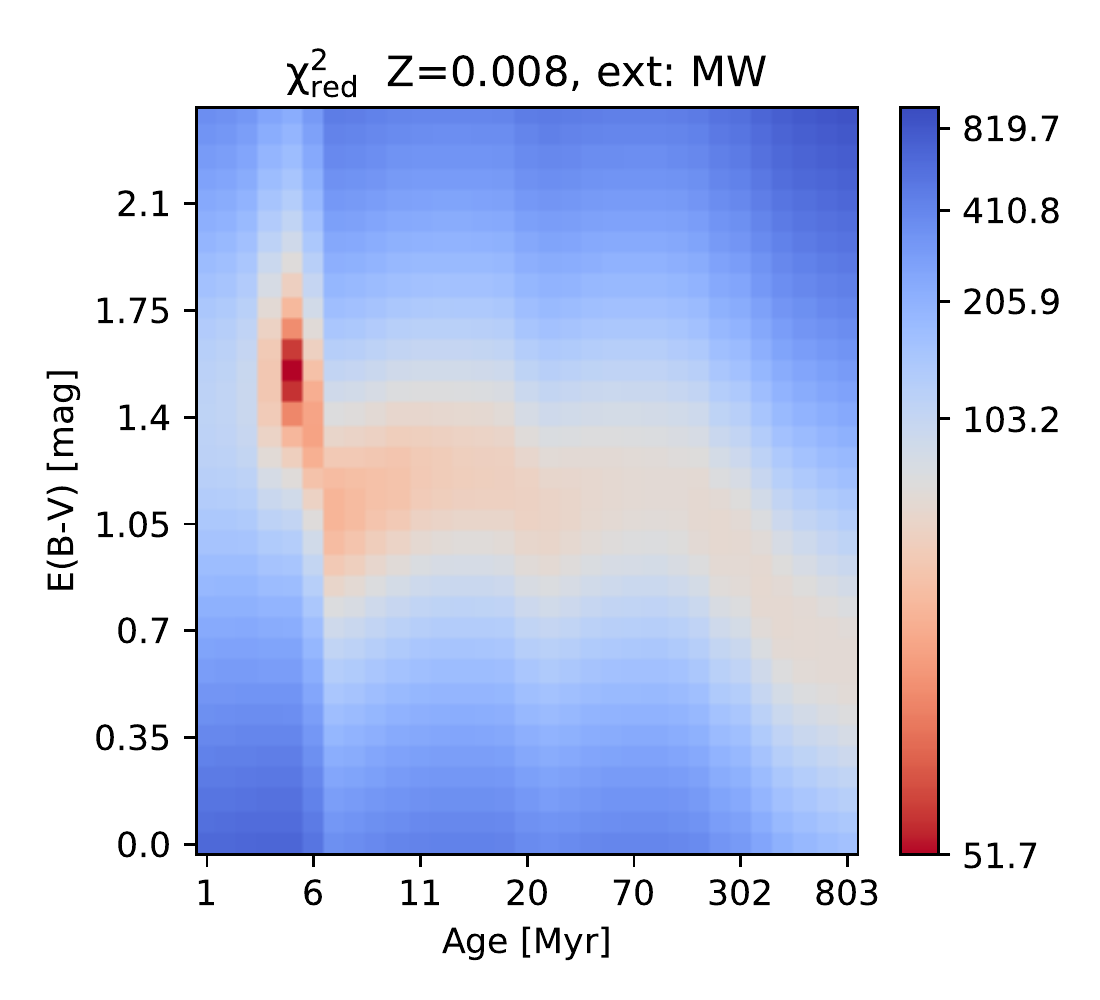}}
\subfigure{\includegraphics[width=0.23\textwidth]{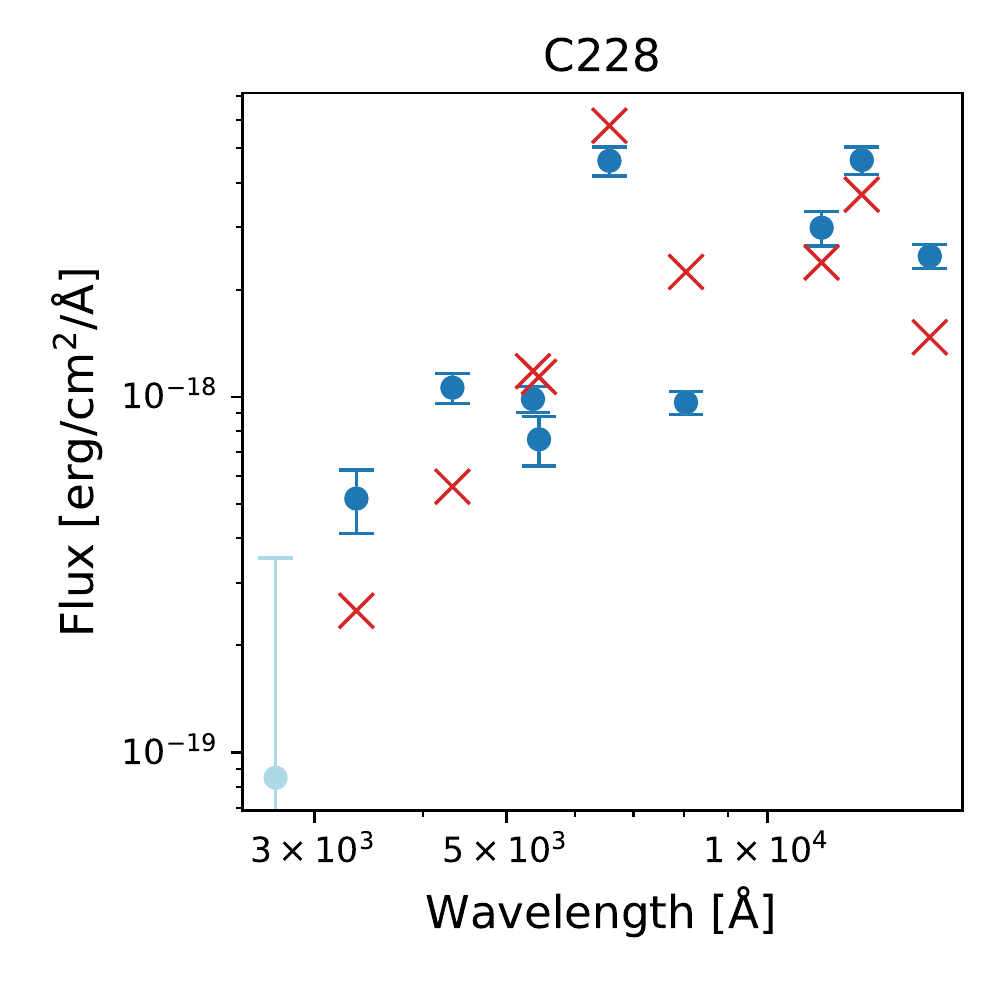}}
\caption{Candidate young and extincted sources from both the \textit{ExtmapCat--final} and \textit{PBcompactCat--final} samples. For each source we show the distributions of \chisq\ values on the age-extinction grid used in the fit process (cut at ages $\sim1$ Gyr) and the fluxes from the best-fit (red crosses) plotted over the photometric data (blue circles). The data not used in the fit because of either large uncertainty or contamination are plotted with shaded blue color. }
\label{fig:individual_ext1}
\end{figure*}

\begin{figure*}
\centering
\subfigure{\includegraphics[width=0.26\textwidth]{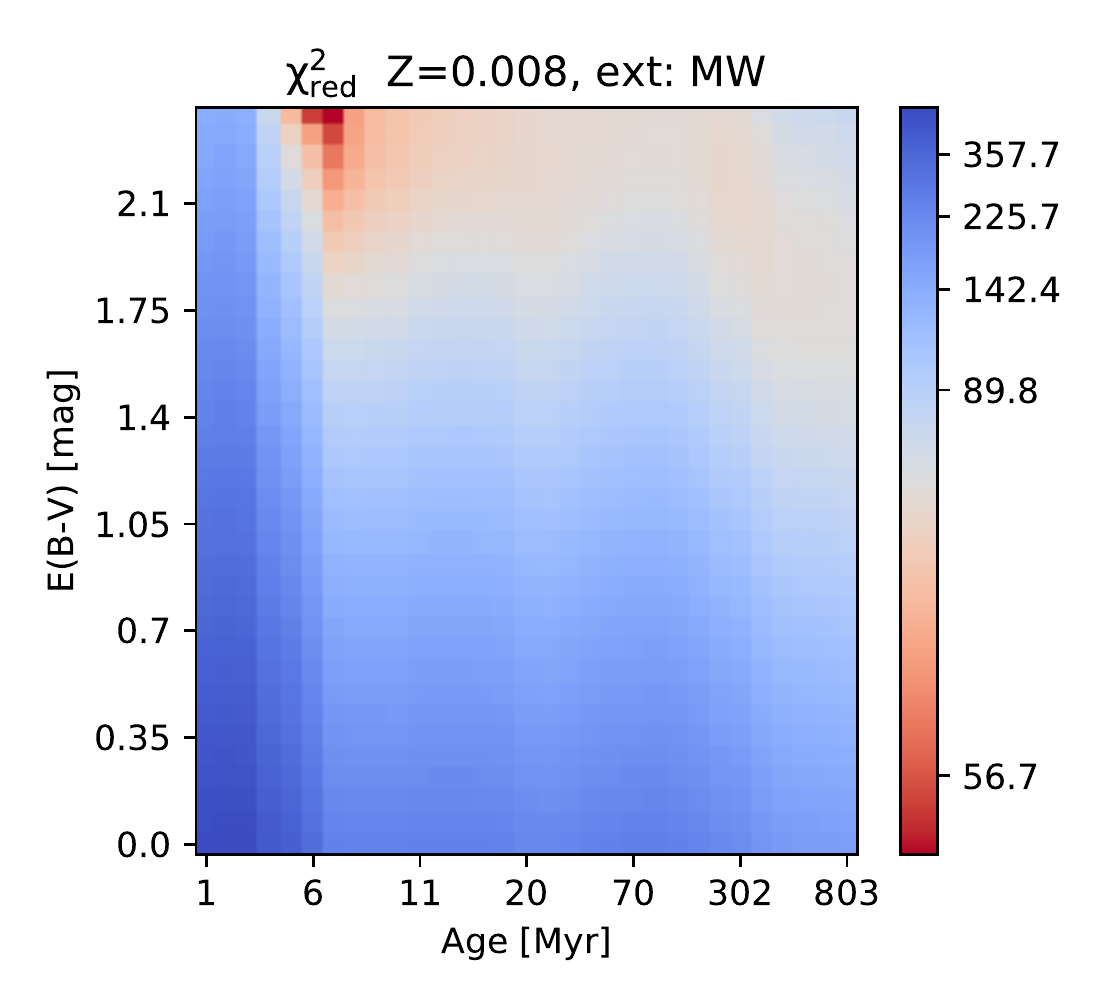}}
\subfigure{\includegraphics[width=0.23\textwidth]{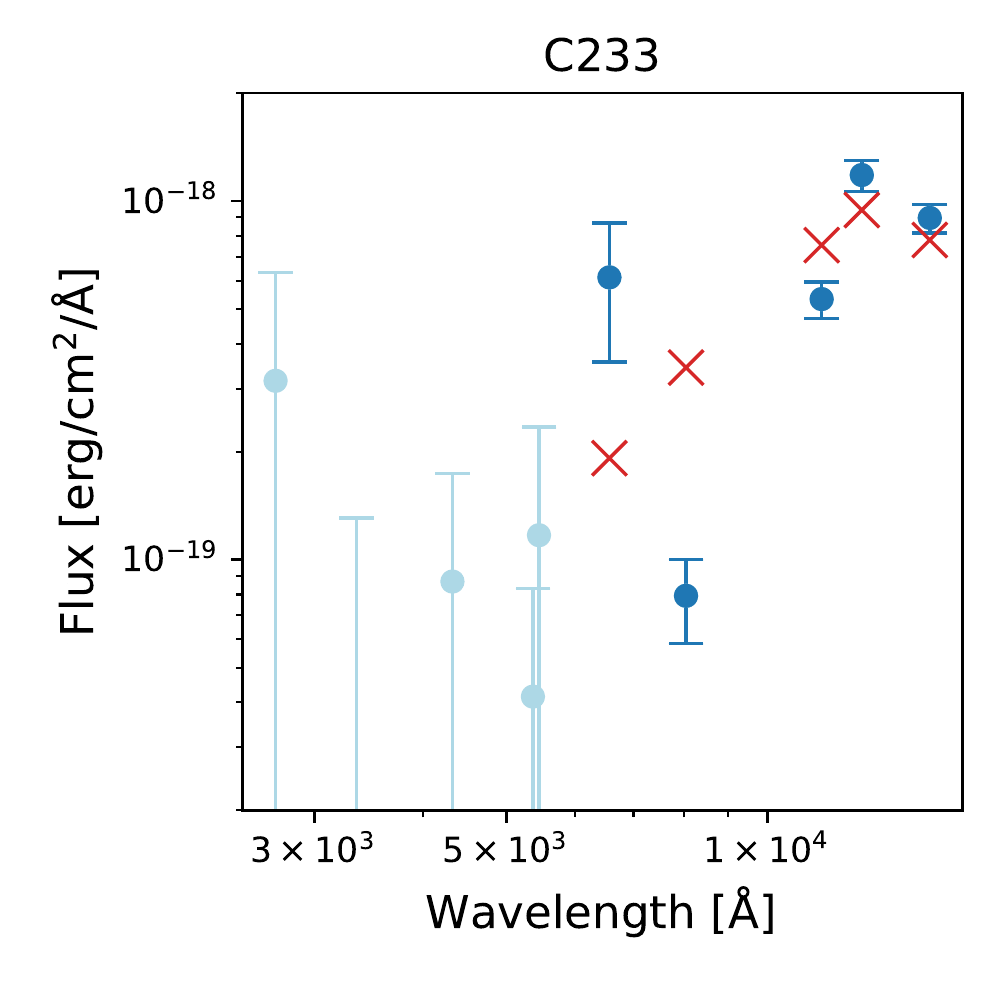}}
\subfigure{\includegraphics[width=0.26\textwidth]{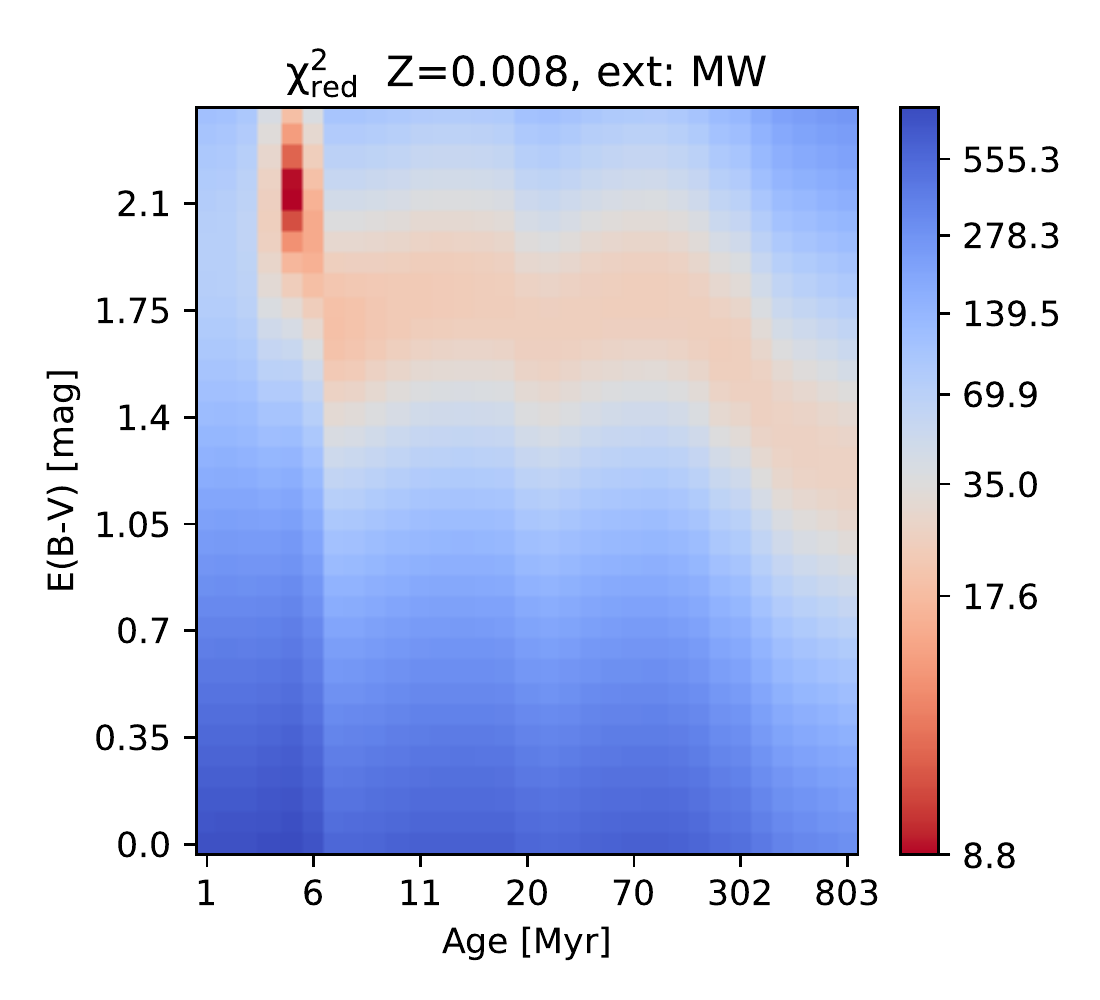}}
\subfigure{\includegraphics[width=0.23\textwidth]{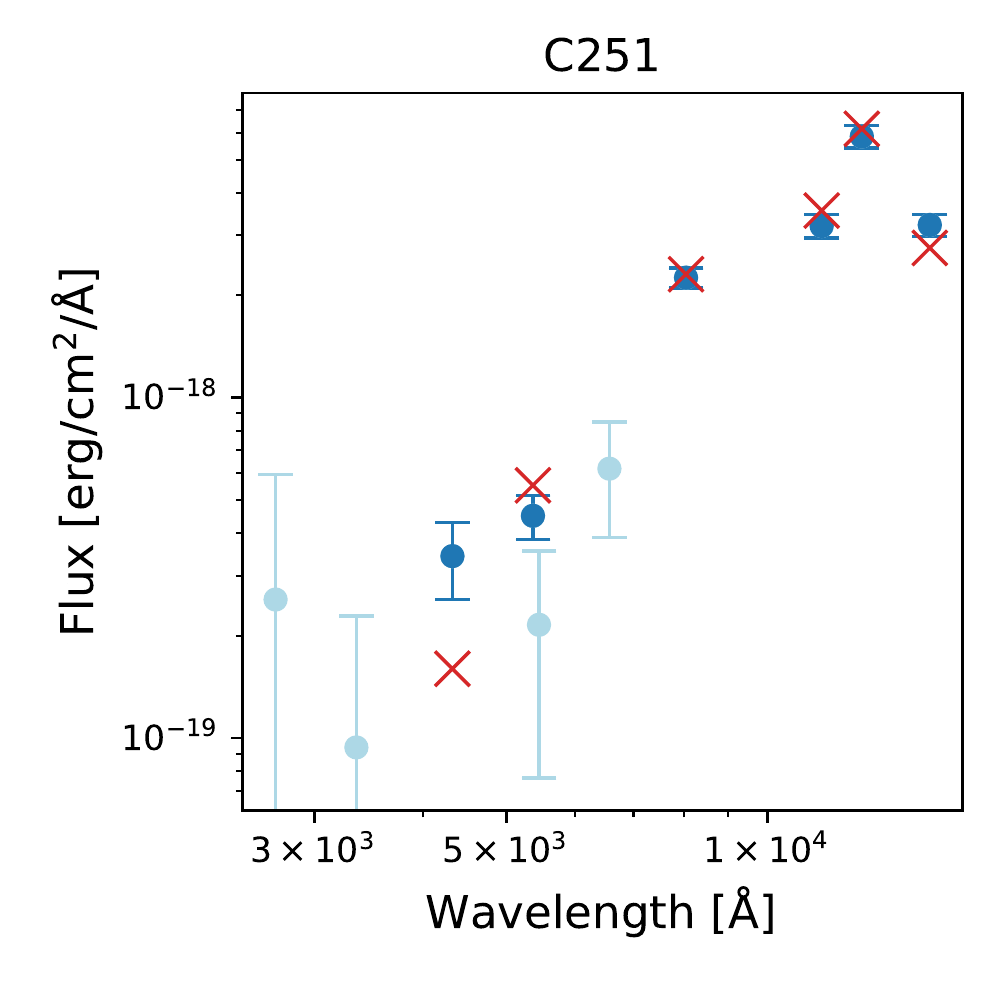}}
\subfigure{\includegraphics[width=0.26\textwidth]{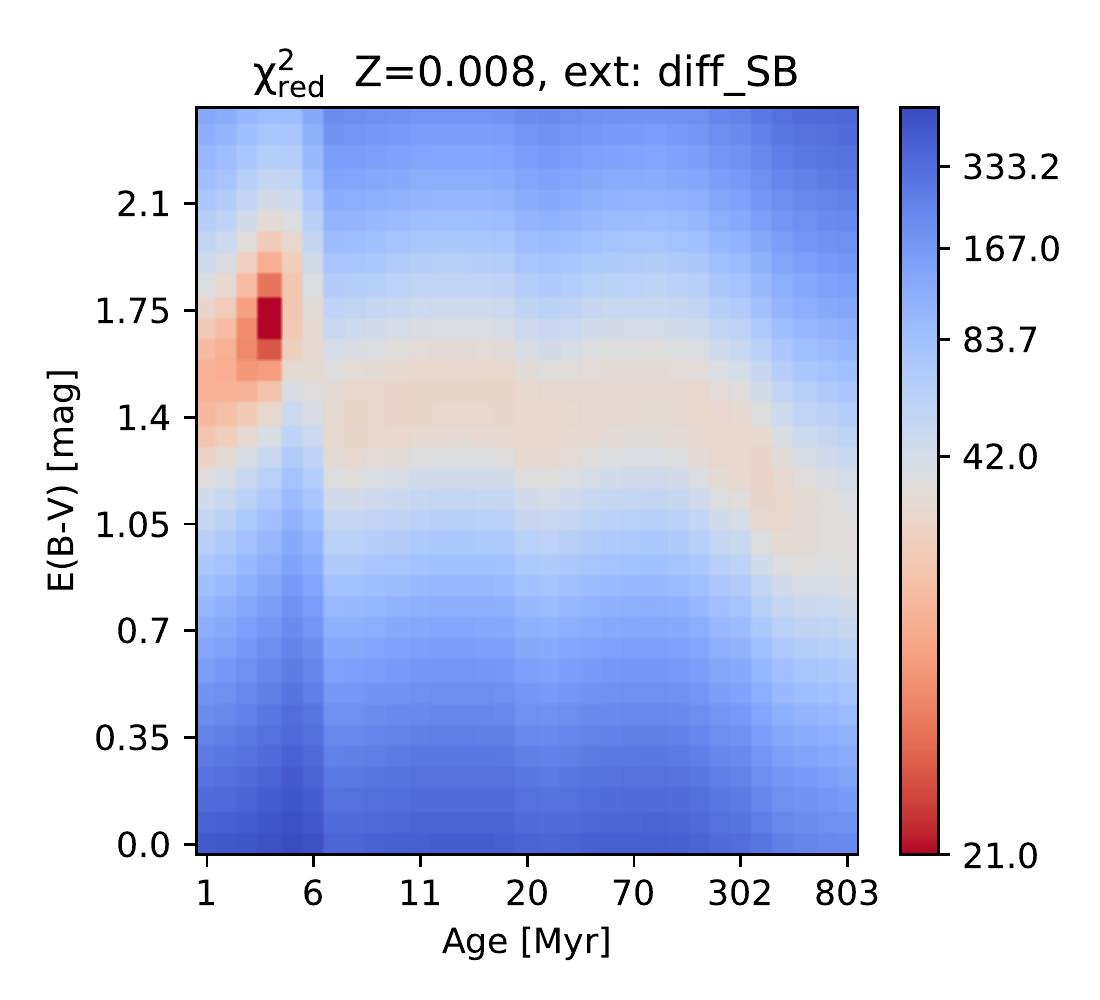}}
\subfigure{\includegraphics[width=0.23\textwidth]{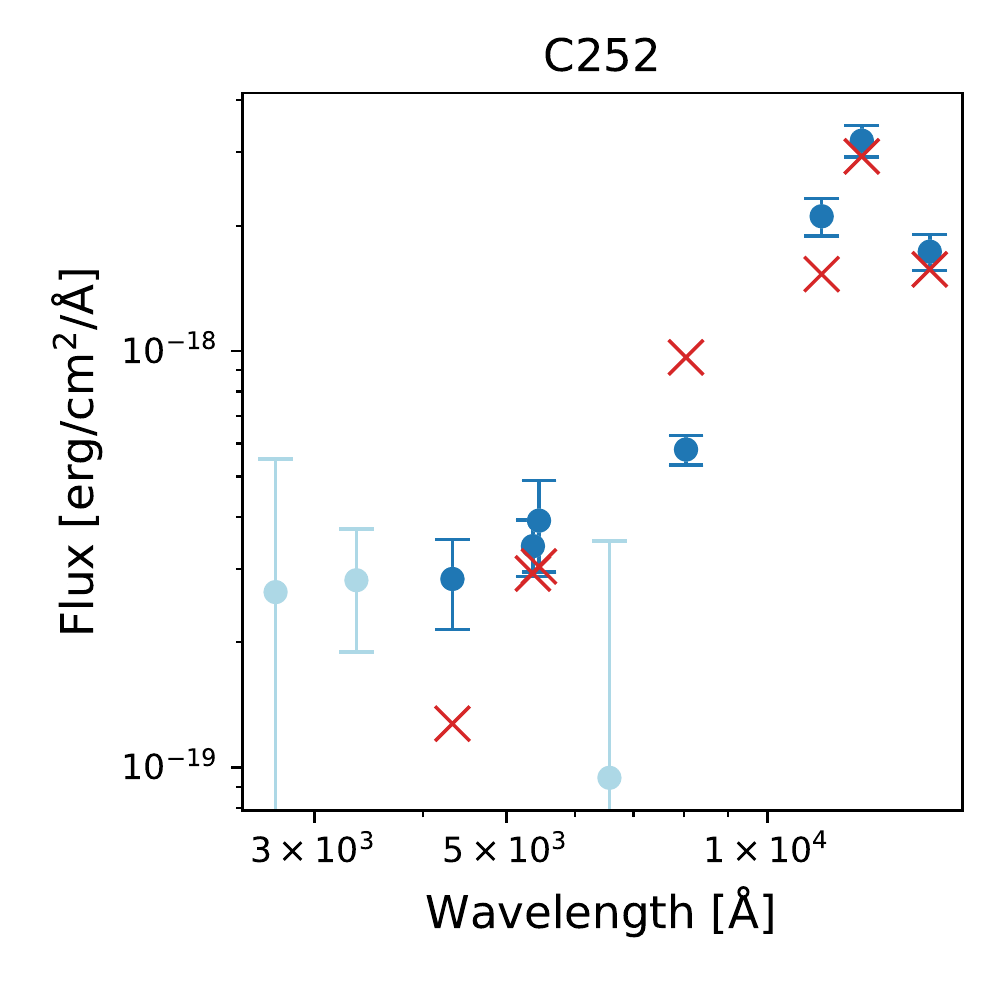}}
\subfigure{\includegraphics[width=0.26\textwidth]{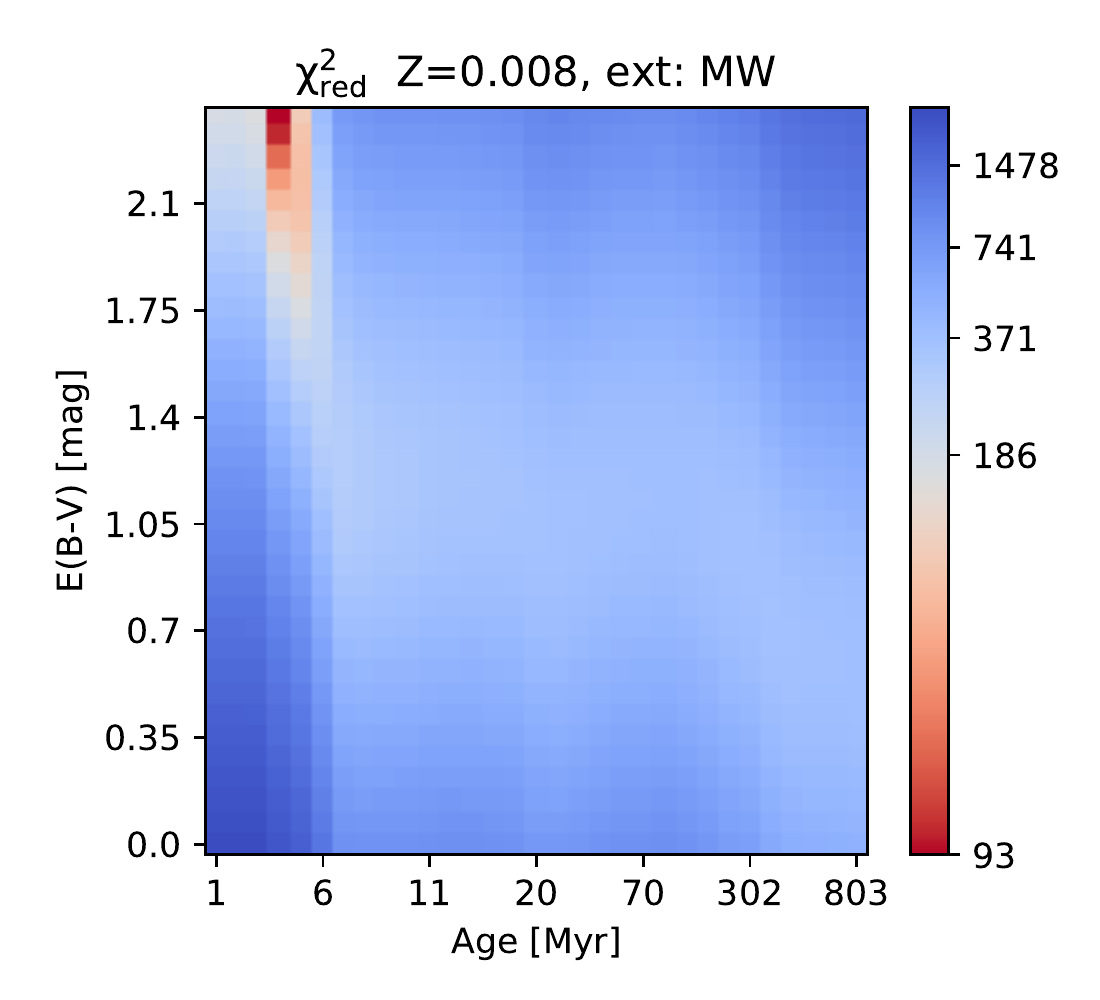}}
\subfigure{\includegraphics[width=0.23\textwidth]{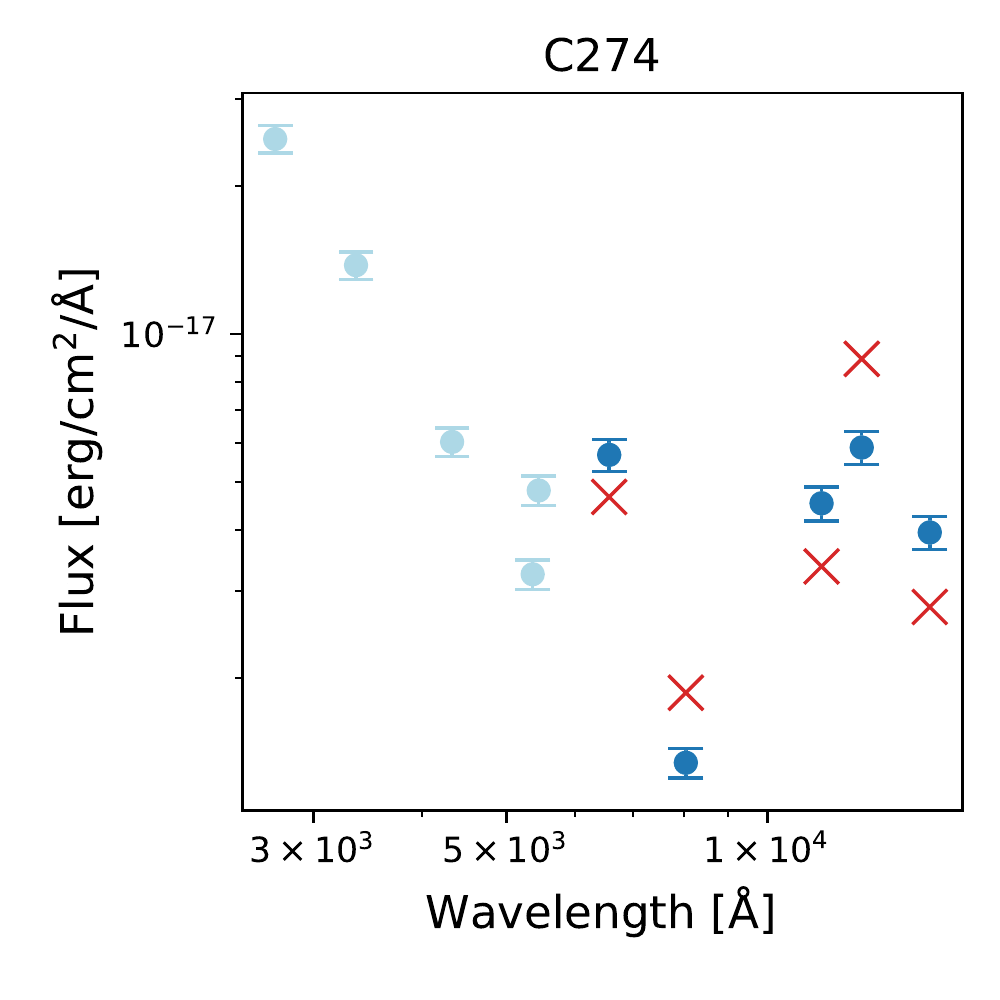}}
\flushleft
\subfigure{\includegraphics[width=0.26\textwidth]{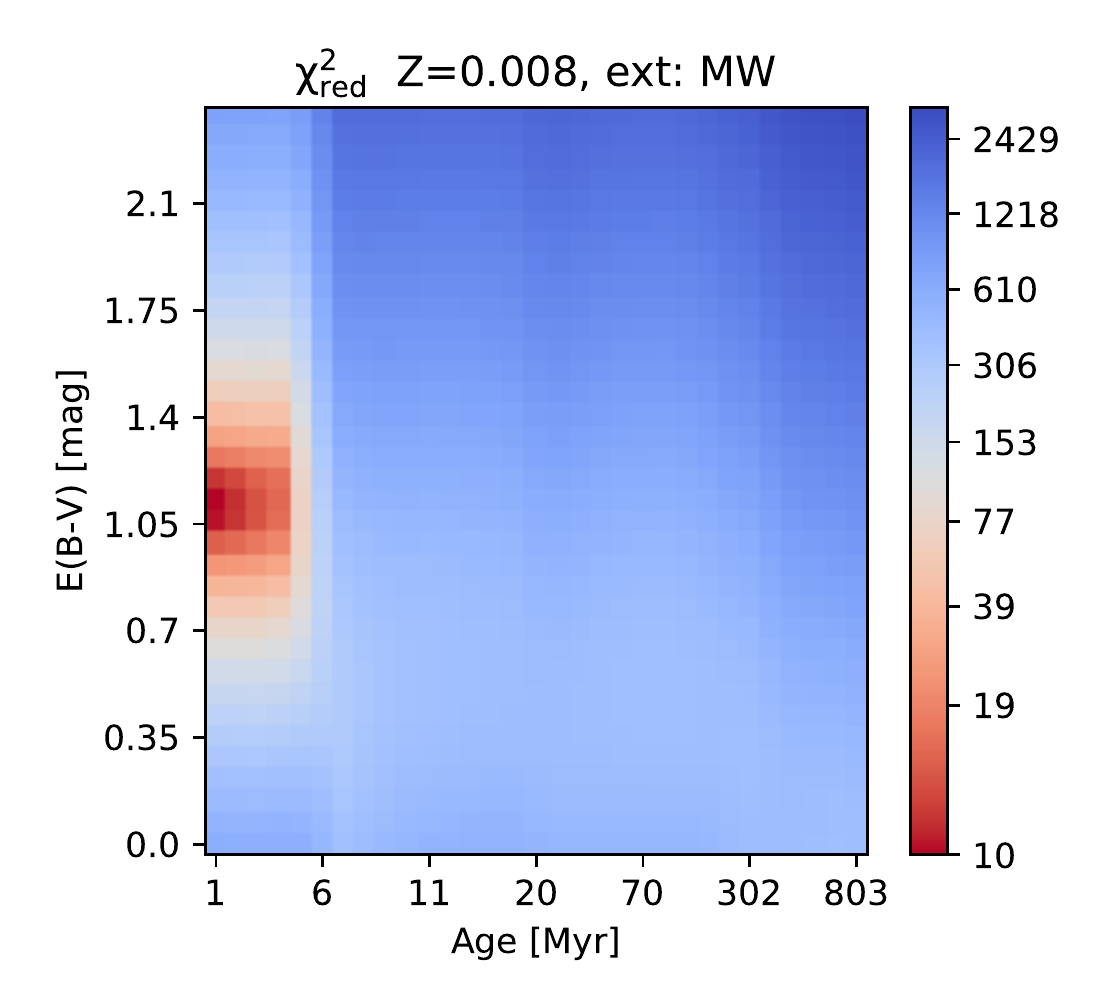}}
\subfigure{\includegraphics[width=0.23\textwidth]{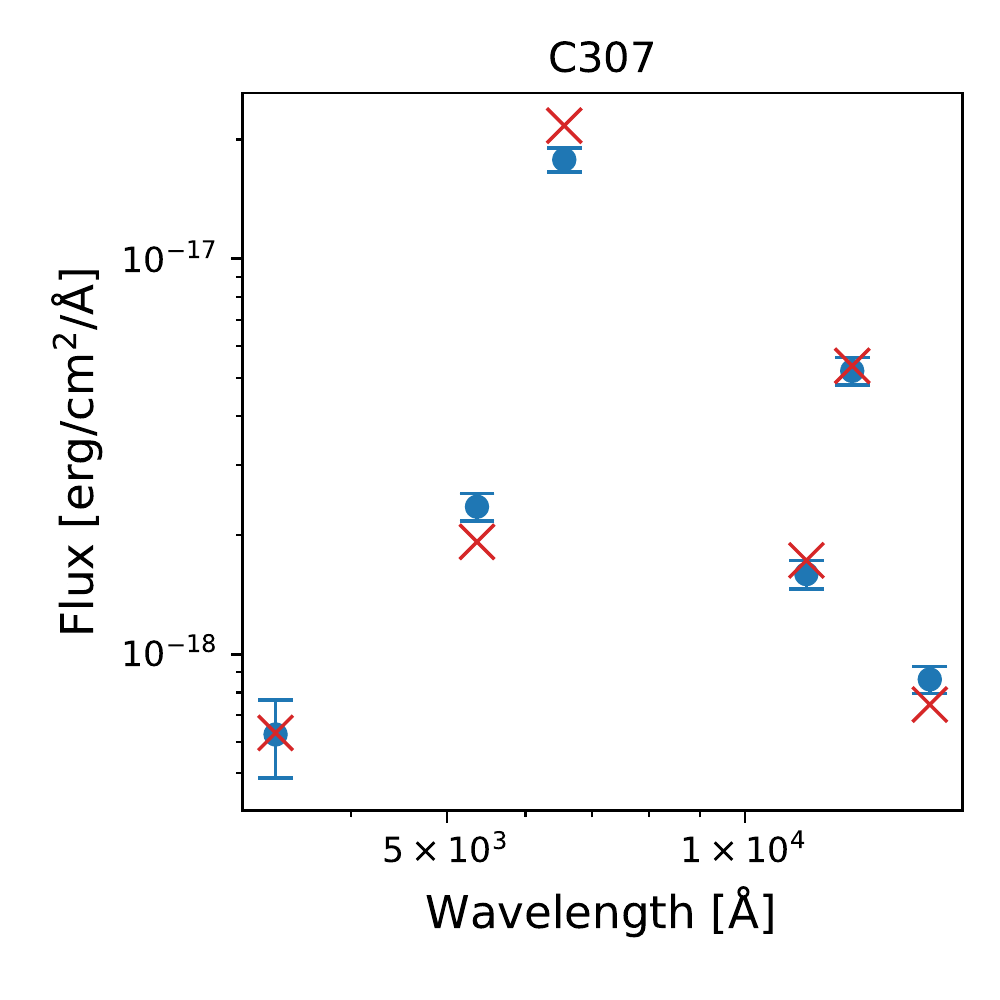}}
\caption{Continuation of Fig.~\ref{fig:individual_ext1}.}
\label{fig:individual_ext2}
\end{figure*}

\end{document}